\documentclass[aps,pre,reprint,superscriptaddress,noeprint,longbibliography,floatfix]{revtex4-2}

\usepackage{placeins}
\usepackage{bm}
\usepackage[normalem]{ulem}
\usepackage{amsmath}
\usepackage{amsbsy}
\usepackage{bm}
\usepackage{hhline}
\usepackage{multirow}
\usepackage{amssymb} 
\usepackage{graphicx}
\usepackage{float}
\usepackage{booktabs}
\usepackage{longtable}
\usepackage{upgreek}
\usepackage[caption=false]{subfig}
\usepackage{hyperref}
\hypersetup{
	colorlinks=true,       
	linkcolor=blue,          
	citecolor=green,       
	filecolor=magenta,      
	urlcolor=blue          
}
\usepackage[capitalise]{cleveref}

\usepackage[dvipsnames]{xcolor}
\usepackage{tikz}

\newcommand{\bx}{{\mathbf x}}
\newcommand{\bq}{{\mathbf q}}
\newcommand{\bphi}{\boldsymbol{\Phi}}
\newcommand{\sgn}{{\rm{sgn}}}

\begin{document}
	
	\title{Nonequilibrium universality of the nonreciprocally coupled \boldmath $O(n_1) \times O(n_2)$ \unboldmath model}
	\author{Jeremy T. Young}
	\email[Corresponding author: ]{j.t.young@uva.nl}
    \affiliation{Institute of Physics, University of Amsterdam, 1098 XH Amsterdam, the Netherlands}
	\affiliation{JILA, University of Colorado and National Institute of Standards and Technology, Boulder, Colorado 80309, USA}
    \affiliation{Center for Theory of Quantum Matter, University of Colorado, Boulder, Colorado 80309, USA}
	
	\author{Alexey V. Gorshkov}
	\affiliation{Joint Quantum Institute and Joint Center for Quantum Information and Computer Science, NIST/University of Maryland, College Park, Maryland 20742 USA}
	
	\author{Mohammad Maghrebi}
	\affiliation{Department of Physics and Astronomy, Michigan State University, East Lansing, Michigan 48824 USA}
	
	\date{\today}
	
	\begin{abstract}
        Nonequilibrium dynamics play an important role in all contexts of physics, both classical and quantum as well as living and non-living, so it is crucial to develop a foundational understanding of nonequilibrium phase transitions. In this work, we investigate an important class of nonequilibrium dynamics in the form of nonreciprocal interactions. In particular, we study how nonreciprocal coupling between two $O(n_i)$  order parameters (with $i = 1, 2$) affects  the universality at a multicritical point, extending the analysis of [J.~T.~Young \emph{et al.}, Phys.~Rev.~X \textbf{10}, 011039 (2020)], which considered the case $n_1 = n_2 = 1$, i.e., a $\mathbb{Z}_2 \times \mathbb{Z}_2$ model. We show that nonequilibrium fixed points (NEFPs) emerge for a broad range of $n_1,n_2$ and exhibit intrinsically nonequilibrium critical phenomena, namely a violation of fluctuation-dissipation relations at \textit{all} scales and underdamped oscillations near criticality in contrast to the  overdamped relaxational dynamics of the corresponding equilibrium models. Furthermore, the NEFPs exhibit an emergent discrete scale invariance in certain physically-relevant regimes of $n_1,n_2$, but not others, depending on whether the critical exponent $\nu$ is real or complex. The boundary between these two regions is described by an exceptional point in the renormalization group (RG) flow, leading to distinctive features in correlation functions and the phase diagram. Another contrast with the previous work is the number and stability of the NEFPs as well as the underlying topology of the RG flow.
        Finally, we investigate an extreme form of nonreciprocity where one order parameter is independent of the other order parameter but not vice versa. Unlike the $\mathbb{Z}_2 \times \mathbb{Z}_2$ model, which becomes non-perturbative in this case, we identify a distinct nonequilibrium universality class whose dependent field similarly violates  fluctuation-dissipation relations but does not exhibit discrete scale invariance or underdamped oscillations near criticality.
	\end{abstract}
	
	
	\maketitle

While the study of phase transitions initially focused on equilibrium or near-equilibrium systems \cite{Wilson1972,Wilson1972a,Hohenberg1977,Chaikin1995, Zinn-Justin2002, SachdevBook}, extensive experimental, theoretical, and numerical efforts have been applied to understanding how the nature of phase transitions are modified when the assumption of equilibrium is broken in a nonequilibrium system \cite{Odor2004,Henkel2008,Zia2010,Henkel2010a,Kamenev2023,Tauber2014,Sieberer2024}.
These efforts encompass both classical and quantum physics as well as living and non-living matter.
In light of this, it is crucial to theoretically identify and classify the manner in which nonequilibrium universality can emerge.

An important class of nonequilibrium dynamics is defined by nonreciprocal interactions, which is also the focus of this work. These are interactions in which the response of one part of the system to the other is not constrained by the reverse process, a feature impossible in equilibrium systems. Nonreciprocal interactions are ubiquitous in many systems, including open or non-Hermitian quantum systems \cite{Pichler2015,Metelmann2015,Lodahl2017,Gong2018,Dykman2018,Ashida2020,Hanai2020,Bergholtz2021, Zhang2022c, Wang2023,Daviet2023,Kawabata2023,Orr2023,Chiacchio2023,Zelle2023,Begg2024,Zhu2024,Lee2024,Fang2017,Dogra2019,Wanjura2023, Liang2022,Han2024,Miri2019}, active matter \cite{Saha2020,You2020,Suchanek2023,Shankar2022,Bowick2022,Osat2023}, flocking models \cite{Dadhichi2020,Loos2023}, and a variety of other contexts \cite{Scheibner2020,Fruchart2023,Hanai2024,Brauns2024,Nassar2020,Meredith2020, Sohn2021,DelPino2022,Orsel2023,Orsel2024,Veenstra2024,Kar2024,Zhou2024,Avni2023,Xue2025}. Due to the intrinsic nonequilibrium nature of these interactions, their study provides an excellent pathway towards understanding new forms of nonequilibrium phases and phase transitions \cite{Fruchart2021}.

In a previous work \cite{Young2020}, we showed how the emergence of nonreciprocal interactions at a multicritical point of two $\mathbb{Z}_2$ Ising order parameters in a driven-dissipative quantum system can give rise to a new form of intrinsically nonequilibrium universality described by a nonequilibrium fixed point (NEFP). These NEFPs exhibit a variety of exotic phenomena such as discrete scale invariance, genuinely  nonequilibrium features of the critical scaling and exponents, the violation of the fluctuation-dissipation theorem at all scales, and underdamped oscillations at criticality in contrast to the overdamped relaxational dynamics in the corresponding  equilibrium models. Crucially, we showed that these phenomena rely on a sign difference in the nonreciprocal interactions. 

For a weaker form of nonreciprocity, with the same sign  but different strengths, effective equilibrium universality emerges. This latter robustness of equilibrium Ising universality to nonequilibrium modifications falls under a common trend in the context of Monte Carlo models \cite{Garrido1989,Wang1988,Marques1989,Marques1990,Tome1991,DeOliveira1993,Achahbar1996,Godoy2002}, field theoretical models \cite{Bassler1994,Tauber1997,Tauber1998,Tauber2002,Santos2002,Akkineni2004,Risler2004,Risler2005}, and open quantum systems 
\cite{Hoening2014,Marcuzzi2014,Chan2015,Sieberer2016,Maghrebi2016,FossFeig2017,Sieberer2024}.
In light of this, our previous work \cite{Young2020} showcases that with the proper form of nonreciprocity (the sign difference in this case), stable NEFPs emerge.
In contrast to our nonreciprocal approach, other avenues towards nonequilibrium criticality include conservation laws \cite{Grinstein90,Garrido90,Cheng91,Katz1984,Zia2010,Zia2010};  violating detailed balance \cite{Tauber2002}, colored \cite{Torre2010} or non-Markovian \cite{Cheung2017} noise; tuning the dissipation strength to tune an effective temperature to zero \cite{Marino2016,Rota2019}; and complex Gross-Pitaevskii equations \cite{Sieberer2016} which can exhibit critical exceptional points \cite{Hanai2020,Zelle2023}.

Thus the utilization of strongly nonreciprocal interactions provides an excellent avenue for identifying new forms of intrinsically nonequilibrium phase transitions. Although the form of nonreciprocity was an emergent feature due to the competition of coherent and incoherent dynamics in Ref.~\cite{Young2020}, nonreciprocity can be engineered through measurement and feedback processes. While this is straightforward in classical systems, in quantum systems it may also be realized via cascaded quantum systems \cite{Gardiner1993,Carmichael1993,Metelmann2015} or, as has been recently shown, via dissipative gauge symmetries \cite{Wang2023}. Beyond the conceptually straightforward approach of feedback, there have been a variety of experimental investigations of nonreciprocity in recent years, both emergent or engineered, likewise spanning a variety of contexts \cite{Fang2017,Meredith2020, Nassar2020,Liang2022,Wanjura2023,Miri2019,Sohn2021,DelPino2022,Orsel2023,Orsel2024,Veenstra2024,Kar2024,Zhou2024,Dogra2019,Han2024}.

The goal of the present work is to understand how the nonequilibrium universality investigated in Ref.~\cite{Young2020} is modified when the underlying symmetries of the system are changed and elucidate the effects that nonreciprocal interactions can have on universality in a generalized context. In particular, we investigate what happens when two $n_i$-component real order parameters, $\bphi_1, \bphi_2$, possess $O(n_1)$ and $O(n_2)$ symmetries, respectively, giving rise to an overall $O(n_1) \times O(n_2)$ symmetry. The corresponding equilibrium model has been studied extensively, giving rise to both bicritical and tetracritical points and exhibiting a rich variety of phenomena \cite{Fisher1965,Bruce1975,Kosterlitz1976,Folk2008a,Folk2008b,Eichhorn2013}. Thus in addition to comparing the universality we identify to the original  nonreciprocal $\mathbb{Z}_2 \times \mathbb{Z}_2$ model (note $O(1)$ is equivalent to $\mathbb{Z}_2$), we may also compare the resulting behaviors to the corresponding equilibrium model. 

We find that this new form of nonreciprocal universality generalizes to more complicated order parameters, and all of the qualitative features observed in the $\mathbb{Z}_2 \times \mathbb{Z}_2$ model can persist in the $O(n_1) \times O(n_2)$ $\phi^4$ models, with some important differences. Although for a certain range of $n_1, n_2$, the corresponding NEFPs exhibit discrete scale invariance, others do not. 
Yet, we show that for any $n_1, n_2$, the effective temperature at the NEFPs only becomes ``hotter'', never ``colder'', at longer wavelengths. Furthermore, we show that, for all the NEFPs, there is a region in the doubly-ordered phase near the multicritical point where the dynamics exhibits underdamped oscillations towards the steady state regardless of the discrete scale invariance.

Furthermore, we investigate the critical phenomena that emerge when one order parameter (dependent field) is affected by the other (independent field) but not the reverse. In Ref.~\cite{Young2020}, this scenario was not considered due to the field theory becoming non-perturbative. For the more general symmetries considered here, we find that the system remains perturbative and identify another new class of fixed points. While one of the fields exhibits typical equilibrium universality, the other exhibits several of the exotic critical behaviors that are present in the fully-coupled models. Moreover, the two effective ``temperatures'' are now no longer the same, with the independent field having a scale-independent temperature and the dependent field having a scale-dependent temperature. 
Moreover, we identify a transient criticality which could emerge when these fixed points become non-perturbative, such as for the $\mathbb{Z}_2 \times \mathbb{Z}_2$ model.

The remainder of this paper is organized as follows. In Sec.~\ref{sec:model}, we introduce the important features of the model and discuss the relevant critical phenomena of interest. In Sec.~\ref{sec:rg}, we present the results of our renormalization-group (RG) analysis. First, we introduce the formalism we use and determine the beta functions to lowest non-trivial order, discussing the various features of the beta functions and identifying the corresponding fixed points that emerge. In Sec.~\ref{sec:scale}, we discuss the forms of critical behavior that can emerge for the NEFPs and present the RG flow equations which give rise to this behavior. Finally, we discuss how the choice of $n_1,n_2$ affects the critical behavior of the NEFPs, first focusing on the case of $n_1 = n_2$ before extending to arbitrary $n_1,n_2$. In Sec.~\ref{sec:halfcouple}, we investigate the criticality which can emerge when one order parameter evolves independently while the other is coupled to the first, identifying an additional class of nonequilibrium fixed points that is distinct from the previous NEFPs. In Sec.~\ref{sec:outlook}, we summarize several possible future directions that emerge from the results of the present work. Finally, in the appendix, we present the one- and two-loop calculations used to derive the RG equations, derive an expression for the maximal underdamping angle $\theta^*$, and present numerical values of the various stable fixed points and their corresponding critical exponents.

\section{Model}

\label{sec:model}
In this section, we introduce the model that is the focus of this work. 
We consider a pair of coupled nonequilibrium $O(n_i)$ order parameters $\bphi_i$. These order parameters are vector fields, $\bphi_i\equiv (\phi_{i,1}, \cdots, \phi_{i,n_i})$, with $n_i$ real components. Common examples of such order parameters include classical spins whose interactions are isotropic ($n=3$) or anisotropic due to an easy axis or external field ($n=2$), commonly described via the paradigmatic classical Heisenberg and XY models. Similar rotational symmetries can emerge via the internal degrees of freedom for quantum spins as well, although the underlying symmetries and realization of $O(n)$ symmetry can be more complex. For example, driven-dissipative condensates in exciton-polariton systems are described by a complex $U(1)$ order parameter, but the critical point can be described by an effective \emph{classical} $O(2)$ model \cite{Tauber2014a,Sieberer2013, Sieberer2014, Tauber2014a}. Furthermore, $\mathbb{Z}_2$ ($n=1$) symmetries are ubiquitous in both classical and quantum contexts, including both intrinsic (e.g., anti-ferromagnetism) and emergent (e.g., liquid-gas critical points) symmetries, while open quantum systems often display effectively classical criticality \cite{Sieberer2016,Maghrebi2016,FossFeig2017,Sieberer2024}. Extending our analysis from Ref.~\cite{Young2020}, we aim to identify the effects that nonreciprocal couplings between a pair of these generalized order parameters exhibit at a multicritical point.

We consider the most representative form of the dynamics for a pair of classical real $O(n_i)$ order parameters in the absence of fine-tuning or conserved quantities. The resulting dynamics are described by the Langevin equations
\begin{subequations}\label{eq:1}
\begin{equation}
    \zeta_1 \partial_t \bphi_1 = - \frac{\delta \mathcal{F}_1}{\delta \bphi_1} + g_{12} \bphi_1 |\bphi_2|^2 + \bm{\xi}_1,
\end{equation}
\begin{equation}
    \zeta_2 \partial_t \bphi_2 = - \frac{\delta \mathcal{F}_2}{\delta \bphi_2} + g_{21} |\bphi_1|^2 \bphi_2 + \bm{\xi}_2,
\end{equation}
where
\begin{equation} 
    \mathcal{F}_i [\bphi_i]\equiv \int_\mathbf{x} \frac{D_i}{2} |\nabla \bphi_i|^2 +  \frac{r_i}{2} |\bphi_i|^2 + \frac{g_i}{4} |\bphi_i|^4,
\end{equation}
\begin{equation}
    \langle \xi_{i,\alpha}(t,\mathbf{x}) \xi_{j,\beta}(t',\mathbf{x}') \rangle = 2 \zeta_i T_i \delta_{ij} \delta_{\alpha \beta} \delta(t-t') \delta(\mathbf{x} - \mathbf{x}'),
\end{equation}
\begin{equation}
    |\bphi_i|^2 \equiv \bphi_i \cdot \bphi_i = \sum_{\alpha} \phi_{i,\alpha}^2.
\end{equation}
\end{subequations}
In these equations, $\zeta_i$ denotes a ``friction'' coefficient, $D_i$ the stiffness, $T_i$ an effective ``temperature'' characterizing the noise level of the Gaussian white noise $\bm{\xi}_i$ whose $n_i$ components $\alpha$ experience otherwise independent noise, $r_i$ the distance from the critical point (which will shift when fluctuations are taken into account), and $g$ the coupling terms. While thermal fluctuations in classical systems are responsible for the Gaussian white noise, in the context of open quantum systems, this noise emerges from the combination of drive and dissipation even at zero temperature \cite{Sieberer2016,Maghrebi2016,FossFeig2017,Sieberer2024}, for example due to spontaneous emission, generically rendering quantum fluctuations irrelevant in the sense of RG. Furthermore, the identify form of the noise correlations is enforced by the $O(n_i)$ symmetry.

In the absence of coupling between the two order parameters, $g_{12}=g_{21}=0$, the dynamics is described by the model A dynamics \cite{Hohenberg1977} for the $O(n_1)$ as well as the $O(n_2)$ models which are nevertheless decoupled.
When $g_{12} = g_{21}$, the model is closely related to the equilibrium  $O(n_1) \times O(n_2)$ model, with the free energy $\mathcal{F} = \mathcal{F}_1 + \mathcal{F}_2 + ({g_{12}}/{2}) \int_\bx |\bphi_1|^2 |\bphi_2|^2$. Note, however, that this free energy does not necessarily describe an equilibrium system unless the temperatures are equal ($T_1 = T_2$), where the fluctuation-dissipation theorem is restored. More generally, we are interested in $g_{12}\ne g_{21}$ and/or $T_1 \neq T_2$ in which case the dynamics is generically nonreciprocal, that is, the coupled dynamics between the two fields $\bphi_1$ and $\bphi_2$ do not originate from a term in the free energy with a single temperature $T$. As we later show, the nonequilibrium dynamics of the nonreciprocally coupled order parameters are crucial for the emergence of exotic, fundamentally nonequilibrium critical dynamics. 

\subsection{Alternative representation}

It may appear that there are two variables that control the nonequilibrium nature of the dynamics: the ratio $g_{12}/g_{21}$ setting the degree of nonreciprocity in the dynamics, and $T_1/T_2$ characterizing the mismatch of the two temperatures. However, owing to a scaling freedom, one can exploit a redundancy to fix one of these variables and focus on the other one. To this end, consider scaling one of the fields, say $\bphi_1$, as $\bphi_1 \to \bphi_1 / c$. The dynamics remains invariant if we simultaneously make the changes 
\begin{equation}
    T_1 \to T_1 / c^2, \quad g_1 \to g_1 c^2, \quad g_{21}\to g_{21} c^2,
\end{equation}
while all the other variables including $g_2, g_{12}, T_2$ and $\zeta_1,\zeta_2$ are unchanged. Using this scaling freedom, we can bring the interaction strengths to a form where $g_{21} = \sigma g_{12}$ with $\sigma=0, \pm 1$.
The factor $\sigma$ cannot be removed since  $g_{21}$ can only be rescaled by a positive number $c^2$; additionally, we have included $\sigma=0$, which represents $g_{21} = 0$. 
Upon this transformation, the temperature ratio $T_1/T_2$ remains free and is determined by the microscopic dynamics. 
A complementary  representation is to fix the ratio $T_1/T_2 =1$ while allowing a general ratio $g_{21}/g_{12}$.
In this work, we find both representations useful for different purposes. To avoid any confusion, we reserve the interactions strengths $g_a$ (where $a = 1,2,12,21$) for a representation where the temperature ratio is fixed to unity, $T_1/T_2=1$, or when $\sigma = 0$. 

The role of $\sigma$ here reveals an important key distinction between our work here and prior works which have investigated the role of multiple temperatures on criticality \cite{Tauber1997,Tauber1998,Tauber2002,Santos2002,Akkineni2004}. In contrast to these previous works, our model does not require conserved quantities or spatially anistropic noise to ensure that equilibrium is not restored at criticality. Instead, $\sigma$ will provide the sole mechanism for preventing the restoration of effective equilibrium behavior at criticality. Here, we note that although $\sigma = 0$ corresponds to an effective temperature ratio $T_1/T_2 \to 0$, which has been shown can lead to new universality in other two-temperature models \cite{Tauber1997,Tauber1998,Tauber2002,Santos2002,Akkineni2004}, $\sigma = -1$ has no counterpart in these prior studies.

We find the representation where the ratio $T_1/T_2$ is free while $|g_{21}/g_{12}|$ is fixed to be the most convenient when $\sigma \neq 0$, so we define a new set of interaction strengths $u_a$ via 
\begin{equation}
    g_1 \equiv u_1/c^2, \quad g_2 \equiv u_2, \quad g_{12}\equiv u_{12}, \quad g_{21} \equiv \sigma u_{12}/c^2,
\end{equation}
where $c=\sqrt{|g_{12}/g_{21}|}$ with $\sigma=\sgn(g_{21}/g_{12})$ assuming $g_{21}\ne 0$.
The dynamics is governed by \cref{eq:1} with the substitution $g_i\to u_i$ and $g_{12} \to u_{12}$, and $g_{21}\to \sigma u_{12}$.
In the $u$ representation, the temperature ratio $T_1/T_2$ can take any value (set by the microscopic dynamics). In an abuse of notation, we shall use the same symbols for temperatures in the two cases, while the distinction will be clear from the context.
Whenever $g_{21} = 0$, we simply set $\sigma = 0$. Since $c$ is introduced to ensure the inter-coupling interaction strengths are equal in magnitude, it is no longer needed in this case.
We now briefly discuss the dynamics in the $u$ representation. 

\textbf{$\sigma=1$.---}In this case, we can write the dynamics in terms of a single free energy $\mathcal{F} = \mathcal{F}_1 + \mathcal{F}_2 + ({u_{12}}/{2}) \int_\bx |\bphi_1|^2 |\bphi_2|^2$, where ${\cal F}_i$ are given by \cref{eq:1} upon the scaling transformation and replacing $g_i \to u_i$ for $i=1,2$. Note however that the two temperatures are generically different, $T_1/T_2\ne 1$, which is then inherently nonequilibrium.  Interestingly, in spite of the nonequilibrium dynamics, we find that the critical behavior in this case is always governed by equilibrium fixed points where $T_1/T_2 \to 1$ under the RG flow to long scales, thus representing another example of the robustness of equilibrium to nonequilibrium perturbations
\cite{Garrido1989,Wang1988,Marques1989,Marques1990,Tome1991,DeOliveira1993,Achahbar1996,Godoy2002,Bassler1994,Tauber1997,Risler2004,Risler2005,Tauber2002,Hoening2014,Marcuzzi2014,Chan2015,Sieberer2016,Maghrebi2016,FossFeig2017,Sieberer2024}. We discuss this behavior in detail in Sec.~\ref{subsec:RGEq}.

\textbf{$\sigma=-1$.---}In this case, the dynamics cannot be described by a free energy, even if the temperatures are identical. 
In fact, nonreciprocity takes an extreme form here where the effect of $\bphi_1$ on $\bphi_2$ is not only different from that of $\bphi_2$ on $\bphi_1$ but it even takes the opposite sign relative to the expected behavior in equilibrium. We show that genuinely nonequilibrium fixed points can emerge in this case in Sec.~\ref{sec:fixedpoints} and discuss the corresponding universal features in Sec.~\ref{sec:scale}.

\textbf{$\sigma=0$.---}%
This case corresponds to a \textit{one-way} coupling where the dynamics of $\bphi_1$ is coupled to $\bphi_2$ but not \textit{vice versa}, so we refer to $\bphi_1$ as the dependent field and $\bphi_2$ as the independent field because the latter evolves independently of the former. 
In terms of the $g_{12}, g_{21}$ parameters, this regime corresponds to the $g_{12} g_{21} = 0$ subspace; here, we have assumed that $g_{12}\ne 0$ without loss of generality. 
The one-way dynamics in this sector cannot be derived from a free energy and gives rise to genuinely nonequilibrium behavior, which we discuss in detail in Sec.~\ref{sec:halfcouple}. Furthermore, 
this sector puts an important constraint on the nonequilibrium critical behavior in the $\sigma=-1$ sector. To see this, we first note that the RG flow is closed in the subspace $g_{12}g_{21}=0$
since one field is decoupled from the other at all scales. 
This implies that the long-distance behavior in the $\sigma = - 1$ sector cannot flow to the $\sigma=1$ sector under RG, hence the dynamics in the former sector must be genuinely nonequilibrium provided the fields remain coupled under RG \cite{Young2020}.

\subsection{Summary of Previous and New Results}
Before proceeding to the RG analysis of the above model, we summarize several of the key features of the nonequilibrium critical phenomena that emerge at the new nonreciprocal fixed points. We  discuss three types of critical exponents: $\nu, z, \gamma_T$. The first, $\nu$, describes the divergence of the correlation length $\xi$ in $r$ as the multicritical point is approached and also plays a role in the structure of the phase diagram. The second, $z$, is the dynamical critical exponent and relates the scaling of the correlation length to the correlation time $\tau$ according to $\tau \propto \xi^z$.
The third exponent, $\gamma_T$, describes the scaling violation of the fluctuation-dissipation theorem (FDT). In a thermal equilibrium setting, the FDT relates the correlation functions $C_i$ and the response functions $\chi_i$ via 
\begin{equation}
    C_i(\mathbf{q},\omega) = \frac{2 T}{\omega}\text{Im} \hspace{.07cm} \chi_i(\mathbf{q},\omega),
\end{equation}
where $\mathbf{q}$ is the momentum, $\omega$ is the frequency, and $T$ is the temperature. At the nonequilibrium fixed points, this relation no longer holds, and the scaling of the correlation and response functions is no longer connected. We quantify this mismatch relative to the FDT via an effective \textit{scale-dependent} temperature, which we quantify via $\gamma_T$ and formally describes the mismatch between the anomalous dimensions of the correlation and response functions compared to equilibrium.
Finally, we investigate a universal feature of the relaxation to the steady-state. Due to nonreciprocity, some regions of the phase diagram can exhibit underdamped oscillations arbitrarily close to the critical point. The underdamped oscillations are described by a complex frequency $\omega_r + i \kappa_r$.
We capture this behavior via a universal constant $\theta^*$, which describes the maximal angle (as a function of the position in the phase diagram)  $\theta \equiv \arctan \frac{\omega_r}{\kappa_r}$ possible near the critical point.
Since this relies on the nonreciprocal interactions, this is not possible for equilibrium universality in our model, which effectively exhibits overdamped dynamics with no oscillations.

Before summarizing the new results of this paper, we discuss the critical phenomena identified in Ref.~\cite{Young2020}, which considered the case of $n_1 = n_2 = 1$ and provides a basis for understanding the more general case. There, we identified a pair of stable NEFPs which emerge for $\sigma = -1$, one for $u_{12} > 0$ and one for $u_{12} < 0$, which we expect to describe criticality in the top left and bottom right quadrants in $g_{12}$-$g_{21}$ space, respectively. In contrast to typical equilibrium systems, $\nu$ takes on a complex value $\nu^{-1} = \nu'^{-1} + i \nu''^{-1}$. In this case, the real part $\nu'$ continues to describe the behavior of the correlation length. However, the imaginary part $\nu''$ heralds the emergence of discrete scale invariance, where scale invariance is preserved only under a preferred scaling factor (in momentum) $b_* \equiv e^{2 \pi \nu''}$. This discrete scale invariance imprints itself both on the form of the correlation and response functions as well as on the  structure of the phase diagrams, which exhibits a discrete scale invariance in the form of spiraling phase boundaries. Furthermore, we showed $\gamma_T < 0$, corresponding to the complete violation of the fluctuation-dissipation theorem with an effective temperature becoming ``hotter'' at larger length scales. Finally, we showed that $\theta^* = \pi/3$, heralding underdamped dynamics in the relaxation of the order parameters to the steady state, which occur within the doubly-ordered phase ($\langle \bphi_1\rangle \neq 0 \neq \langle \bphi_2\rangle$).

In the generalized model, i.e., the $O(n_1) \times O(n_2)$ model, the NEFPs do not necessarily  appear as a pair. Depending on $n_1, n_2$, there can be no, one, or two NEFPs. Moreover, two NEFPs, one stable and one unstable, can be present in the same quadrant (top left or bottom right) of $g_{12}$-$g_{21}$ space.
Depending on the values of $n_1,n_2$, the exponent $\nu$ can be either complex, as in Ref.~\cite{Young2020}, or purely real. By analytically continuing the model in $n_1,n_2$, an exceptional point occurs in the RG at the boundary of complex and real $\nu$, resulting in logarithmic features in the correlation and response functions as well as the phase diagram.
The NEFPs always violate the FDT with an effective temperature which always appears to get ``hotter'' at longer wavelengths. Similarly, $\theta^*$ is always nonzero (unlike the discrete scale invariance which may or may not emerge).

In the case of one-way coupling where $\bphi_2$ is independent of $\bphi_1$, we identify a single coupled fixed point which is stable when both $n_1 < n_2$ and the decoupled fixed point is unstable.
Naturally, the independent field always exhibits the corresponding equilibrium criticality. In contrast to the NEFPs, $\nu$ is always real and $\theta^* = 0$,
both of which intrinsically rely on $\sigma = -1$. Nevertheless, the dependent field still exhibits a scale-dependent temperature which gets ``hotter'' at large length scales, and the FDT is violated for the dependent field. Finally, we investigate one-way coupling for $n_1 = n_2$, where the fixed points become non-perturbative. This form of coupling was not considered for the $\mathbb{Z}_2 \times \mathbb{Z}_2$ model of Ref.~\cite{Young2020} due to this non-perturbative behavior. We find that 
a transient criticality relevant to physical systems at intermediate scales can potentially emerge before the system can flow to a non-perturbative regime, allowing us to identify potential transient critical phenomena.

\section{Field theory and RG Analysis}
\label{sec:rg}
To investigate the dynamics due to the nonreciprocal coupling, we shall use the response-function formalism. This allows us to investigate the critical phenomena of this model by extending standard techniques of RG analysis to a dynamical setting.

\subsection{Formalism}
We define the nonequilibrium partition function $Z = \int \mathcal{D}[i \tilde{\bphi}_1, \bphi_1; i \tilde{\bphi}_2, \bphi_2] e^{- \mathcal{A}[ \tilde{\bphi}_1, \bphi_1; \tilde{\bphi}_2, \bphi_2]}$, where we have introduced the functional integral measure $\mathcal{D}$ (composed of a product of four measures, one for each field) and the ``action'' $\mathcal{A}$, which involve both the fields $\bphi_{1,2}$ and their corresponding ``response'' fields $\tilde{\bphi}_{1,2}$ \cite{Tauber2014}. Note that the measure integrates the response fields over the imaginary axis.
The statistical weight of $\bphi_{1,2}(t,\mathbf{x})$ can be obtained by integrating out both response fields as $P[\bphi_1,\bphi_2] = \int \mathcal{D}[i \tilde{\bphi}_1]\mathcal{D}[i \tilde{\bphi}_2] e^{-\mathcal{A}[ \tilde{\bphi}_1, \bphi_1; \tilde{\bphi}_2, \bphi_2]}$. While the partition function $Z=1$ by construction, the expectation value of any quantity---the fields themselves or their correlations---can be determined by computing a weighted average in the partition function.
To this end, we shall use the representation in terms of $u_a$ where the ratio of the inter-coupling  coefficients is fixed (up to the constant $\sigma=0,\pm 1$) rather than that of $g_a$ where the temperature ratio is set to 1.  
The technical reason is that, although all the coupling terms $g_a$ are renormalized at one loop, the ratio $g_{12}/g_{21}$ is not renormalized until two loops. Moreover, the effective temperatures $T_i$ are similarly not renormalized until two loops. Thus by considering the renormalization of $u$ at one loop and $T_i$ at two loops, we may determine both the RG fixed points and the critical exponents to lowest non-trivial order. 

\begin{figure}[]
\centering
\subfloat[]
{
\centering
\includegraphics[scale=1]{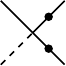}
}
\hspace{.7cm}
\subfloat[]
{
\centering
\includegraphics[scale=1]{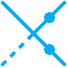}
}
\hspace{.7cm}
\subfloat[]
{
\centering
\includegraphics[scale=1]{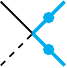}
}
\hspace{.7cm}
\subfloat[]
{
\centering
\includegraphics[scale=1]{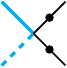}
}
\caption{Interaction vertices. Thin black (thick cyan) lines correspond to the first (second) field and solid (dashed) lines correspond to the classical (response) field. The inclusion of the circles is to illustrate the pairing of legs, i.e., the two legs without circles are involved in one dot product while the two legs with circles are involved in the other. For each vertex, we must sum over all of the  components contributing to the dot product for each leg pair. The vertices correspond to (a) $u_1 |\bphi_1|^2 \bphi_1 \cdot \tilde{\bphi}_1$, (b) $u_2 |\bphi_2|^2 \bphi_2 \cdot \tilde{\bphi}_2$, (c) $u_{12} |\bphi_2|^2 \bphi_1 \cdot \tilde{\bphi}_1$, and (d) $\sigma u_{12} |\bphi_1|^2 \bphi_2 \cdot \tilde{\bphi}_2$. \label{fig:ints}}
\end{figure}

Next, we write the explicit form of the action corresponding to the Langevin equations after rescaling to the  $u_a$ variables \cite{Tauber2014}:
\begin{subequations}
\label{VectorAction} %
\begin{equation}
\mathcal{A}[ \tilde{\bphi}_1, \bphi_1; \tilde{\bphi}_2, \bphi_2] = \mathcal{A}_0 + \mathcal{A}_{\text{int}},
\end{equation}
\begin{equation}
\mathcal{A}_0 =  \int_{t,\mathbf{x}} \sum_i  \tilde{\bphi}_i \cdot (\zeta_i \partial_t  - D_i \nabla^2  + r_i )\bphi_i - \zeta_i T_i |\tilde{\bphi}_i|^2,
\end{equation}
\begin{equation}
\begin{aligned}
\mathcal{A}_{\text{int}} &= \int_{t,\mathbf{x}} u_1 |\bphi_1|^2 (\bphi_1 \cdot \tilde{\bphi}_1) \\
&+ \int_{t,\mathbf{x}} u_2 |\bphi_2|^2 (\bphi_2 \cdot \tilde{\bphi}_2) \\
&+ \int_{t,\mathbf{x}} u_{12} |\bphi_2|^2 (\bphi_1 \cdot \tilde{\bphi}_1) \\
&+ \int_{t,\mathbf{x}} \sigma u_{12} |\bphi_1|^2 (\bphi_2 \cdot \tilde{\bphi}_2).
\end{aligned}
\end{equation}
\end{subequations} %
In Fig.~\ref{fig:ints}, we illustrate the resulting interaction vertices that are considered in our perturbative RG analysis.

\subsection{RG Equations}
\label{subsec:RGEq}

\begin{figure*}
\centering
\subfloat[]{
\centering
\includegraphics[scale=1]{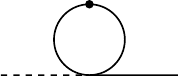}
}
\hspace{1cm}
\subfloat[]{
\centering
\includegraphics[scale=1]{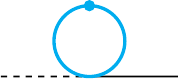}
}
\hspace{1cm}
\subfloat[]{
\centering
\includegraphics[scale=1]{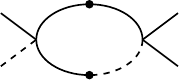}
}
\hspace{1cm}
\subfloat[]{
\centering
\includegraphics[scale=1]{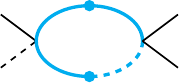}
}

\subfloat[]{
\includegraphics[scale=1]{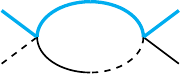}
}
\hspace{1cm}
\subfloat[]{
\centering
\includegraphics[scale=1]{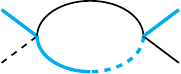}
}
\hspace{1cm}
\subfloat[]{
\centering
\includegraphics[scale=1]{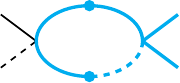}
}
\hspace{1cm}
\subfloat[]{
\centering
\includegraphics[scale=1]{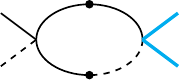}
}

\subfloat[]{
\centering
\includegraphics[scale=1]{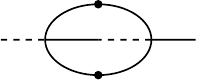}
}
\hspace{.7cm}
\subfloat[]{
\centering
\includegraphics[scale=1]{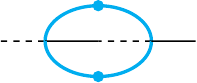}
}
\hspace{.7cm}
\subfloat[]{
\centering
\includegraphics[scale=1]{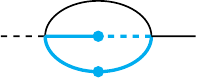}
}

\subfloat[]{
\centering
\includegraphics[scale=1]{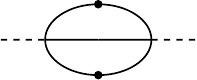}
}
\hspace{.7cm}
\subfloat[]{
\centering
\includegraphics[scale=1]{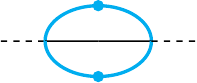}
}

\caption{One-loop corrections to (a,b) $r_1 \bphi_1 \cdot \tilde{\bphi}_1$, (c,d) $u_1 |\bphi_1|^2 \bphi_1 \cdot \tilde{\bphi}_1$, and (e-h) $u_{12} |\bphi_2|^2 \bphi_1 \cdot \tilde{\bphi}_1$. Analogous diagrams for $r_2 \bphi_2 \cdot \tilde{\bphi}_2$, $u_2 |\bphi_2|^2 \bphi_2 \cdot \tilde{\bphi}_2$ and $\sigma u_{12} |\bphi_1|^2 \bphi_2 \cdot \tilde{\bphi}_2$ can be obtained by switching thin black and thick cyan lines. Two-loop corrections to (i-k) $\zeta_1$ and $D_1$ as well as (l,m) $\zeta_1 T_1$. Analogous diagrams for $\zeta_2,D_2$, and $\zeta_2 T_2$ can be obtained by switching thin black and thick cyan lines. In these diagrams, the circles indicate propagators corresponding to the same component of a given field involved in a dot product and which must be summed over. \label{fig:loops}}
\end{figure*}

To study the RG flow, we first define renormalized parameters
\begin{equation}
\begin{array}{ccc}
D_{i_R} = Z_{D_i} D_i, & &r_{i_R} = Z_{r_i} r_i \mu^{-2}, \\ \\
u_{i_R} = Z_{u_i} u_i A_d \mu^{-\epsilon}, & &u_{12_R} = Z_{u_{12}} u_{12} A_d  \mu^{- \epsilon},\\ \\
\zeta_{i_R} = Z_{\zeta_i} \zeta_i, & & T_{i_R} = Z_{T_i} T_i ,
\end{array}
\end{equation}
where $A_d = \Gamma(3-d/2)/(2^{d-1} \pi^{d/2}) $ is a geometrical factor, $\Gamma(x)$ is Euler's Gamma function, $\mu$ is an arbitrary small momentum scale (compared to the lattice spacing), and $\epsilon = 4 - d$ defines the small parameter of the epsilon expansion. The effect of renormalization is captured in the $Z$ factors that contain the divergences according to the minimal subtraction procedure. Here, we note that in contrast to other approaches, the renormalization has been defined entirely in terms of the parameters, while the fields themselves have no additional non-trivial renormalization beyond their bare scaling.

We determine these $Z$ factors perturbatively to the lowest nontrivial order in $\epsilon$ or loops. The lowest-order corrections to $Z_{r}$ and $Z_{u}$ occur at one loop ($\sim\epsilon$), while those of $Z_\zeta, Z_T, Z_D$ appear at two loops ($\sim\epsilon^2$). The corresponding diagrams are illustrated in Fig.~\ref{fig:loops}. As in equilibrium, these corrections are modified from the case of $n_1 = n_2 = 1$ entirely through the inclusion of $n_i$-dependent combinatorial factors, and the integrals associated with the diagrams are otherwise unchanged.  While the diagrams are similar to those of model A, which have been calculated up to five loops \cite{Adzhemyan2022,Adzhemyan2022a}, the presence of two order parameters modify the associated integrals, leading to more complex forms which cannot be directly mapped to model A results, which is further complicated by the nonequilibrium features of our model. Expressions for the resulting $Z$ factors are presented in Appendix \ref{sec:Z}. The corresponding RG flow equations are
\begin{equation}
\label{gammaflow}
\gamma_p = \mu \partial_\mu \ln(p_R/p),
\end{equation}
where $p \in \{r_i,\zeta_i,D_i,T_i\}$. To identify the fixed points, we define the parameters
\begin{equation}
\begin{gathered}
v \equiv \frac{T_2}{T_1}, \qquad w \equiv \frac{\tilde{D}_2}{\tilde{D}_1}, \\
\tilde{u}_i \equiv \frac{T_i}{D_i^2} u_i, \qquad \tilde{u}_{12} \equiv \frac{T_1}{D_1 D_2} u_{12},
\end{gathered}
\end{equation}
where we have defined $\tilde{D}_i \equiv D_i/\zeta_i$.
The corresponding beta functions are
\begin{equation}
\beta_{s_a} = \mu \partial_\mu s_{a_R},
\end{equation}
where $s_a\in\{\tilde{u}_1,\tilde{u}_2,\tilde{u}_{12}, v, w\}$. By introducing these five parameters, the resulting beta functions are closed. These are given by
\begin{subequations}
\label{eq:beta}
\begin{equation}
\beta_{\tilde{u}_1} = \tilde{u}_{1_R} [-\epsilon + (n_1+8) \tilde{u}_{1_R}] + \sigma v_R n_2 \tilde{u}_{12_R}^2,
\end{equation}
\begin{equation}
\beta_{\tilde{u}_2} = \tilde{u}_{2_R} [-\epsilon + (n_2+8) \tilde{u}_{2_R}] + \sigma v_R n_1 \tilde{u}_{12_R}^2,
\end{equation}
\begin{multline}
\beta_{\tilde{u}_{12}} = \tilde{u}_{12_R}\bigg[ -\epsilon + 4 \frac{v_R+\sigma w_R}{1+w_R}\tilde{u}_{12_R}  \\
+ (n_1 +2)\tilde{u}_{1_R} + (n_2+2) \tilde{u}_{2_R} \bigg],
\end{multline}
\begin{equation}
\beta_v = - n_2 v_R F(w_R) \tilde{u}_{12_R}^2 [v_R - \sigma] \left[v_R + \sigma \frac{n_1}{n_2}\frac{F(w_R^{-1})}{F(w_R)}\right],
\label{betav}
\end{equation}
\begin{equation}
\begin{aligned}
\beta_w = -w_R\Big\{ & C'\big[(n_1+2)\tilde{u}_{1_R}^2 - (n_2+2)\tilde{u}_{2_R}^2\big] \\
   &+ \tilde{u}_{12_R}^2 \big[n_2 v_R^2 G(w_R) - n_1 G(w_R^{-1})\big] \\
   &+ 2 \sigma v_R \tilde{u}_{12_R}^2 \big[n_2 H(w_R)-n_1 H(w_R^{-1})\big]\Big\},
 \end{aligned}
\end{equation}
\end{subequations}
where we have defined $C' = 3\log(4/3) - 1/2 $ and the functions
\begin{subequations}
\begin{equation}
F(w) = -\frac{2}{w} \log \left(\frac{2+2 w}{2+w} \right),
\end{equation}
\begin{equation}
 G(w) = \log \left( \frac{(1+w)^2}{w(2+w)} \right)  - \frac{1}{2+3 w + w^2},
\end{equation}
\begin{equation} 
H(w) = \frac{1}{w} \log \left( \frac{2+2 w}{2+w}\right) - \frac{3 w + w^2}{8 + 12 w + 4 w^2}.
\end{equation}
\end{subequations}

These equations exhibit several important features. First, we note that, for $n_1 = n_2$, the beta functions are invariant under the transformation
$\tilde{u}_{1_R} \leftrightarrow \tilde{u}_{2_R}$, $\tilde{u}_{12_R} \to \sigma v_R \tilde{u}_{12_R}, v_R \to v_R^{-1}$, and $w_R \to w_R^{-1}$. This implies that the fixed points with either $\sigma = - 1, v_R \neq 1$, or $w_R \neq 1$ come in equivalent pairs. However, when $n_1 \neq n_2$, there is no such invariance since the two fields have different symmetries, so the fixed points no longer emerge in pairs.

Furthermore, under equilibrium conditions with $\sigma = v_R = 1$, we immediately see that $\beta_v = 0$, so the two temperatures remain identical at all scales. Indeed, one can further show that $\gamma_T = 0$ in this case, so the temperature itself does not flow, which is a consequence of the fact that the temperature of the system is fixed and scale invariant in equilibrium. Moreover, the beta functions for $\tilde{u}$ become independent of $w_R$, which underscores how statics are independent of the dynamics in equilibrium. Similarly, in Refs.~\cite{Sieberer2013,Sieberer2014,Tauber2014a}, the critical exponent associated with the transient nonequilibrium features represented an additional hierarchy of this form, where the nonequilibrium features were dependent on the statics and dynamics but not the reverse. This illustrates a key difference between an (effective) equilibrium setting and our nonequilibrium model, where statics and dynamics are inherently intertwined and there is no such distinction.

When considering the beta function $\beta_w$, there are two distinct scenarios we must discuss, noting that the beta function for the ratio $p/q$ is $\beta_{p/q} = \mu \partial_\mu (p_R/q_R) =  \frac{p_R}{q_R} (\gamma_p - \gamma_q)$. In the first scenario, $\beta_w$ vanishes when $\gamma_{\tilde{D}_1} = \gamma_{\tilde{D}_2}$. Since the dynamical critical exponent is related to the flow of the $\tilde{D}_i$ parameters through $z_i = 2 + \gamma_{\tilde{D}_i}$, we see that, in this scenario, both fields are governed by the same dynamical critical exponent, which is known as ``strong dynamic scaling''\cite{Halperin1969,DeDominicis1977,Dohm1977,Tauber2014}. In the second scenario, $\gamma_{\tilde{D}_1} \neq \gamma_{\tilde{D}_2}$, which takes $w_R$ to 0 or $\infty$ depending on the sign of $\gamma_{\tilde{D}_1} - \gamma_{\tilde{D}_2}$. As a result, the two fields are governed by different dynamical critical exponents, which is known as ``weak dynamic scaling'' \cite{Halperin1969,DeDominicis1977,Dohm1977}; see also \cite{Tauber2014}. Similarly, one can also consider the beta function $\beta_v$ with a similar perspective. Either the two effective temperatures $T_i$ realize a finite ratio or, depending on the sign of $\gamma_{T_1} - \gamma_{T_2}$, the system flows to $v_R = 0$ or $v_R = \infty$, both of which correspond to one of the two $u_{12}$ coupling terms vanishing, corresponding to $\sigma = 0$.
We investigate this scenario in \cref{sec:halfcouple}.

\subsection{Fixed points and stability}
\label{sec:fixedpoints}
In this section, we identify the fixed points of the RG flow in different regimes. To this end, we find it useful to consider separately the cases where the number of components $n_1$ and $n_2$ are identical or different. As noted above, we restrict our focus for now to $\sigma = \pm 1$, while the fixed points when $\sigma = 0$ are considered later in Sec.~\ref{subsec:OneWayBeta} since they require extra care due to non-perturbative behaviors.

\subsubsection{Equal number of components: $n_1=n_2$}

For an equal number of components $n_1=n_2$, the fixed points can be identified analytically (similar to the coupled Ising models with $n_1=n_2=1$ studied in previous work \cite{Young2020}).  We first consider $\sigma = 1$, in which case we immediately find that the roots of $\beta_v = 0$ are given by $v_R = 0, 1, - F(w_R^{-1})/F(w_R)$. Setting aside the case of $v_R = 0, \infty$ (the latter can be understood as a fixed point for $1/v_R$) and noting that $- F(w_R^{-1})/F(w_R) < 0$ for $w_R >0$ ($w_R < 0$ results in unbounded dynamics), the only physically sensible fixed point value is $v_R^* = 1$, implying that the two temperatures are identical at the fixed point. As a result, only equilibrium fixed points are possible for $\sigma = 1$, illustrating the robustness of emergent equilibrium descriptions of nonequilibrium Ising models. Moreover, this means that an effective FDT emerges, so $\gamma_T = 0$ and the $T_1 = T_2$ are constant, as in equilibrium. 

Thus for $\sigma = 1$, there are three types of fixed points, one of which is stable and two of which are unstable, where the type of the stable fixed point depends on the value of $n$. The first is the decoupled fixed point $\mathcal{D}$, at which both fields become effectively decoupled. The second is the Heisenberg fixed point $\mathcal{H}$, associated with the emergence of an $O(n_1+n_2)$ symmetry at the critical point, where $\tilde{u}_{1_R} = \tilde{u}_{2_R} = \tilde{u}_{12_R}$. Finally, the third fixed point is the biconical fixed point $\mathcal{B}$ which remains coupled but with no emergent symmetry at the fixed point. The biconical fixed point is commonly associated with its ability to host a doubly-ordered phase at the critical point \cite{Fisher1965,Bruce1975,Kosterlitz1976,Folk2008a,Eichhorn2013}.

Next we consider the $\sigma = -1$ sector; the resulting NEFPs are given by 
\begin{equation}
\begin{gathered}
v_R^* = 1, \qquad w_R^* = 1, \\
\tilde{u}_{1_R}^* = \frac{1}{2(n+2)}\epsilon, \qquad \tilde{u}_{2_R}^* = \frac{1}{2(n+2)} \epsilon, \\
 \tilde{u}_{12_R}^* = \pm \frac{\sqrt{4/n - 1}}{2(n+2)} \epsilon.
\end{gathered}
\label{eq:equalfixed}
\end{equation}
There are several interesting features to note about these fixed points. First, although $v_R^* = 1$ at this point, indicating that the two temperatures are the same, this is not the same as an effective equilibrium description. This is because there are two ingredients necessary for an equilibrium description: (1) The two effective temperatures are the same; (2) There is an effective Hamiltonian which describes the dynamics. Since $\sigma = -1$, the latter requirement is violated, and the fixed points are indeed nonequilibrium.

Second, we see that the two NEFPs merge at $\tilde{u}_{12_R}^* = 0$ when $n=4$. For $n>4$, $\tilde{u}_{12_R}^*$ becomes imaginary and thus unphysical, so $n=4$ represents the crossover from criticality governed by NEFPs to an equilibrium criticality governed by decoupled fixed points. Interestingly, this is also the value of $n$ at which the stable equilibrium fixed points change from biconical fixed points to decoupled fixed points. As we show in the following section, there is a similar relation for $n_1 \neq n_2$ to all orders in $\epsilon$. 

Third, since the merging/splitting of fixed points coincides with changes in stability, this means that all the NEFPs for $0 < n < 4$ have the same stability as for $n=1$. 
In previous work for $n=1$ \cite{Young2020}, we showed that three of the five eigenvalues of the stability matrix are stable at $\mathcal{O}(\epsilon)$, while two are marginal at this order and would require considering two-loop corrections to $\tilde{u}$ and three-loop corrections to $v, w$ to determine their stability to $\mathcal{O}(\epsilon^2)$.

Before closing this subsection, we 
conclude with a couple of additional remarks. 
First, if we allow ourselves to consider the $n \to 0$ limit, $\tilde{u}_{12_R}$ diverges, and the theory becomes non-perturbative. In equilibrium, analytic continuation of the  $O(n)$ Ising model in the limit $n \to 0$ describes the behavior of self-avoiding walks \cite{DeGennes1979,Pelissetto2002}, so a multicritical, nonequilibrium generalization of self-avoiding walks may lead to new, non-perturbative nonequilibrium criticality, although identifying such a generalization is by no means straightforward. Second, we note that the two fixed point values of $\tilde{u}_{12_R}^*$ are purely imaginary for $n > 4$. While nominally non-physical, these complex fixed points can give rise to approximate conformality and weakly first-order transitions near (in $n_i$) the crossover from real to complex fixed points \cite{Kaplan2009,Wang2017,Gorbenko2018,Ma2019}.

\begin{figure*}[t!]
\centering
\includegraphics[scale=1]{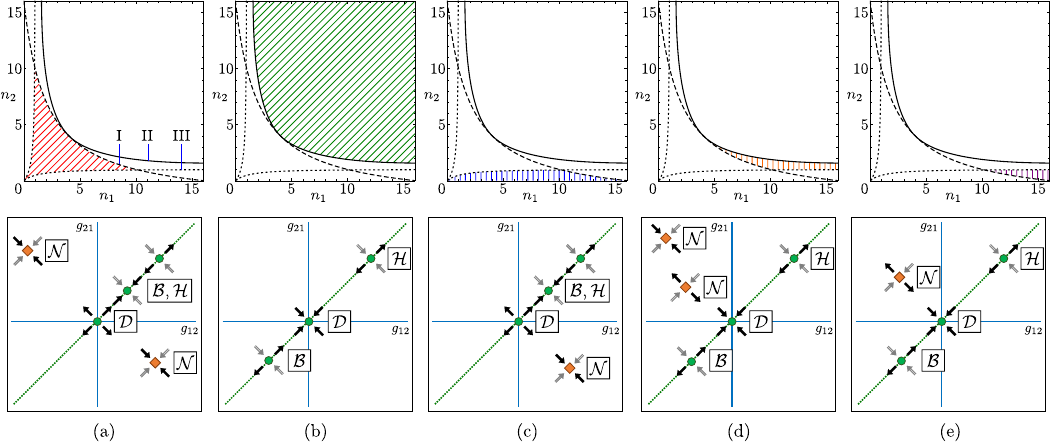}
\caption{Fixed point behavior as a function of $n_1, n_2$ with qualitative stability to $O(\epsilon)$ aside from $\tilde{u}_{i_R}^* < 0$ fixed points. 
The top plots indicate different regions of fixed point and stability behavior, separated by three types of boundaries, labeled I, II, and III, indicated by dashed, solid, and dotted lines, respectively. Shading indicates which region of $n_1, n_2$ we consider. The corresponding bottom plots illustrate qualitatively the RG flow diagrams in the $g_{12}$-$g_{21}$ plane, although the full flow occurs in a five-dimensional space. $\mathcal{N}$ denotes the NEFPs, $\mathcal{H}$ the Heisenberg fixed points with $O(n_1+n_2)$ symmetry, $\mathcal{B}$ the biconical fixed points, and $\mathcal{D}$ the decoupled fixed points. The dotted green lines in the bottom plots indicate parameters corresponding to effective equilibrium behavior, so $\mathcal{H},\mathcal{B}, \mathcal{D}$ lie along these lines. Since stability is only known to first-order in $\epsilon$ along directions which preserve $g_{21}/g_{21}$, we use filled black arrows to indicate known stability at this order and gray arrows to indicate the anticipated stability at higher orders.
Given the fact that the system cannot flow through $g_{12} = 0$ or $g_{21} = 0$, we expect each quadrant (minus the axes) to broadly define the region of attraction for the corresponding stable fixed point, provided one exists and the system remains in the perturbative regime.
(a) In this region, both NEFPs are present and stable at $\mathcal{O}(\epsilon)$. Whether $\mathcal{B}$ or $\mathcal{H}$ is stable is determined by the values of $n_1, n_2$. (b) In this region, there are no physically valid NEFPs, and criticality is described by $\mathcal{D}$. (c) In this region, one of the two NEFPs has diverged, leaving only the other NEFP, which is stable at $\mathcal{O}(\epsilon)$. Whether $\mathcal{B}$ or $\mathcal{H}$ is stable is determined by the values of $n_1, n_2$. (d) In this region, both NEFPs are in the same quadrant, with one stable and one unstable at $\mathcal{O}(\epsilon)$. (e) In this region, one of the two NEFPs has diverged, leaving only the other NEFP, which is unstable at $\mathcal{O}(\epsilon)$. \label{NEFPplot}}
\end{figure*}

\subsubsection{Arbitrary numbers of components}

In this section, we 
investigate the fixed points in a more general scenario where $n_1$ and $ n_2$ are not necessarily equal. 
Let us consider the $\sigma = 1$ sector first. Once again, we immediately find that $v_R^* = 1$ is the only physically valid solution when the two fields are coupled and we set aside the case of $v_R = 0, \infty$ for now. Thus the robustness of the equilibrium fixed points in the $\sigma = 1$ sector continues to hold. Moreover, the same classes of fixed points ($\mathcal{D},\mathcal{H},\mathcal{B}$) continue to describe the criticality in this sector as for $n_1=n_2$.

Next, consider the $\sigma = -1$ sector. Although the fixed points at this order can be identified analytically for both $n_1 = n_2$ and when $n_1 \neq n_2$ in equilibrium to $\mathcal{O}(\epsilon)$, there is no analytic solution here. This is largely a consequence of the fact that there is no separation of ``statics'' ($\tilde{u}$) and ``dynamics'' ($v,w$) in the nonequilibrium setting. As a result, the beta functions for $\tilde{u}$ become coupled to those for $v,w$, and finding the fixed points no longer corresponds to finding the roots of simple polynomials. For $n_1 = n_2$, this was not an issue because $v_R^* = w_R^* = 1$, and a simple analytic solution was possible. Instead, the nonequilibrium fixed points are identified numerically. 
Interestingly, we can still determine a simple analytical  expression for the temperature ratio as
\begin{equation}
v_R^* = \frac{n_1}{n_2} \frac{F(w_R^{-1})}{F(w_R)},
\end{equation}
which reduces the number of beta functions we need to consider to four.

Away from $n_1 = n_2$, we numerically find a pair of additional fixed points that are not connected to the NEFPs found for $n_1 = n_2$. These two fixed points are characterized by a negative value of either $u_{1_R}^*$ or $u_{2_R}^*$. Thus, for these fixed points, the higher-order terms (equivalent to $\phi^6$ terms in the equilibrium free energy) become dangerously irrelevant since they are necessary for a finite expectation value of the order parameter in the ordered phase. In the limit $n_1 \to n_2$, the $\tilde{u}$ parameters diverge, hence their absence in the previous subsection. Moreover, stability analysis indicates that these fixed points are unstable to order $\epsilon$. In light of these factors, we anticipate these unstable fixed points to have minimal relevance in a physical system, so we only focus on the NEFPs which are present for $n_1 = n_2$ in the $\sigma = -1$ sector.

In Fig.~\ref{NEFPplot}, we illustrate the regions where the NEFPs exist and the qualitative behavior of their stability. In total, there are eight different regions, six of which are composed of three equivalent pairs under $\bphi_1  \leftrightarrow \bphi_2$. These different regions are divided by three different types of boundaries. Each boundary corresponds to a qualitative change in the behavior/stability of one or more fixed points.

The first (I) boundary we consider is the dashed curve in \cref{NEFPplot} which is defined by
\begin{equation}
\frac{n_1+2}{n_1+8} + \frac{n_2 + 2}{n_2+8} = 1.
\label{eq:boundary1}
\end{equation} 
This boundary corresponds to one (or both for $n_1 = n_2 = 4$) of the NEFPs passing through the decoupled fixed point. As a result, the latter two fixed points exchange their stability in the nonequilibrium direction (i.e., towards the upper left and bottom right quadrants), leading to one stable NEFP and one unstable NEFP. Additionally, beyond the behavior of the NEFPs, this boundary is also associated with a change in the behavior of the equilibrium fixed points. Like the NEFP, the biconical fixed point also passes through the decoupled fixed point, which exchange stability in the equilibrium direction (i.e., along the equilibrium line) with each other. 

The fact that these two boundaries are equivalent is not merely a coincidence of perturbation theory at this order; indeed, it extends to all orders and is exact. This can be understood by considering the stability matrix
\begin{equation}
\Lambda_{ab} \equiv \frac{\partial \beta_{s_a}}{\partial s_{b_R}},
\end{equation}
and considering $\tilde{g}_{12}, \tilde{g}_{21}$ rather than $\tilde{u}_{12}, v$ since $v$ becomes 0 or $\infty$ at the decoupled fixed point when $n_1 \neq n_2$ due to weak dynamic scaling. Moreover, note
\begin{equation}
\beta_{\tilde{g}_{12}} = \tilde{g}_{12} f_{12} (s_R), \qquad \beta_{\tilde{g}_{21}} = \tilde{g}_{21} f_{21} (s_R),
\end{equation}
which is a consequence of the fact that $g_{12}$ (similarly, $g_{21}$) cannot be generated if it is zero. For the decoupled fixed point, $\frac{\partial \beta_{\tilde{g}_{ij}}}{\partial s_{b_R}} = \tilde{g}_{ij_R} \frac{\partial f_{ij}}{\partial s_{b_R}} = 0$ for $s_{b} \neq \tilde{g}_{ij}$ and $i \neq j$. As a result, $\Lambda$ becomes block triangular, so we can consider the stability in the $\tilde{g}_{12}, \tilde{g}_{21}$ subspace on its own. For the same reasons, the corresponding $2 \times 2$ submatrix is diagonal with eigenvalues $\lambda_{12}, \lambda_{21}$. 
Moreover, since the stability is marginal when any fixed point passes through the decoupled fixed point, at least one of the two eigenvalues $\lambda_{12}, \lambda_{21}$ must be zero. 
Now we consider the equilibrium boundary where the biconical fixed point passes through the decoupled fixed point. Since equilibrium perturbations to the model cannot give rise to nonequilibrium dynamics, the stability matrix can be put into a different basis and retain a triangular form. This means that (in)stability in the equilibrium case extends to the more general model. Since the  stability in the equilibrium direction is marginal when the biconical fixed point passes through the decoupled fixed point, we find that the corresponding eigenvalue is $\lambda_{12} + \lambda_{21} = 0$. Combined with the fact that one of $\lambda_{12}, \lambda_{21}$ must be zero when the decoupled fixed point exhibits marginal stability, it thus follows that $\lambda_{12} = \lambda_{21} = 0$, i.e., the stability is marginal to both equilibrium and nonequilibrium perturbations. Since the equilibrium model only exhibits marginality in this decoupled fixed point when the biconical fixed point passes through it, a NEFP must also pass through the decoupled fixed point here; otherwise, it would contradict the established behavior of the equilibrium model. This means that conclusions based on the more sophisticated approaches to determining the biconical/decoupled fixed point boundary in equilibrium also apply to one of the two NEFPs. Note that when $n_1 = n_2 = n$, we anticipate that the decoupled fixed point describes the phase transition for $n \geq 2$ based on higher-order results for the equilibrium model in 3D \cite{Calabrese2003,Folk2008a,Eichhorn2013}. 

An interesting consequence of this relation between the NEFP and the biconical fixed point is that the stability of one gives insight into the stability of the other. Since the stability of the $n_1 = 1, n_2 = 2$ biconical fixed point in three dimensions has been the subject of extensive research using both numerical and diagrammatic approaches, with some results indicating stability \cite{Calabrese2003,Folk2008a,Eichhorn2013} and others indicating instability \cite{Selke2011,Selke2013,Hu2014,Xu2019}, investigating the stability of the related NEFP could provide an alternative means of understanding the nature of the equilibrium multicritical point.

The second (II) boundary we consider is the solid boundary in \cref{NEFPplot}. This boundary corresponds to the two NEFPs merging in the same quadrant (with the exception of $n_1 = n_2$, which occurs at $\tilde{u}_{12_R} = 0$) and becoming complex, which can lead to approximate conformality and weakly first-order transitions \cite{Kaplan2009,Wang2017,Gorbenko2018,Ma2019}, as discussed for $n_1 = n_2$. 

The third (III) boundary is in fact a pair of boundaries denoted by the dotted lines. These boundaries are approximately defined by
\begin{equation}
\pm 2 (n_1 - n_2) \approx n_2(n_1+2) \sqrt{\frac{n_1}{n_1+8}} + n_1(n_2+2) \sqrt{\frac{n_2}{n_2+8}}.
\end{equation}
These boundaries are different from the previous two in that they do not involve a pair of the previously-discussed fixed points merging or passing through one another. Rather, these two boundaries correspond to a breakdown in perturbation theory, expressed by the divergence of coupling terms, hence why there is no stable fixed point for the upper left quadrant in Figs.~\ref{NEFPplot}(c,e). Since this divergence is quite sharp, it is important to ensure that the model is still perturbative in the vicinity of this boundary. In light of this non-perturbative feature, functional/exact RG techniques \cite{Dupuis2021,Boettcher2015} could give valuable insight into the behavior of the field theory in this regime.

Note that the asymptotic behaviors of boundaries II and III indicate that the two NEFPs persist for $n_2 = 1$ in the limit of $n_1 \to \infty$, which occurs in the region shown in Fig.~\ref{NEFPplot}(d), where boundary II approaches $n_2 \approx 1.141$ and boundary III approaches $n_2 \approx 0.9707$, both of which may be sensitive to higher-order terms in $\epsilon$. In this limit, the two fixed points neither merge nor diverge. We identify the stable fixed point for $n_2 = 1$ in this limit numerically as 
\begin{equation}
\begin{gathered}
v_R^* \approx 0.4885 n_1 , \qquad w_R^* \approx 0.4302, \\
\tilde{u}_{1_R}^* \approx \frac{7.4336}{n_1} \epsilon, \qquad \tilde{u}_{2_R}^* \approx 2.361 \epsilon, \\
 \tilde{u}_{12_R}^* \approx -\frac{9.895}{n_1} \epsilon, 
\end{gathered}
\end{equation} 
and the unstable fixed point exactly as
\begin{equation}
\begin{gathered}
v_R^* = n_1, \qquad w_R^* = 1, \\
\tilde{u}_{1_R}^* = \frac{3 + 2 \sqrt{3}}{6 n_1} \epsilon, \qquad \tilde{u}_{2_R}^* = \frac{\epsilon}{6}, \\
 \tilde{u}_{12_R}^* = -\frac{\sqrt{3}}{6 n_1} \epsilon.
\end{gathered}
\end{equation} 
An interesting future direction would be to investigate whether these fixed points continue to persist along $n_2 = 1$ when higher-order corrections are taken into account. 
Finally, we remark that there exist unstable fixed points with unbounded dynamics in the absence of higher-order terms (i.e., $\tilde{u}_{1_R}^* < 0$ or $\tilde{u}_{2_R}^* < 0$) which we do not investigate.

\section{Universal scaling behavior}
\label{sec:scale}

In this section, we investigate the universal scaling behavior that emerges due to the NEFPs. Near criticality, the correlation and response functions can be generically expressed
\begin{subequations}\label{scaling functions}
\begin{equation}
\begin{aligned}
C_i(\mathbf{q},\omega,\{r_j\}) &= \mathcal{F} \langle \phi_i(\mathbf{0},0) \phi_i(\mathbf{r},t) \rangle\\
&\propto |\mathbf{q}|^{-2 + \eta_i - z} \hat{C}_i\left(\frac{\omega}{|\mathbf{q}|^z},\left\{\frac{r_j}{|\mathbf{q}|^{1/\nu_j}}\right\}\right),
\end{aligned}
\end{equation}
\begin{equation}
\begin{aligned}
\chi_i(\mathbf{q},\omega,\{r_j\}) &= \mathcal{F} \langle \tilde{\phi}_i(\mathbf{0},0) \phi_i(\mathbf{r},t) \rangle\\
&\propto |\mathbf{q}|^{-2 + \eta_i'}\hat{\chi}_i\left(\frac{\omega}{|\mathbf{q}|^z},\left\{\frac{r_j}{|\mathbf{q}|^{1/\nu_j}}\right\} \right),
\end{aligned}
\end{equation}
\end{subequations}
where we have introduced general scaling functions $\hat C_i, \hat \chi_i$ and several critical exponents. Here, $\eta_i, \eta_i'$ define the anomalous dimensions of the correlation and response functions, respectively, which are equal in equilibrium systems. The dynamical critical exponent $z$ describes the relative scaling of time compared to spatial coordinates, where we have dropped the subscript $i$ since in nearly every case we consider, the criticality exhibits strong dynamic scaling where $z_1 = z_2$, with the one exception only weakly violating this relation. In contrast, the various anomalous dimensions are typically different for  the nonequilibrium fixed points we study. Furthermore, we have expressed $r_j$ for $j=a,b$ to reflect the fact that the RG equations couple $r_1, r_2$ in a non-trivial way, while the values of $\nu_j$ determine the scaling of the correlation length and the crossover exponent, which describes the scaling of the phase boundaries, near criticality.

Unlike equilibrium, there is no longer a single anomalous dimension which is the same for both the correlation and response functions. This represents a breakdown in the fluctuation-dissipation theorem and the absence of an effective temperature. Nevertheless, through an appropriate modification of the usual fluctuation-dissipation theorem, we may define an effective scale-dependent temperature by roughly mimicking the fluctuation-dissipation relation as $T_i^{\rm eff} (\mathbf{q},\omega)\sim \omega C_i(\mathbf{q}, \omega)/{\text{Im}} \chi_i (\mathbf{q},\omega)$. A similar relation is considered in the context of ageing at criticality, where the violation of the fluctuation-dissipation relation is instead described by a universal amplitude ratio \cite{Godreche2000,Calabrese2005}. In contrast, here the analogous amplitude ratio is no longer a universal constant but instead exhibits universal scaling.
This effective temperature is then related to the anomalous dimensions according to $T_i^{\rm eff} \sim |\mathbf{q}|^{\eta_i-\eta_i'}$ at long wavelengths and fixed $\omega/|\mathbf{q}|^z$, and hence $\gamma_T = \eta_i - \eta_i'$, thereby reflecting the mismatch in anomalous dimensions between the fields and their corresponding response fields. For the fully-coupled NEFPs, the value of $\gamma_T$ is the same for both fields. In the following analysis, we always find that $\eta_i' \geq \eta_i$, so the effective temperatures always get ``hotter'' or remain invariant at increasing wavelengths.

Note that although the NEFPs do not require $\eta_1 = \eta_2$ or $\eta_1' = \eta_2'$, we find $\eta_1 - \eta_1' = \eta_2 - \eta_2'$, indicating that the effective temperatures for each field scale in the same way, realizing a constant temperature ratio $T_2^{\rm eff}/T_1^{\rm eff}$. Since there is only a single dynamical critical exponent $z$, this indicates the system is described by strong dynamic scaling \cite{Halperin1969,DeDominicis1977,Dohm1977,Tauber2014}. 
Additionally, depending on the values of $n_i$, the value of $\nu$ can take on complex values, corresponding to the emergence of discrete scale invariance \cite{Sornette1998} rather than the usual continuous scale invariance associated with phase transitions.

This section is arranged as follows. In Sec.~\ref{subsec:anomdynam}, we discuss the flow equations which define the anomalous dimensions $\eta$ and dynamical critical exponent $z$. In Sec.~\ref{subsec:massrenorm}, we discuss three different scenarios for the mass renormalization, concluding with a discussion of the implications for the phase diagram in all three scenarios. In Sec.~\ref{subsec:critexp}, we discuss the qualitative features of the critical exponents as a function of $n_i$. Finally, in Sec.~\ref{subsec:hyper}, we discuss how the usual hyperscaling relations are modified at the NEFPs due to complex $\nu$ and $\eta_i \neq \eta_i'$.

\subsection{Anomalous and dynamic exponents}
\label{subsec:anomdynam}
We utilize the method of characteristics to identify the critical exponents \cite{Tauber2014, Young2020}, which we relate to the flow functions as
\begin{equation}
\begin{aligned}
\eta_i &= \gamma_T - \gamma_{D_i}, & \eta_i' &= - \gamma_{D_i}, & z_i &= 2 + \gamma_{D_i} - \gamma_{\zeta_i}.
\end{aligned}\label{exponents from gamma}
\end{equation}

Using the $Z$ factors presented in Appendix \ref{sec:Z}, we can determine expressions for these flow functions in terms of $\tilde{u}_R, v_R, w_R$:
\begin{subequations}
\begin{equation}
\begin{aligned}
\gamma_{\zeta_1} =&~ -C_\zeta' (n_1+2) \tilde{u}_{1_R}^2 -  n_2 v_R^2 \tilde{u}_{12_R}^2 G_\zeta (w_R) \\
&~ - 2 \sigma n_2 v_R \tilde{u}_{12_R}^2 H_\zeta (w_R),
\end{aligned}
\end{equation}
\begin{equation}
\begin{aligned}
\gamma_{\zeta_2} =&~ -C_\zeta' (n_2+2) \tilde{u}_{2_R}^2 -  n_1 \tilde{u}_{12_R}^2 G_\zeta (w_R^{-1}) \\
&~ - 2 \sigma  n_1 v_R \tilde{u}_{12_R}^2 H_\zeta (w_R^{-1}),
\end{aligned}
\end{equation}
\begin{equation}
\begin{aligned}
\gamma_{D_1} =&~ -C'_D (n_1+2)\tilde{u}_{1_R}^2 - n_2 v_R^2 \tilde{u}_{12_R}^2 G_D(w_R) \\
&~ - 2 \sigma n_2 v_R \tilde{u}_{12_R}^2 H_D(w_R),
\end{aligned}
\end{equation}
\begin{equation}
\begin{aligned}
\gamma_{D_2} =&~ -C'_D (n_2+2)\tilde{u}_{2_R}^2 - n_1 \tilde{u}_{12_R}^2 G_D(w_R^{-1}) \\
&~ - 2 \sigma n_1 v_R \tilde{u}_{12_R}^2 H_D(w_R^{-1}),
\end{aligned}
\end{equation}
\begin{equation}
\gamma_{T_1} = n_2 F(w_R) \tilde{u}_{12_R}^2 v_R (v_R - \sigma),
\end{equation}
\begin{equation}
\gamma_{T_2} = - \sigma n_1 F(w_R^{-1}) \tilde{u}_{12_R}^2 (v_R - \sigma),
\end{equation}
\end{subequations}
where we have defined the constants and functions
\begin{subequations}
\begin{align}
C'_\zeta &= 3 \log(4/3), & C'_D &=1/2, \\
G_\zeta(w) &= \log \left(\frac{(1+w)^2}{w(2+w)}\right), & G_D(w) &= \frac{1}{2 + 3w + w^2}, \\
H_\zeta(w) & = \frac{1}{w} \log\left(\frac{2+2w}{2+w}\right), & H_D(w) &= \frac{3w + w^2}{8+12w + 4w^2}.
\end{align}
\end{subequations}
Note that these constants and functions are related to those used in $\beta_w$ via $C' = C'_\zeta - C'_D$, $G(w) = G_\zeta(w) - G_D(w)$, $H(w) = H_\zeta(w) - H_D(w)$. By inserting the fixed points derived in the previous section in the above equations, one can readily determine the exponents $\eta_i,\eta_i', z$. The reader can skip to \cref{sec:critical_exp} to see a summary of the qualitative behaviors of these critical exponents in different cases or Table \ref{tab:exp} in the Appendix for their numerical values. First, however, we must devote special care to the criticality associated with the mass ($r_i$) renormalization, which is more complex and is further responsible for several nonequilibrium features of note, so that we may provide a full summary of the universality of the NEFPs in various regimes.

\subsection{Mass renormalization}
\label{subsec:massrenorm}
In this section, we consider the renormalization of the mass terms $r_i$. Because their renormalization is intertwined with one another due to the nonreciprocal coupling, there are several qualitatively different scaling behaviors which are possible, each of which require different treatments. Additionally, we  focus specifically on the renormalization of $r_i/D_i$, which is the parameter associated with the scaling of the correlation length $\xi$, so that we only need to consider two flow equations, similar to our consideration of redefinition of $u$ for the beta functions, and we replace $r_i \to D_i r_i$ in a slight abuse of notation. Defining the flowing parameters $r_i(l)$ and the corresponding momentum scale $\mu (l) = \mu l$ with $r_i(1) = r_{i_R}$, the flow equations take the general form
\begin{subequations}
\begin{equation}
l \frac{d}{dl} \left( \begin{array}{c}
r_1(l) \\
r_2(l)
\end{array} \right) = \left(\begin{array}{cc}
\mathcal{R}_{11} & \mathcal{R}_{12} \\
\mathcal{R}_{21} & \mathcal{R}_{22} \end{array} \right)
\left( \begin{array}{c}
r_1(l) \\
r_2(l)
\end{array} \right),
\end{equation}
\begin{equation}
\begin{aligned}
\mathcal{R}_{11} &= -2 + (n_1 + 2) \tilde{u}_{1_R}, & \mathcal{R}_{12} &= n_2 v_R w_R^{-1} \tilde{u}_{12_R}, \\
\mathcal{R}_{22} &= -2 + (n_2 + 2) \tilde{u}_{2_R}, &  \mathcal{R}_{21} &= \sigma n_1 w_R \tilde{u}_{12_R},
\end{aligned}
\end{equation}
\end{subequations}
which have been determined using the $Z$ factors presented in Appendix \ref{sec:Z}. We shall denote the above $2\times 2$ matrix by $\mathcal{R}$.
The values of $\nu_i^{-1}$ are determined by the eigenvalues of $\mathcal{R}$:
\begin{equation}
-\nu^{-1} = \frac{\mathcal{R}_{11} + \mathcal{R}_{22}}{2} \pm \sqrt{ \left(\frac{\mathcal{R}_{11} - \mathcal{R}_{22} }{2}\right)^2 + \mathcal{R}_{12} \mathcal{R}_{21} }.
\end{equation} 
There are three separate scenarios we consider with regards to the matrix $\mathcal{R}$ depending on the behavior of its eigenvalues and eigenvectors. The first scenario is when the eigenvalues are real. This is the usual behavior that occurs for equilibrium models. The second scenario is when the eigenvalues are complex-valued. As we discuss, this corresponds to the emergence of discrete scale invariance and can only occur when $\sigma = -1$. Finally, the third scenario corresponds to the transition between these two cases (as a function of $n_i$ for $\sigma = -1$) where an exceptional point occurs. In this case, there is only a single eigenvalue and eigenvector. Since this occurs at the boundary of two regions, we expect that it may not apply directly to a physical system with integer $n_i$. Nevertheless, if the values of $n_i$ are sufficiently close to this boundary, the resulting critical behavior may be nearly indistinguishable from an exceptional point for practical purposes. Finally, we discuss the implications for all three scenarios on the structure of the phase diagram.

\subsubsection{Real eigenvalues}
For the equilibrium fixed points, one always has $\mathcal{R}_{12} \mathcal{R}_{21} \geq 0$, which implies that the eigenvalues $\nu_a^{-1}, \nu_b^{-1}$ are purely real. For the NEFPs, this also holds for some values of $n_1, n_2$, as we later show in Sec.~\ref{subsec:critexp}. 
To determine the scaling functions, we diagonalize $\mathcal{R}$: 
\begin{subequations}
\begin{equation}
    \mathcal{P}^{-1} \mathcal{R} \mathcal{P} =
    -\left(
    \begin{array}{cc}
    \nu_a^{-1} & 0 \\
    0 & \nu_b^{-1}
    \end{array}
    \right).
\end{equation}
\begin{equation}
\label{eq:P}
    \mathcal{P}^{-1} = \left(
    \begin{array}{cc}
    \mathcal{R}_{11} - \mathcal{R}_{22} + \nu_a^{-1} - \nu_b^{-1} & 2 \mathcal{R}_{12} \\
     \mathcal{R}_{11} - \mathcal{R}_{22} - \nu_a^{-1} + \nu_b^{-1} &  2 \mathcal{R}_{12}
    \end{array}
    \right),
\end{equation}
\end{subequations}
where we assume $\nu_a \geq \nu_b$. 
To describe the eigenvectors, we introduce the vector $\mathbf{r}$,
which represents the two-dimensional parameter space defined by $(r_1, r_2)$.
The right and left eigenvectors of the matrix $\mathcal R$ are now denoted by $\mathbf{v}_{a,b}$ and $\mathbf{u}_{a,b}$, respectively, corresponding to the eigenvalues $\nu_{a,b}$. Given the non-symmetric form of $\mathcal R$, these right and left eigenvectors need not be identical; however, they still satisfy the orthogonality relation $\mathbf{u}_i \cdot \mathbf{v}_j = \delta_{ij}$. Note that $\mathbf{u}$ are row vectors of $\mathcal{P}^{-1}$ and $\mathbf{v}$ are column vectors of $\mathcal{P}$. While $\mathbf{u}$ can be read off directly from $\mathcal{P}^{-1}$ above, $\mathbf{v}$ can be determined up to a constant factor (which is determined through the orthogonality relation) by replacing $\mathcal{R}_{12}$ with $\mathcal{R}_{21}$ or directly determining $\mathcal{P}$.

\begin{figure}
    \centering
    \includegraphics[scale=.9]{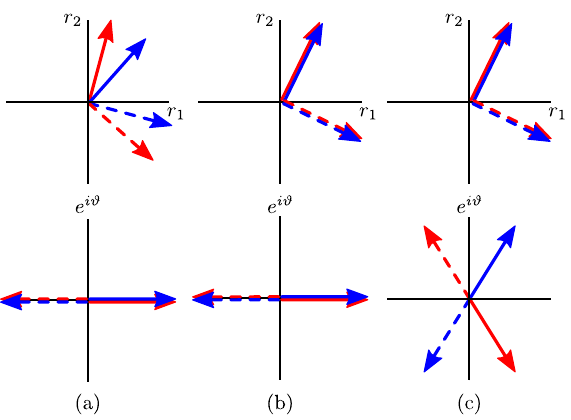}
    \caption{Behavior of right (solid) and left (dashed) eigenvectors 
    of the flow defined by $\mathcal{R}$ for each of the three possible cases at a NEFP.
    The top plots illustrate the behavior of 
    the magnitude of the $r_i$ components while the bottom plots illustrate the behavior of the relative complex phase $\vartheta$ of the $r_i$ components. (a) Real-valued $\nu$: two different real eigenvectors. The analogous plots for equilibrium fixed points can be obtained by applying a reflection to the eigenvectors (both left and right) associated with one eigenvalue, corresponding to a shift of $\pi$ in $\vartheta$. (b) Exceptional point which occurs between real- and complex-valued $\nu$ regions: both eigenvectors coalesce into a single eigenvector. (c) Complex-valued $\nu$: two complex eigenvectors which are conjugate to one another.}
    \label{fig:eig}
\end{figure}

The above diagonalization motivates casting $\mathbf r$ in the new basis as  
\begin{equation}
    \mathbf{r} = r_a \mathbf{v}_a + r_b \mathbf{v}_b,
\end{equation}
with the coefficients $r_{a,b}$ determined by 
\begin{equation}
    r_{a,b} = \mathbf{u}_{a,b} \cdot \mathbf{r},
\end{equation}
or, alternatively, $(r_a, r_b)^T \equiv \mathcal{P}^{-1} (r_1, r_2)^T$. 
Indeed, we find that the scaling functions are naturally described in terms of these parameters: Eq.~(\ref{scaling functions}) can be now cast as
\begin{subequations}\label{scaling functions real}
\begin{equation}
\hat{C}_i = \tilde{C}_i\left(\frac{\omega}{|\mathbf{q}|^z},\frac{r_{a_R}}{|\mathbf{q}|^{1/\nu}}, \frac{r_{b_R}}{|r_{a_R}|^\phi}\right),
\end{equation}
\begin{equation}
\hat{\chi}_i = \tilde{\chi}_i\left(\frac{\omega}{|\mathbf{q}|^z},\frac{r_{a_R}}{|\mathbf{q}|^{1/\nu}}, \frac{r_{b_R}}{|r_{a_R}|^\phi}\right),
\end{equation}
\end{subequations}
where $\nu = \nu_a$ and $\phi = \nu_a/\nu_b$. 
The behavior of the eigenvectors in this region is illustrated in Fig.~\ref{fig:eig}(a). There is a qualitative change in these eigenvectors compared to the equilibrium case, which provides a geometrical reason for why exceptional points and complex $\nu$ are only possible with a nonreciprocal coupling. We can understand this by considering the role that $\mathcal{R}_{12},\mathcal{R}_{21}$ play in the structure of $\mathcal{P}^{-1}$ in Eq.~(\ref{eq:P}), $\mathcal{P}$, and the related eigenvectors. First, the sign of the second component must swap for either the left or the right eigenvectors due to the sign change in one of $\mathcal{R}_{12}, \mathcal{R}_{21}$ when considering an equilibrium model. Second, the (unnormalized) first component can be re-expressed as $\mathcal{R}_{11} - \mathcal{R}_{22} \pm \sqrt{(\mathcal{R}_{11} - \mathcal{R}_{22})^2 + 4 \mathcal{R}_{12} \mathcal{R}_{21}}$, so the sign flip also results in a sign flip for the first component of either both $\nu_a$ eigenvectors or both $\nu_b$ eigenvectors since the relative magnitude of the two terms changes. Hence, the eigenvector structure for the equilibrium flow can be obtained by applying a reflection (in $r_1$ or $r_2$) to one orthogonal pair of left/right eigenvectors from the nonequilibrium flow. This precludes an exceptional point, and thus complex $\nu$, because tuning $n_1, n_2$ can only rotate the orthogonal pairs, so it is impossible for both the left eigenvectors and the right eigenvectors to become simultaneously degenerate [cf.~exceptional point in Fig.~\ref{fig:eig}(b)]. 

\subsubsection{Complex eigenvalues}

Interestingly, the NEFPs can exhibit a new regime where   $\mathcal{R}_{12} \mathcal{R}_{21} < 0$, so the eigenvalues of the matrix $\cal R$ can assume complex values as $\nu^{-1} = {\nu'}^{-1} \pm i {\nu''}^{-1}$, hence the two eigenvalues $\nu_a=\nu_b^*$ are a pair of complex valued numbers; our motivation for this notation becomes clear shortly. This condition, and the emergence of a complex-conjugate pair, is realized if 
\begin{equation}
\left(\frac{\mathcal{R}_{11} - \mathcal{R}_{22} }{2}\right)^2 < - \mathcal{R}_{12} \mathcal{R}_{21}.
\end{equation} 
In this case,
it is convenient to consider a new basis defined by 
\begin{subequations}
\label{eq:tildeS}
\begin{equation}
\tilde{\mathcal{S}} = \tilde{\mathcal{M}}^{-1} \mathcal{R} \tilde{\mathcal{M}} = -\left( \begin{array}{cc}
\nu'^{-1} &  \nu''^{-1} \\
-\nu''^{-1} & \nu'^{-1} 
\end{array}\right),
\end{equation}
\begin{equation}
\tilde{\mathcal{M}}^{-1} = \left( \begin{array}{cc}
\mathcal{R}_{21} & \frac{\mathcal{R}_{22} - \mathcal{R}_{11}}{2} \\
0 & \nu''^{-1} 
\end{array}\right).
\end{equation}
\end{subequations}
This transformation brings the form of the matrix to the same form as the  $\mathbb{Z}_2 \times \mathbb{Z}_2$ ($n_1 = n_2 = 1$) model \cite{Young2020}. 
Based on the structure of $\tilde{\mathcal{M}}^{-1}$, we see that this corresponds to applying a skew transformation to the coordinate system. Similar to the case of real $\nu$, this corresponds to a basis transformation
\begin{subequations}
\begin{equation}
    \mathbf{r} = \tilde{s}_1 \tilde{\mathbf{v}}_1 + \tilde{s}_2 \tilde{\mathbf{v}}_2,
\end{equation}
\begin{equation}
    \tilde{s}_{1,2} = \tilde{\mathbf{u}}_{1,2} \cdot \mathbf{r},
\end{equation}
\end{subequations}
where the tildes denote that these are not right/left eigenvectors, but we still have $\tilde{\mathbf{u}}_i \cdot \tilde{\mathbf{v}}_j = \delta_{ij}$, where $\tilde{\mathbf{u}}$ ($\tilde{\mathbf{v}}$) are the row (column) vectors of $\tilde{\mathcal{M}}^{-1}$ ($\tilde{\mathcal{M}}$).

If we define a complex variable $\tilde{s} \equiv \tilde{s}_1 + i \tilde{s}_2$, then we have $\tilde{s}(l) = l^{-1/\nu} s_R$. In this basis, it is easy to consider the solutions for the flowing parameters $\tilde{s}_1(l), \tilde{s}_2(l)$ separately as
\begin{subequations}
\begin{equation}
\tilde{s}_1(l) = l^{-1/\nu'} \left[ \tilde{s}_{1_R} \cos \frac{\log l}{\nu''} + \tilde{s}_{2_R} \sin\frac{\log l}{\nu''} \right],
\end{equation}
\begin{equation}
\tilde{s}_2(l) = l^{-1/\nu'} \left[ \tilde{s}_{2_R} \cos\frac{\log l}{\nu''} - \tilde{s}_{1_R} \sin\frac{\log l}{\nu''} \right],
\end{equation}
\end{subequations}
from which we can get the corresponding equations for $r_i$ via $\tilde{\mathcal{M}}$. Note that the skewed basis we find here is connected to the fact that, as we shall later show, at the boundary between these two regions in $n_1, n_2$, the matrix $\mathcal{R}$ exhibits an exceptional point, and the eigenvectors become identical. Thus, as this line of exceptional points is approached, the flow equations must become more and more skewed until they only affect one direction. Given the structure of $\tilde{\mathcal{S}}$, the corresponding left/right eigenvectors are complex conjugate to one another, and we may readily identify the  eigenvectors of $\mathcal{R}$ as 
\begin{equation}
    \mathbf{v}_a = \mathbf{v}_b^* = \tilde{\mathbf{v}}_1 - i \tilde{\mathbf{v}}_2,  \quad \mathbf{u}_a = \mathbf{u}_b^* = \tilde{\mathbf{u}}_1 + i \tilde{\mathbf{u}}_2.
\end{equation}
Note that $\mathbf{v} = \tilde{s} \mathbf{v}_a + \tilde{s}^* \mathbf{v}_b = \tilde{s} \mathbf{v}_a + c.c.$ The behavior of $\mathbf{v}_a,\mathbf{v}_b$ in this region is illustrated in Fig.~\ref{fig:eig}(c). 

We can now express the scaling functions in Eq.~(\ref{scaling functions}) as
\begin{subequations}\label{scaling functions 2}
\begin{equation}
\hat{C}_i = \tilde{C}_i\left(\frac{\omega}{|\mathbf{q}|^z},\frac{|\tilde{s}_R|}{|\mathbf{q}|^{1/\nu'}}, P\Big(\frac{\log|\mathbf{q}|}{\nu''}-\arg( \tilde{s}_R) \Big)\right),
\end{equation}
\begin{equation}
\hat{\chi}_i = \tilde{\chi}_i\left(\frac{\omega}{|\mathbf{q}|^z},\frac{|\tilde{s}_R|}{|\mathbf{q}|^{1/\nu'}}, P\Big(\frac{\log|\mathbf{q}|}{\nu''}-\arg( \tilde{s}_R) \Big)\right),
\end{equation}
\end{subequations}
where 
\begin{equation}
|\tilde{s}_R|^2 =  \left(\mathcal{R}_{21} r_{1_R} +  \frac{\mathcal{R}_{22} - \mathcal{R}_{11}}{2}r_{2_R} \right)^2 + \nu''^{-2} r_{2_R}^2,
\end{equation}
while $\arg( \tilde{s}_R)$ denotes the polar angle in the $\tilde{s}_R$ plane and $P$ is a $2 \pi$-periodic function. 
Here, we have rewritten $\tilde{s}/l^{1/\nu'+i/\nu''}$ as a function of $|\tilde{s}|/l^{1/\nu'}$ and $e^{i(\log l)/\nu'' - i \arg( \tilde{s})}$. The first of these two is the usual scaling form that characterizes the correlation length, while the second results in a log-periodic function $P$, since the transformation $\log l \to \log l + 2 \pi \nu''$ leaves the exponential invariant.
The appearance of log-periodic functions corresponds to the emergence of a discrete scale invariance rather than the characteristic continuous scale invariance that is typically found at criticality. Thus a preferred scaling factor emerges as
\begin{equation}
    b_* = e^{2 \pi \nu''},
\end{equation}
rescaling by which, or any integer power thereof, leaves the system scale invariant, mimicking a fractal-like structure. However, in contrast to fractals where this emerges in the discrete microscopic structure, here the discrete scale invariance emerges only at long length scales in the continuum. Additionally, we note that the effect of a physical momentum cutoff $\Lambda$ determines the phase of the oscillations by entering the functions $P$ as a phase shift.

Similar phenomena appear to arise in earthquakes \cite{Sornette1995}, equilibrium models on fractals \cite{Karevski1996}, driven-dissipative quantum criticality \cite{Marino2016,Marino16-2}, the dynamics of strongly interacting nonequilibrium systems \cite{Maki2018}, the behavior of Efimov states \cite{Braaten2006,Efimov1970}, Berezinskii-Kosterlitz-Thouless phase transitions \cite{LeClair2003,Kaplan2009}, disordered classical \cite{Aharony75,Chen77,Khmelnitskii78,Weinrib83,Boyanovsky82} and quantum \cite{Hartnoll-16} systems, artifacts of position-space RG in the early development of renormalization group theory \cite{Jona-Lasinio75,Nauenberg75,Niemeijer:1976bs}, and several other systems \cite{Sornette1998}. 
However, the discrete scale invariance in the current work is distinct from these examples since it arises in a classical nonequilibrium model without disorder.
For a more detailed discussion of the examples discussed above, see Ref.~\cite{Young2020}.

As the upper critical dimension $d_c = 4$ is approached, the discrete scale invariant approaches a continuous one. Thus, in three dimensions, perturbative values at the NEFPs can result in a very large scaling factor $b_*$ (e.g., $b_* \sim 10^9$ for the $\mathbb{Z}_2 \times \mathbb{Z}_2$ model), although these are very sensitive to even small corrections beyond lowest-order perturbation theory. Additionally, higher harmonics in the periodic function $P$ can be significant in principle, which could be observed over smaller variations in the physical scale.

\subsubsection{Exceptional point}

As mentioned above, when moving from a region of real $\nu$ to complex $\nu$, an exceptional point in the flow occurs, leading to qualitatively different behavior. This occurs when 
\begin{equation}
\left(\frac{\mathcal{R}_{11} - \mathcal{R}_{22} }{2}\right)^2 + \mathcal{R}_{12} \mathcal{R}_{21} = 0.
\end{equation} 
In this case, the matrix $\mathcal{R}$ is no longer diagonalizable, possessing only a single eigenvalue and eigenvector, where we define $\mathbf{u}_{e}, \mathbf{v}_{e}$ as the left and right eigenvectors, respectively, which satisfy $\mathbf{u}_{e} \cdot \mathbf{v}_{e} = 0$. Instead of diagonalizing the matrix, we may instead express it in lower-triangular form according to
\begin{subequations}
\begin{equation}
    \check{\mathcal{S}} = \check{\mathcal{M}}^{-1} \mathcal{R} \check{\mathcal{M}} = \left(\begin{array}{cc}
    -\nu^{-1} & 0 \\
    1 & - \nu^{-1}
    \end{array}\right),
\end{equation}
\begin{equation}
    \check{\mathcal{M}}^{-1} = \left( \begin{array}{cc}
    \frac{\mathcal{R}_{11} - \mathcal{R}_{22}}{2} & \mathcal{R}_{12} \\
    1 & 0
    \end{array} \right).
\end{equation}
\end{subequations}

As before, we define new coordinates using the row (column) vectors $\check{\mathbf{u}}$ ($\check{\mathbf{v}}$) of $\check{\mathcal{M}}^{-1}$ ($\check{\mathcal{M}}$) with $\mathbf{r} = \check{s}_1 \check{\mathbf{v}}_1 + \check{s}_2 \check{\mathbf{v}}_2$ and coefficients $\check{s}_i = \check{\mathbf{u}}_i \cdot \mathbf{r}$ (where we have chosen the convention $\check{\mathbf{v}}_2 = \mathbf{v}_{e}, \check{\mathbf{u}}_1 = \mathbf{u}_{e}$), and we solve the new flow equation $l \frac{d}{dl} \check{\mathbf{s}} = \check{\mathcal{S}} \check{\mathbf{s}}$:
\begin{subequations}
\begin{equation}
    \check{s}_1(l) = l^{-1/\nu} \check{s}_{1_R},
\end{equation}
\begin{equation}
    \check{s}_2(l) = l^{-1/\nu} (\check{s}_{2_R} + \check{s}_{1_R} \log l),
    \label{eq:exceptionalflow}
\end{equation}
\end{subequations}
where the $l$ argument denotes that we are considering the flowing parameters.
Here, we see that the effect of the exceptional point is to add a logarithmic correction to the flow. This logarithmic correction is similar to the logarithmic corrections which emerge due to a parameter becoming marginal in the RG flow (e.g., the quartic couplings at $d_c=4$ \cite{Tauber2014}). However, in this case it is more correct to say that one direction is marginal relative to the other since $\check{s}_1$ and $\check{s}_2$ themselves are both relevant parameters.

The resulting scaling functions are now
\begin{subequations}\label{scalingexceptional}
\begin{equation}
\hat{C}_i = \tilde{C}_i \left(\frac{\omega}{|\mathbf{q}|^z},\frac{ \check{s}_{1_R}}{|\mathbf{q}|^{1/\nu}}, \frac{r_{e_R}}{ \check{s}_{1_R}} + \log |\mathbf{q}| \right),
\end{equation}
\begin{equation}
\hat{\chi}_i = \tilde{\chi}_i\left(\frac{\omega}{|\mathbf{q}|^z},\frac{ \check{s}_{1_R}}{|\mathbf{q}|^{1/\nu}}, \frac{r_{e_R}}{ \check{s}_{1_R}} + \log |\mathbf{q}| \right),
\end{equation}
\end{subequations}
where we have identified $r_e \equiv \check{s}_2$.
Note that we may write $\mathbf{v}_1 = a \mathbf{v}_{e} + b \mathbf{u}_{e}$, and thus $\mathbf{r} = (a \check{s}_1 + r_e) \mathbf{v}_e + b \check{s}_1 \mathbf{u}_e$ in a convenient orthogonal basis. We see that $\check{s}_1$ can be associated with the distance from the axis defined by $\mathbf{v}_e$, so $\check{s}_1$ determines the correlation length. However, even if $\check{s}_1$ is 0, we see that the logarithmic corrections in Eq.~(\ref{eq:exceptionalflow}) disappear, and the flow of $\check{s}_2$ is determined by $\nu$ alone, so the critical exponent for the correlation length is still $\nu$ with no logarithmic corrections.
The coalescense of $\mathbf{v}_{a,b}$ ($\mathbf{u}_{a,b}$) into a single eigenvector $\mathbf{v}_e$ ($\mathbf{u}_e$) in this region is illustrated in Fig.~\ref{fig:eig}(b). We see that, on the one hand, this behavior is qualitatively similar to the case of real $\nu$ via the third argument's dependence on $r_{e_R}/\check{s}_{1_R}$, corresponding to a crossover exponent $\phi = 1$ (up to logarithmic corrections) due to the existence of a single eigenvalue. On the other hand, it is qualitatively similar to the case of complex $\nu$ via the dependence on $\log |\mathbf{q}|$. Hence the fact that the exceptional point corresponds to a crossover between the two scenarios is reflected in the resulting scaling behavior.

\subsubsection{Phase diagram}

The NEFPs also exhibit qualitative differences in the structure of the phase diagram compared to their equilibrium counterparts. While bicritical points with only two ordered phases (either $\langle \bphi_1 \rangle \neq 0$ or $\langle \bphi_2 \rangle \neq 0$, but not both) are possible for the equilibrium fixed points, the NEFPs only exhibit tetracritical points involving four phases: The disordered phase where $\langle \bphi_1 \rangle = \langle \bphi_2 \rangle = 0$, the two singly-ordered phases present in the bicritical case, and a fourth doubly-ordered phase where $\langle \bphi_1 \rangle \neq 0 \neq \langle \bphi_2 \rangle$.
As we discussed earlier in the text, the stability of the equilibrium fixed points is related to that of the NEFPs. As a result, investigating the stability of the NEFPs can give insight into open questions about whether bicriticality or tetracriticality emerges in certain equilibrium systems.

\begin{figure*}
\centering
\includegraphics[scale=.3]{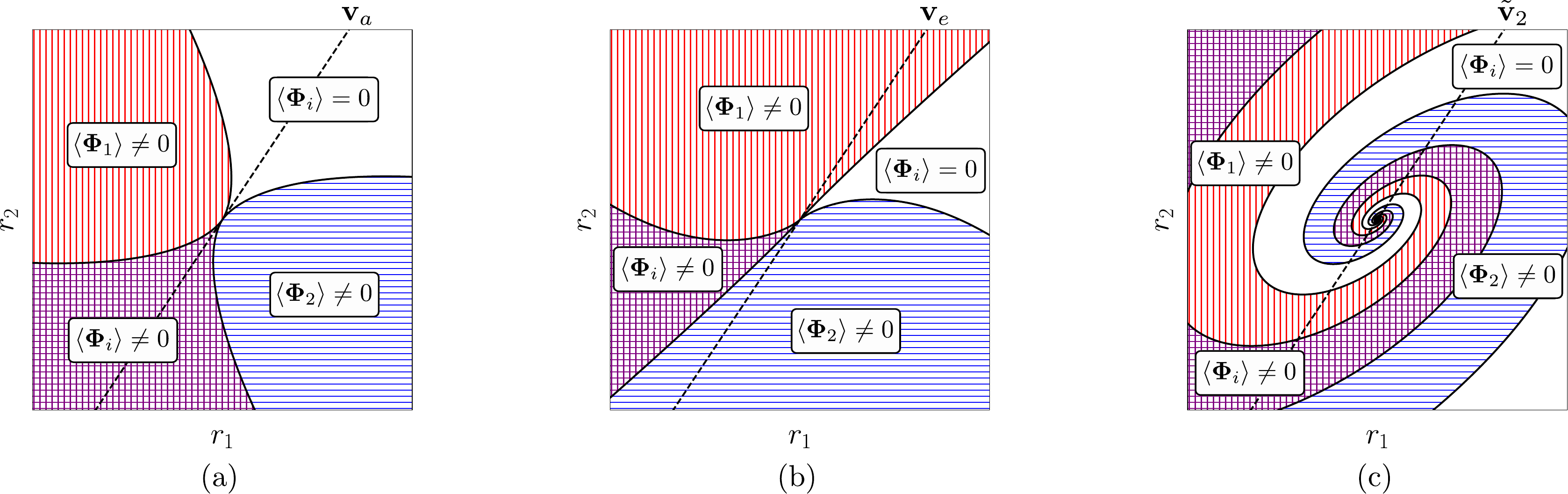}
\caption{Phase diagrams associated with the NEFPs. The white region indicates the disordered phase with $\langle \bphi_1 \rangle = \langle \bphi_2\rangle = 0$, the red vertically shaded region (blue horizontally shaded region) one of the singly-ordered phases with $\langle \bphi_1 \rangle \neq 0$ ($\langle \bphi_2 \rangle \neq 0$), and the purple square shaded region the doubly-ordered phase with both $\langle \bphi_1 \rangle, \langle \bphi_2 \rangle \neq 0$. The behavior of the phase diagram depends on the exponent $\nu$. (a) When $\nu$ is real, the phase diagram generally behaves like for the equilibrium coupled fixed points, where the phase boundaries approach the multicritical point tangentially to $\mathbf{v}_{a}$, the right eigenvector of $\mathcal{R}$ associated with the correlation length exponent $\nu$, like a polynomial. (b) In the transition between these two scenarios as a function of $n_1$, $n_2$, the flow of $r$ undergoes an exceptional point. The corresponding phase boundaries approach the multicritical point like $r \log r$. Note that the phase boundaries approach from the same side of 
$\mathbf{v}_e$, the only right eigenvector of $\mathcal{R}$, a consequence of the coalescence of the eigenvectors. (c) When $\nu$ is complex, the eigenvectors of $\mathcal{R}$ are as well. As a result, the phase diagram exhibits logarithmic spirals with discrete scale invariance. In general, these spirals are skewed along a vector $\tilde{\mathbf{v}}_2$, which is defined by a matrix $\tilde{\mathcal{M}}$ [cf.~Eq.~(\ref{eq:tildeS})] that transforms the skewed spirals into isotropic spirals via a basis change. 
\label{fig:phasediag}}
\end{figure*}

The behavior of the phase boundaries is determined by the scaling behavior of $r_i$. In light of this, we expect three types of behavior of the phase boundaries. To identify this behavior, we consider the scaling functions in Eq.~(\ref{scaling functions}) in the limit $\omega, \bq \to 0$.
In the region of real $\nu$, the simplest case most similar to equilibrium fixed points, we expect the phase boundaries to approach the multicritical point tangentially to $\mathbf{v}_a$ 
with power-law scaling $r_{b_R} \propto |r_{a_R}|^\phi$. This is illustrated in Fig.~\ref{fig:phasediag}(a).

In the case of complex $\nu$, 
in the $\omega, \bq \to 0$ limit the scaling functions are determined purely by the $2\pi$-periodic function of $\frac{\nu'}{\nu''} \log\left(|s_R|\right)-\arg( s_R)$, which can be seen by switching the momentum scale in Eq.~(\ref{scaling functions 2}) from $|\bq|$ to $s_R$. The phase boundaries, characterized by a divergence of correlations, thus occurs at fixed values of the periodic function, and the shape of the phase boundaries is set by
\begin{equation}
\label{eq:spiral}
\frac{\nu'}{\nu''} \log(|s_R|)-\arg(s_R) = \mbox{const},
\end{equation}
which defines a spiral in $s_R$. Recall that in general, $s_R$ describes a skewed basis in terms of $r_{i_R}$, and so the corresponding spirals are skewed. In the $\tilde{s}$ basis, these spirals are perfectly isotropic, so in order to determine the phase diagram in terms of $r_i$, we can apply a skew to these spirals. The resulting phase diagram in $r_i$ is acquired in this fashion and illustrated in Fig.~\ref{fig:phasediag}(c).

Similar to the discrete scale invariance in the correlation and response functions, the perturbative values of the exponents at $\epsilon = 1$ generally require large variations in $r$ to observe a full spiral, although these scales are very sensitive to small corrections. However, partial spirals can still be observed for reasonable scales. Since the phase boundaries all spiral in the same direction, it may be possible to distinguish them from equilibrium critical points even for very weak spiraling. Finally, when two NEFPs are present, they can be distinguished via the form of the phase diagram. For example, if both have complex $\nu$, then they can be distinguished by the direction of the spirals.

The phase diagram at an exceptional point in the flow of $\mathcal{R}$ also leads to logarithmic corrections, although not in the form of spirals. 
Fixing the second and third arguments of Eq.~(\ref{scalingexceptional}) to $c_1, c_2$ and eliminating the momentum scale in favor of $r_{e_R}$, we have 
\begin{equation}
    r_{e_R} = \left(c_2 - \nu \log \frac{\check{s}_{1_R}}{c_1} \right) \check{s}_{1_R},
\end{equation}
which corresponds to the phase boundaries approaching $\mathbf{v}_e$ 
logarithmically. Again, this illustrates qualitative similarities to both cases of real and complex $\nu$. If we ignore the logarithmic contribution, it corresponds to linear phase boundaries with $\phi = 1$ and real $\nu$. On the other hand, the equation defining the phase boundary for the exceptional point is very similar in form to the case of complex $\nu$ [cf.~Eq.~(\ref{eq:spiral})].
Additionally, unlike for the usual case of real $\nu$, the phase boundaries for the disordered phase approach $\mathbf{v}_e$ from the same side rather than opposite sides.
The corresponding phase diagram at an exceptional point is illustrated in Fig.~\ref{fig:phasediag}(b).

\subsection{Relaxational behavior}
Finally, we remark on the relaxational behavior in the doubly-ordered phase. For the equilibrium fixed points, it is equivalent to overdamped dynamics with no oscillations, which is captured by the real-valued poles of the propagators. Indeed, even for emergent equilibrium criticality in nonequilibrium systems, oscillatory effects tend towards zero as criticality is approached \cite{Tauber2014a,Torre2013}. As a result, the poles rapidly become real-valued near criticality, corresponding to overdamped dynamics. However, for some regions of the doubly-ordered phase, the pole can take on complex values arbitrarily close to the multicritical point, leading to underdamped dynamics in the form of oscillatory relaxation to the steady state.
This gives rise to a new universal quantity, which is the maximum angle that the pole can take compared to overdamped dynamics. Physically, this corresponds the maximum ratio of the frequency of the oscillations to the rate of the exponential decay near the steady state. 

To understand the origin of this behavior, it is helpful to utilize a mean-field picture. In the doubly-ordered phase, the two order parameters realize nonzero values, and for each order parameter we can identify an ordered mode $\varphi_i$ and $n_i-1$ corresponding Goldstone modes. Since the Goldstone modes are massless, we are primarily interested in the behavior associated with the relaxation of the ordered mode. Hence, we consider the linearized dynamics of $\varphi_i$ about the mean-field steady-state values $M_i$. Making the change of variables $\varphi_i \to \varphi_i + M_i$ and defining $\bm{\varphi} = (\varphi_1, \varphi_2)^T$, we find
\begin{subequations}
\label{eq:relaxeig}
\begin{equation}
    \partial_t \bm{\varphi} \approx R \bm{\varphi},
\end{equation}
\begin{equation}
    R = -\left( \begin{array}{cc}
    2 u_1 M_1^2 & 2 u_{12} M_1 M_2 \\
    2 \sigma u_{12} M_1 M_2 & 2 u_2 M_2^2 
    \end{array}\right),
\end{equation}
\end{subequations}
where $R$ is the matrix that defines the relaxational behavior. Hence, if the eigenvalues of $R$ are complex, the system exhibits underdamped oscillations. This occurs when $-4\sigma u_{12}^2 M_1^2 M_2^2 > (u_1 M_1^2 - u_2 M_2^2)^2$, so we see complex eigenvalues are only possible when $\sigma \neq 0$, indicating the underlying nonequilibrium origin of this behavior. Under an appropriate rescaling of $\varphi_1, \varphi_2$ such that the diagonal entries are equal, we see that the dynamics map to a damped oscillator, where the off-diagonal nonreciprocal elements define the frequency and the diagonal elements define the damping.

Near criticality, the maximal ratio (as a function of $r_1, r_2$) of the imaginary to real parts of the eigenvalues of Eq.~(\ref{eq:relaxeig}) approaches a universal value. We describe this ratio via the angle that the complex eigenvalues form with respect to the real axis. However, given the scaling freedom we utilized previously as well as the redefinition of $u$ couplings in terms of $\tilde{u}$ couplings, this limit requires care. The detailed derivation of $\theta^*$ via the action is presented in Appendix ~\ref{sec:relax}. 
The value of the corresponding maximal complex angle is defined by
\begin{equation}
\theta^* = \arctan \left( \frac{\sqrt{- \sigma v_R^*} \tilde{u}_{12_R}^*}{\sqrt{\tilde{u}_{1_R}^* \tilde{u}_{2_R}^*}} \right).
\end{equation}
Qualitatively, this expression is best reproduced in mean-field theory by shifting all features of the nonreciprocity from $T_1, T_2$ (by rescaling such that $T_1 = T_2$) to $\tilde{g}_{12} \equiv v \tilde{u}_{12},~\tilde{g}_{21} \equiv \sigma \tilde{u}_{12}$, $\tilde{g}_i \equiv \tilde{u}_i$, i.e., by treating the strength of the noise for each field on equal grounds. In this case,
\begin{equation}
\theta^* = \arctan \left( \sqrt{\frac{ -\tilde{g}_{12_R}^* \tilde{g}_{21_R}^*}{\tilde{g}_{1_R}^* \tilde{g}_{2_R}^*}} \right).
\end{equation} 
Note that there are underdamped oscillations only in the doubly-ordered phase away from the phase boundaries (but arbitrarily close to the multicritical point). 
Near the phase boundaries, the dynamics become overdamped  due to the restoration of effective equilibrium criticality. From a mean-field theory perspective, we see that as one of the magnetizations $M_i$ goes to 0, thereby approaching a phase boundary, $R$ can no longer possess complex eigenvalues, so the underdamped dynamics can disappear even before criticality from one field dominates. This is analogous to the crossover from underdamped dynamics to overdamped dynamics, where critical damping occurs between the two. For a damped harmonic oscillator $x(t)$, the damping ratio $\zeta_\text{d}$ describes the damping strength relative to the harmonic oscillator frequency $\omega_\text{ho}$ via $\ddot{x} + 2 \zeta_\text{d} \omega_\text{ho} \dot{x} + \omega_\text{ho}^2 x = 0$, where $\zeta_\text{d} < 1$ corresponds to underdamping. Thus, we may relate $\theta^*$ to a minimal damping ratio $\zeta_{\text{d},\text{min}} = \cos \theta^*$.

\subsection{Critical exponents}\label{sec:critical_exp}
\label{subsec:critexp}
\subsubsection{Equal number of components}

We now proceed to identify the critical exponents and $\theta^*$ for the case of $n_1 = n_2 = n$, which are the same for both fields for all $n$ to lowest non-trivial order:
\begin{subequations}
\label{eq:equalexp}
\begin{equation}
\eta = \frac{1+2n + 12 (n-4) \log(4/3)}{12(2+n)^2} \epsilon^2,
\end{equation}
\begin{equation}
\eta' = \frac{1+2n}{12(2+n)^2} \epsilon^2,
\end{equation}
\begin{equation}
z - 2 = (6 \log(4/3) -1 ) \eta'
\end{equation}
\begin{equation}
\nu^{-1} = \nu'^{-1} + i \nu''^{-1} = \left( 2 - \frac{\epsilon}{2} \right) \pm i \frac{n \sqrt{4/n - 1}}{2(n+2)} \epsilon,
\end{equation}
\begin{equation}
\theta^* = \arctan \left(\sqrt{4/n-1} \right).
\end{equation}
\end{subequations}
There are a variety of interesting features to note concerning the behavior of these exponents at this order: (1) The relationship between $z$ and $\eta'$ is the same as in equilibrium and independent of $n$. (2) The exponent $\eta$ changes sign at $n \approx 2.35$. However, for all $n < 4$, $\gamma_T = \eta - \eta' < 0$, which indicates that the effective temperatures always get ``hotter'' at large length scales. (3) The real part of $\nu^{-1}$ does not depend on $n$. This is a consequence of the fact that at this order, $\tilde{u}_{1_R}^* = \tilde{u}_{2_R}^*$, which likely changes at higher orders. (4) The maximal value of $\nu''^{-1}$ is realized at $n=1$. (5) Although $u_{12_R}^*$ is divergent as $n \to 0$, the critical exponents themselves do not diverge, and $\nu''^{-1}$ goes to zero. (6) The value of $\theta^*$ takes on simple values: $\pi/2, \pi/3, \pi/4, \pi/6$ for $n=0, 1, 2, 3$, respectively. Interestingly, the $n \to 0$ limit indicates that the damping can go to 0, leaving purely oscillatory dynamics, which may have connections to critical exceptional points or nonequilibrium rotating phases \cite{Zelle2023}.

\subsubsection{Arbitrary number of field components}

\begin{figure*}
\centering
\includegraphics[scale=1]{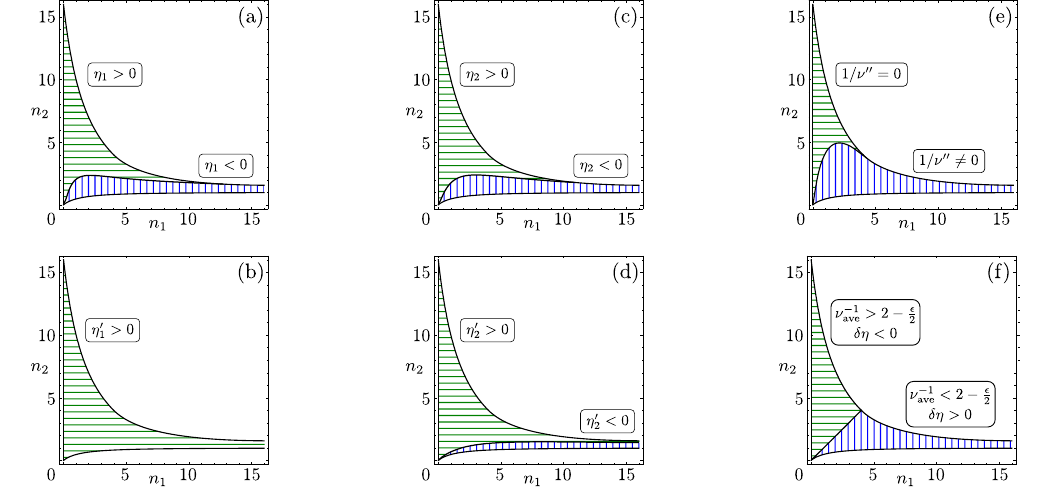}
\caption{Scaling behavior for different values of $n_1, n_2$ for the NEFPs with $\tilde{u}_{12_R} < 0$. Equivalent figures exist for $\tilde{u}_{12_R} > 0$ but with the field indices swapped. (a) Sign of $\eta_1$. The anomalous dimension $\eta_1$ can be both positive and negative, taking negative values for smaller $n_2$. (b) Sign of $\eta_1'$. The anomalous dimension $\eta_1'$ can only take positive values, unlike the other three anomalous dimensions. (c) Sign of $\eta_2$. Like $\eta_1$, the anomalous dimension $\eta_2$ can be both positive and negative, taking negative values for smaller $n_2$. Although the regions of negative $\eta_1$ and $\eta_2$ are similar, they are distinct, so the signs are not necessarily the same. (d) Sign of $\eta_2'$. Unlike $\eta_1'$, the anomalous dimension $\eta_2'$ can be both positive and negative, taking negative values for smaller $n_2$. (e) Regions of real- and complex-valued $\nu^{-1}$. The region of complex-valued $\nu^{-1}$ is qualitatively similar to the regions where negative anomalous dimensions can occur, although it extends to larger $n_2$. At the boundary between real and complex $\nu^{-1}$, there is an exceptional point in the flow of $r_i$. (f) Sign of $\delta \eta = \eta_1 - \eta_2 = \eta_1' - \eta_2'$ and behavior of $\nu^{-1}_{\text{ave}}\equiv (\nu_a^{-1} + \nu_b^{-1})/2$. The change in behavior at this order occurs at $n_1 = n_2$ since $v_R^* = w_R^* = 1$ here, which causes the flow for the two fields to become equivalent.  \label{fig:scaling}}
\end{figure*}

Next, we consider how the behavior of the critical exponents changes for arbitrary $n_1$ and $n_2$. In particular, we  focus on some of the most distinguishing features of the fixed points: the signs of $\eta_i, \eta_i'$; the asymmetry of the anomalous scaling dimensions, defined as $\delta \eta \equiv \eta_1 - \eta_2 = \eta_1' -\eta_2'$; the imaginary component of $\nu^{-1}$; the scaling of the effective temperatures ($\gamma_T$); and the relaxation angle $\theta^*$. Aside from $\theta^*$, which always takes on nonzero values for the NEFPs, these results are illustrated in Fig.~\ref{fig:scaling}. Numerical values of all fixed point values, critical exponents, and features discussed to lowest nontrivial order in $\epsilon$ are presented in Tables~\ref{tab:fixed}, \ref{tab:exp} of Appendix~\ref{sec:tables}. In the following discussion, we shall consider the behavior of only the stable fixed point with $\tilde{u}_{12_R}^* < 0$.

For both $\eta_1,\eta_2$, the region where these exponents realize negative values are largely present around the area associated with $n_1 \to \infty$. While the regions of negative $\eta_1, \eta_2$ are similar and appear for $n_2 < 3$, they  differ slightly for smaller values of $n_1$. While $\eta_1'$ remains positive everywhere, there is a small region of negative $\eta_2'$ when $n_2 < 2$ that is also associated with the $n_1 \to \infty$ limit. 
For $n_1 = n_2$ at this order, we find $\delta \eta = 0$ and $\nu_\text{ave}^{-1} \equiv (\nu_a^{-1} + \nu_b^{-1})/2 = 2 - \frac{\epsilon}{2}$. For $\delta \eta$, this is because $v_R^* = w_R^* = 1$, and the flows for the two fields become equivalent (aside from the asymmetric part of $\mathcal{R}$), although this is likely to change at higher orders. When $n_1 > n_2$, $ \delta \eta > 0$ and $\nu_\text{ave}^{-1} < 2-\epsilon/2$, while when $n_1 < n_2$, $ \delta \eta < 0$ and $\nu_\text{ave}^{-1} > 2-\epsilon/2$. Additionally, we see that the region of complex $\nu$ is qualitatively similar to the regions where the anomalous dimensions become negative, although it extends as high as $n_2 = 5$. Finally, we remark that the NEFP which persists in the $n_1 \to \infty$ limit exhibits 3 negative anomalous dimensions as well as complex-valued $\nu$.

For all values of $n_1,n_2$, we find $\gamma_T < 0$, indicating that the effective temperature always gets ``hotter''; whether this is generally the case for nonreciprocal models remains an open question. In the $n_1 \to \infty$ region, $\gamma_T$ diverges to $-\infty$ at the bottom boundary and approaches $\gamma_T \approx -0.1618 \epsilon^2$ at the top boundary. 

The relaxation angle $\theta^*$ realizes all values between 0 and $\pi/2$ and is generically nonzero everywhere, which is a consequence of the fact that $\sqrt{v_R^*} \tilde{u}_{12_R}^*/\sqrt{\tilde{u}_{1_R}^* \tilde{u}_{2_R}^*} \neq 0$, which determines $\theta^*$ (see Appendix \ref{sec:relax}). 
This emphasizes the fact that, although both nonzero $\theta^*$ and complex $\nu$ require $\sigma = - 1$, $\theta^*$ is generically nonzero while $\nu$ can be either complex or real. 
The generically nonzero value of $\theta^*$ is a consequence of the fact that the relaxation depends on $\langle \bphi_i\rangle$, which can be varied. In contrast, $\nu$ solely depends on the flow of $r_i$, which is entirely determined by the fixed point values of the couplings. Additionally, the region of $n_1 \to \infty$ varies from approximately $\theta^* \approx \pi/3$ at the bottom boundary to $\theta^* \approx \pi/4$ at the top boundary. 

For the stable fixed point in the limit of $n_1 \to \infty$ for $n_2 = 1$, we find the critical exponents
\begin{equation}
\begin{gathered}
\nu^{-1} \approx 2 - (7.259 \pm 6.913i)\epsilon, \\
\eta_1 \approx -10.98 \epsilon^2, \qquad \eta_2 \approx -29.92 \epsilon^2, \\
\eta_1' \approx 6.721 \epsilon^2, \qquad \eta_2' \approx -12.22 \epsilon^2, \\
\gamma_T \approx -17.70 \epsilon^2, \qquad z -2 \approx 8.957 \epsilon^2, \\
\theta^* \approx 0.6532 \frac{\pi}{2}.
\end{gathered}
\end{equation}
Given these values, particularly for $\nu^{-1}$, we can see that this fixed point can no longer be considered perturbative for $\epsilon = 1$, hinting at the breakdown of perturbation theory in this limit. 

\subsection{Hyperscaling relations}
\label{subsec:hyper}
Next, let us investigate how the above critical phenomena can modify the usual hyperscaling relations. We shall focus on the case where $\nu$ is complex and we have four anomalous dimensions $\eta_i, \eta_i'$. The case of real $\nu$ simply corresponds to replacing all instances of $\nu'$ with $\nu$ and removing any functions which depend on $\nu''$. This holds for the exceptional points as well. 
Here, we consider the scaling of the order parameter and magnetic susceptibility
	\begin{equation}
	    \label{hyperdef}
	    \langle \bphi_i \rangle \propto |r|^{\beta_i}, \hspace{1cm}  \frac{\partial \langle \bphi_i \rangle}{\partial h_i}  \propto |r|^{-\gamma_i}.
	\end{equation}
For the equilibrium model, these are related to the other exponents via
	\begin{equation}
	    \beta_i = \nu_i(d-2+\eta_i)/2, \hspace{1cm} \gamma_i = \nu_i(2-\eta_i).
	\end{equation}
	\begin{subequations}
	\begin{equation}
	    \langle \bphi_i^2 \rangle = \lim_{|\bx|,t \to \infty} C_i(\mathbf{x},t,\{r_j\}),
    \end{equation} 
    \begin{equation}
	    \frac{\partial \langle \bphi_i \rangle}{\partial h_i} \bigg|_{h_i = 0} = \lim_{\mathbf{q},\omega \to 0} \chi_i(\mathbf{q}, \omega,\{r_j\}),
	\end{equation}
	\end{subequations}
	where $h_i$ is the field conjugate to $\bphi_i$, and that in the ordered phases, $\langle \bphi_i^2 \rangle$ and $\langle \bphi_i \rangle^2$  scale the same way. 
 To identify the hyperscaling relations, we take $|r_R|$ to define the small momentum scale and consider the limits $\omega,|\bq| \to 0$ for the correlation function and $t, |\mathbf{x}| \to 0$ for the response function, resulting in the following scaling behavior
	\begin{subequations}
	\label{periodicbg}
	\begin{equation}
	    \langle \bphi_i^2 \rangle \propto |r_R|^{\nu'(d-2+\eta_i)} P_{C_i}\left(\frac{\nu'}{\nu''} \log(|r_R|)-\arg(r_R)\right),
	\end{equation}
	\begin{equation}
	    \frac{\partial \langle \bphi_i \rangle}{\partial h_i} \propto |r_R|^{\nu'(2-\eta_i')}P_{\chi_i}\left(\frac{\nu'}{\nu''} \log(|r_R|)-\arg( r_R) \right),
	\end{equation}
	\end{subequations}
where $P_{C_i}, P_{\chi_i}$ are $2\pi$-periodic functions. 
For convenience, we fix the argument of the periodic functions, which correspond to the presence of discrete scale invariance in the phase diagram, thereby restricting the discrete scale invariance to $\nu$. This results in the generalized hyperscaling relations.
	\begin{equation}
	    \beta_i = \nu'(d-2+\eta_i)/2, \hspace{1cm} \gamma_i = \nu'(2-\eta_i'),
	\end{equation}
which depend only on the real part $\nu'$. 
Similar analysis can be used to relate the critical exponent characterizing the order parameter's dependence on the magnetic field
\begin{equation}
\langle \bphi_i\rangle \propto |h|^{1/\delta_i},
\end{equation}
as
\begin{equation}
\frac{\delta_i - 1}{\delta_i+1} \tilde{d} = 2 - \eta_i', \qquad \tilde{d} = d + \gamma_T,
\end{equation}
which involves a mix of $\eta_i, \eta_i'$ since $\gamma_T = \eta_i - \eta_i'$. Interestingly, the scale-dependent temperature leads to a reduced effective dimension $\tilde{d}$ in this expression.

\section{One-way coupling}
\label{sec:halfcouple}
Finally, let us consider the case where only one of the two fields is affected by the other, corresponding to the $g_{12} g_{21} = 0$, $\sigma = 0$, or $v = 0,\infty$ subspaces, all of which are equivalent. In Ref.~\cite{Young2020}, which studied the case $n_1 = n_2 = 1$, this scenario was not considered due to the non-perturbative behavior that emerged when $n_1 = n_2 = 1$. In the more general case of $n_1\ne n_2$, this is not always the case, and we find
a new set of fixed points, although we further find that non-perturbative behavior persists for $n_1 = n_2$. In addition to elucidating the behavior of these new fixed points for $n_1 \neq n_2$, we can better understand the resulting behavior at $n_1 = n_2$ by investigating potential transient phenomena through our perturbative treatment. Similar to generic nonreciprocal coupling, feedback in both classical and quantum \cite{Gardiner1993,Carmichael1993,Metelmann2015} systems or dissipative gauge symmetries in quantum systems \cite{Wang2023} can be utilized to realize one-way coupling between two order parameters.

This section is arranged as follows. In Sec.~\ref{subsec:OneWayBeta}, we discuss the modified beta functions in this subspace and identify the corresponding fixed points. In Sec.~\ref{subsec:OneWayExp}, we present the resulting critical exponents for these fixed points. In Sec.~\ref{subsec:OneWayMarginal}, we investigate the behavior of these fixed points at $n_1 = n_2$, where they simultaneously become non-perturbative and marginal, identifying potential transient critical phenomena and exponents. 

\subsection{Beta functions, fixed points, and stability}
\label{subsec:OneWayBeta}

First, we consider the behavior of the beta functions. While there were five parameters whose fixed point values needed to be identified in the case of full coupling, since one of the original coupling terms has been set to 0, we may anticipate that only four beta functions should be considered. In particular, we see that only one term remains in $\beta_v$ [cf.~Eq.~(\ref{betav})] when $\sigma = 0$, indicating that there can be no finite nonzero value for $v_R^*$. This is a natural result given that this subspace is defined by $v = 0, \infty$. In order to remove the $v$ dependence in the remaining four beta functions, we absorb $v$ into $\tilde{u}_{12}$ via $\tilde{g}_{12} \equiv v \tilde{u}_{12} = u_{12} T_2/D_1 D_2$. The resulting beta functions are 
\begin{subequations}
\begin{equation}
\beta_{\tilde{u}_1} = \tilde{u}_{1_R} [-\epsilon + (n_1+8) \tilde{u}_{1_R}],
\end{equation}
\begin{equation}
\beta_{\tilde{u}_2} = \tilde{u}_{2_R} [-\epsilon + (n_2+8) \tilde{u}_{2_R}],
\end{equation}
\begin{multline}
\beta_{\tilde{g}_{12}} = \tilde{g}_{12_R}\bigg[ -\epsilon + 4 \frac{1}{1+w_R}\tilde{g}_{12_R} \\
~+(n_1 +2)\tilde{u}_{1_R} + (n_2+2) \tilde{u}_{2_R} \bigg],
\end{multline}
\begin{multline}
\beta_w = -w_R\Big\{C'\big[(n_1+2)\tilde{u}_{1_R}^2 - (n_2+2)\tilde{u}_{2_R}^2\big]  \\ 
~+ n_2 {\tilde{g}}^2_{12_R}  G(w_R)\Big\}.
\end{multline}
\end{subequations}

There are several features worth noting in the above beta functions as compared to those of the fully-coupled model. First, as one would expect, the beta function for $\tilde{u}_2$ becomes independent of the other three parameters as $\bphi_2$ is decoupled from $\bphi_1$. Interestingly, this is also the case for $\tilde{u}_1$, meaning that, at this order, both $u_i$ take on their equilibrium value. In the case of $n_1=n_2$, this has important implications on the flow of $w$, where the first pair of terms in $\beta_w$ cancel, leading to a flow which is determined by only one term, which precludes strong dynamic scaling. 

We may readily identify the fixed point values of the coupling terms as functions of $w_R^*$ as
\begin{subequations}
\begin{equation}
    \tilde{u}_{i_R}^* = \frac{\epsilon}{n_i+8},
\end{equation}
\begin{equation}
    \tilde{g}^*_{12_R} = \frac{1+ w_R^*}{4} \left(1 - \frac{n_1+2}{n_1+8} - \frac{n_2+2}{n_2+8} \right) \epsilon.
    \label{betauv}
\end{equation}
\end{subequations}
Using these values, the fixed point value of $w_R^*$ can in general be found self-consistently according to
\begin{widetext}
\begin{equation}
    0 = -w_R^* \left\{ C' \left[\frac{n_1+2}{(n_1+8)^2} - \frac{n_2+2}{(n_2+8)^2}\right] + n_2 \left[ \frac{1+w_R^*}{4} \left( 1 - \frac{n_1+2}{n_1+8} -\frac{n_2+2}{n_2+8}\right) \right]^2 G(w_R^*) \right\}.
\end{equation}
\end{widetext}
We find that a new fixed point exists when $n_2 > n_1$. While solutions exist for $n_2 < n_1$, we find $w_R^* < 0$, which results in unbounded dynamics and is thus not physically relevant. 

\begin{figure}[t!]
\centering
\includegraphics[scale=1]{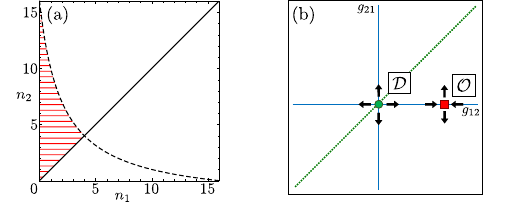}
\caption{Stable fixed point behavior as a function of $n_1, n_2$ for the one-way coupled fixed points when $\bphi_2$ is the independent field (i.e., $g_{21} = 0$). 
(a) Red shading denotes the region where the fixed point is stable, which is enclosed by two boundaries. Boundary I from the fully-coupled fixed points shows up again as the dashed line, cf.~Eq.~(\ref{eq:boundary1}), although this may not hold at higher-orders unlike the fully coupled model, while the second boundary occurs at $n_1 = n_2$ and is associated with a change in the sign of $w_R^*$ and non-perturbative behavior.
(b) Qualitative illustration of the RG flow diagram in the $g_{12}$-$g_{21}$ plane, although the full flow occurs in a five-dimensional space. Arrows indicate the stability of the $\sigma = 0$ fixed points in different directions.
$\mathcal{O}$ denotes the one-way coupled fixed points and $\mathcal{D}$ the decoupled fixed points; other fixed points (e.g., $\sigma \neq 0$) are not shown.
\label{HFCPplot}}
\end{figure}

With this in mind, let us consider more fully the stability of this new type of fixed point to lowest order. Given that $\tilde{u}_i$ are not affected by $\tilde{g}_{12}, w$, the stability matrix takes a block-triangular form. This allows us to consider each of the coupling terms separately. For $\tilde{u}_i$, this just reduces to the usual $O(n)$ stability, and thus we are only concerned with $\tilde{g}_{12}$, whose stability is determined by
\begin{equation}
    \partial_{\tilde{g}_{12_R}} \beta_{\tilde{g}_{12}} = \beta_{\tilde{g}_{12}}/\tilde{g}_{12_R}  + \left(1 - \frac{n_1+2}{n_1+8} - \frac{n_2+2}{n_2+8} \right) \epsilon. 
\end{equation}
At the fixed point, the first term is 0, and thus the stability is entirely determined by the sign of the second term, corresponding to the lower region below the dashed line in Fig.~\ref{HFCPplot}(a), where only the red shaded region has $w_R^* > 0$.

Finally, we investigate the stability of the stable fixed point to perturbations towards full coupling. For convenience, we  work in terms of $v$ with $\sigma \neq 0$, where the above fixed points occur at $v_R^* = \infty$ and a perturbation towards full coupling corresponds to that towards a finite value of $v_R$, allowing us to utilize the beta equations in Eq.~(\ref{eq:beta}). To simplify this analysis, we  consider the stability in terms of $V \equiv 1/v$. While we could instead consider perturbations in the coupling term directly, this would require the full two-loop corrections to the coupling terms, which $v, V$ allow us to avoid as usual. 
Here, we rely on the fact that $\beta_V$ vanishes  with $V_R$ since $V_R = 0$ is a fixed point. As a result,  when $V_R \to 0$, $\partial_{s_{b_R}} \beta_V = 0$ for all $s_{b_R} \neq V_R$, so the stability matrix becomes block triangular and the stability of $V_R$ depends entirely on $\beta_V = - V_R^2 \beta_v$. Hence, we consider
\begin{subequations}
\begin{equation}
    \beta_V =  n_2 V_R F(w_R) \tilde{g}_{12_R}^2 [1 - \sigma V_R] \left[1 + \sigma \frac{n_1}{n_2}\frac{F(w_R^{-1})}{F(w_R)} V_R\right],
    \end{equation}
    \begin{equation}
    \lim_{V_R \to 0} \partial_{V_R} \beta_V  = n_2 {\tilde{g}}^2_{12_R} F(w_R),
\end{equation}
\end{subequations}
and we see that these fixed points are unstable in this direction since $F(w_R) < 0$ for $w_R>0$. 
This means that these fixed points cannot describe systems in which both fields are coupled to one another due to this instability, and the system will flow to a fully-coupled or decoupled fixed point instead. It is only when the system itself exhibits one-way coupling that these new fixed points describe the critical point. 

\subsection{Critical exponents}
\label{subsec:OneWayExp}

Next, we turn our focus to the critical behavior of the one-way coupled fixed points. Since the second field is independent of the first, the criticality associated purely with $\bphi_2$ is the same as the usual equilibrium $O(n)$ model for $n = n_2$, which we  report for completeness:
\begin{subequations}
\begin{equation}
    \eta_{O(n)} = \eta'_{O(n)} = \frac{n+ 2}{2 (n+8)^2} \epsilon^2,
\end{equation}
\begin{equation}
    z_{O(n)} = 2 + \eta_{O(n)} \left( 6 \log 4/3 - 1 \right).
\end{equation}
\end{subequations}
The critical exponents for the one-way coupled fixed points thus take the form
\begin{subequations}
\begin{equation}
    \eta_1 = \eta_{O(n_1)} + n_2 [F(w_R^*) + G_D(w_R^*)]  {\tilde{g}}^{*2}_{12_R},
\end{equation}
\begin{equation}
    \eta_1' = \eta_{O(n_1)} + n_2 G_D(w_R^*) {\tilde{g}}^{*2}_{12_R},
\end{equation}
\begin{equation}
    \eta_2 = \eta_2' = \eta_{O(n_2)},
\end{equation}
\begin{equation}
    z_1 = z_{O(n_1)} + n_2 G(w_R^*) {\tilde{g}}^{*2}_{12_R} = z_{O(n_2)} = z_2,
\end{equation}
\end{subequations}
where we have expressed the nonequilibrium exponents relative to the decoupled equilibrium exponents, using $n_1$ to calculate the corresponding equilibrium exponent. Note that 
$\eta'_1 \neq \eta'_2$, so $z_1 \neq 2 + \eta'_1 \left( 6 \log 4/3 - 1 \right)$ at lowest non-trivial order. 
Additionally, we see that the dynamical critical exponent is defined by the independent field.
This is a consequence of the fact that the fixed points exhibit strong dynamic scaling. Since the second field is independent of the first, its scaling cannot be modified from equilibrium, so the first field can only match it in the region of multicriticality.
Furthermore, we naturally have $\gamma_{T_1} \neq \gamma_{T_2}=0$, and thus $\eta_1' - \eta_1 \neq \eta_2' - \eta_2 =0$, unlike the fully-coupled fixed points. In Table \ref{tab:oneway}, we report the stable fixed point values of $g_{12_R}^*, w_R^*$ as well as the critical exponents $\eta_1 - \eta_{O(n_1)}, \eta_1' - \eta_{O(n_1)}, \gamma_{T_1}$.

Finally, let us turn our attention to the behavior of the flowing parameters $r_i(l)$, which evolve according to
\begin{subequations}
\label{eq:Rflow}
\begin{equation}
l \frac{d}{dl} \left( \begin{array}{c}
r_1(l) \\
r_2(l)
\end{array} \right) = \left(\begin{array}{cc}
\mathcal{R}_{11} & \mathcal{R}_{12} \\
0 & \mathcal{R}_{22} \end{array} \right)
\left( \begin{array}{c}
r_1(l) \\
r_2(l)
\end{array} \right),
\end{equation}
\begin{equation}
\begin{aligned}
\mathcal{R}_{11} &= -2 + (n_1 + 2) \tilde{u}_{1_R}, & \mathcal{R}_{12} &= n_2  \tilde{g}_{12_R}/w_R, \\
\mathcal{R}_{22} &= -2 + (n_2 + 2) \tilde{u}_{2_R}.
\end{aligned}
\end{equation}
\end{subequations}
Here, we see that the flow equations have block-triangular form. This means that the eigenvalues are given by the diagonal entries and the eigenvectors by $r_i(l)$, and we have $-\nu_i^{-1} = -2 + \frac{n_i+2}{n_i+8} \epsilon$. 

Solving the flow equations with $r_i(1) = r_{i_R}$, we find
\begin{subequations}
\begin{equation}
    \begin{aligned}
        r_1(l) =&~ \frac{\nu_1 l^{-1/\nu_1} - \nu_2 l^{-1/\nu_2}}{\nu_1-\nu_2} r_{1_R} + \\
        &~\frac{\nu_1 \nu_2 \mathcal{R}_{12}}{\nu_1-\nu_2} (l^{-1/\nu_1} - l^{-1/\nu_2}) r_{2_R},
    \end{aligned}
    \end{equation}
    \begin{equation}
        r_2(l) = l^{-1/\nu_2} r_{2_R}.
    \end{equation}
\end{subequations}
Fixing the LHS for both,
    \begin{equation}
        \label{eq:rnu}
        r_{1_R} = c_1 l^{1/\nu_1} + c_{12} l^{1/\nu_2}, \qquad r_{2_R} = c_2 l^{1/\nu_2},
    \end{equation}
from which we see that in the limit $l \to 0$, the scaling is governed by the exponent $\nu_2$ since $\nu_2 > \nu_1$. Hence, the correlation length for both fields is defined by $\nu_2$, and the exponent $\nu$ and the crossover exponent $\phi$ are given by
\begin{subequations}
\begin{equation}
    \nu^{-1} = 2 - \frac{n_2+2}{n_2+8} \epsilon,
\end{equation}
\begin{equation}
    \phi = \frac{\nu_1}{\nu_2}  \approx 1 + \left( \frac{n_2+2}{n_2+8} - \frac{n_1+2}{n_1+8} \right) \frac{\epsilon}{2}.
\end{equation}
\end{subequations}
Since $n_1 < n_2$, we find the crossover exponent $\phi > 1$, indicating that the phase diagram is similar to Fig.~\ref{fig:phasediag}(a), with the caveat that the transition associated with $\bphi_2$ becomes a straight line, as it occurs independently of $\bphi_1$. However, unlike the case of a fully-coupled multicritical point, there is no regime with a complex-valued $\nu$ or a non-zero value of $\theta^*$, both of which require $\sigma = -1$.

An interesting question is whether there are any physically-relevant fixed points which realize $\nu_1 > \nu_2$. In this case, there would be two correlation length exponents because the scaling of the dependent field would be dominated by $\nu_1$ in Eq.~(\ref{eq:rnu}) and the independent field by definition is described only by $\nu_2$. In such a case, even if strong dynamic scaling were to hold with $z_1 = z_2$, the scaling of the correlation times would differ. While the $n_1 > n_2$ fixed points have this property, their instability and negative $w_R^*$ values means they are unlikely to play a role in physical systems.

\subsection{Marginality at \boldmath $n_1=n_2$ \unboldmath}
\label{subsec:OneWayMarginal}
As discussed above, for $n_1 > n_2$, there are fixed points with $w_R^* < 0$, indicating that a sign change occurs. Indeed, we find that as the relative values of $n_1, n_2$ are varied, the value of $1/w_R^*$ passes through 0 at $n_1 = n_2$, and $1/w$ becomes marginal. 
Although the fixed point value of $\tilde{g}_{12}^*$ here is divergent and thus unphysical,
the flow towards this divergence is much slower due to the marginality of $1/w$, allowing for a treatment of transient critical phenomena which emerge prior to entering the non-perturbative regime. In this subsection, we  summarize the key transient features that arise due to the marginality, the details of which may be found in Appendix \ref{app:marginal}.

The flow to non-perturbative regimes is a consequence of $w_R$ flowing to infinity, which in turn results in the divergence of $\tilde{g}_{12_R}$ because it rapidly flows to a value proportional to $1+w_R$ [cf.~Eq.~(\ref{betauv})]. To better understand the flow in this regime, we recast the flow in terms of $W \equiv 1/w$:
\begin{equation}
    \beta_W \propto W_R^2 + \mathcal{O}(W_R^3),
\end{equation}
which possesses no linear terms in $W_R$. As pointed out earlier, this is because two fixed points merge along the $W_R$ direction at $n_1 = n_2$, and the two roots of $\beta_W$ combine into a double root,
which in turn renders $W_R$ marginal in terms of its stability. Since this occurs here at $n_1 = n_2$, it might be tempting to view this as a consequence of a symmetry of the model. Nevertheless, we anticipate that the locus of marginality shifts once higher-order corrections in $\epsilon$ are included. Crucially, the structure of $\beta_W$ means that $W(l)$ flows to 0 logarithmically in $l$ as opposed to the typical algebraic flow, and significantly larger length scales are necessary for the model to reach the non-perturbative regime. In light of this, the perturbative expressions may  exhibit transient critical phenomena in real systems.

The independent field exhibits equilibrium criticality as usual, so we summarize the critical behavior of the first field in this transient regime as
\begin{subequations}
\begin{equation}
    \eta_1 = \eta_{O(n)} + n \tilde{G}^{* 2}_{12_R} - (\log 2) n^2 \tilde{G}^{*4}_{12_R} |\log (\mu/\mu^*)|,
\end{equation}
\begin{equation}
    \eta_1' = \eta_{O(n)} + n \tilde{G}_{12_R}^{*2},
\end{equation}
\begin{equation}
    \tau_1 \sim \xi_1^{z_{O(n)}}/|\log \xi_1|,
\end{equation}
\begin{equation}
    \nu = \nu_{O(n)},
\end{equation}
\begin{equation}
    \tilde{G}_{12_R}^* = \frac{1-n/4}{n+8} \epsilon.
\end{equation}
\end{subequations}
where $\mu$ is some relevant momentum scale (e.g., $q$ or $1/\xi$). 
Here, $\tilde{G}_{12_R}^*$ is the fixed point value of $\tilde{G}_{12_R} \equiv W_R \tilde{g}_{12_R}$ and describes the coefficient of the divergence of $\tilde{g}_{12_R}$ with $W_R$, evaluated at $W_R \to 0$. Additionally, $\mu^* > \mu$ is a momentum scale which we have introduced to capture non-universal behavior of $\eta_1$. Specifically, because $W$ continues to flow, $\eta_1$ exhibits scale-dependence via $\mu$, so we must put in a momentum scale $\mu^*$ by hand. Here, $\mu^*$ captures effects from the approach to transient universality which would normally be absent at a true fixed point. Scale-dependent critical exponents also appear in other contexts \cite{Middleton2002,DWang2001,Chen2000,Talocia1995}, 
although the underlying origin of the scale-dependent exponent in the one-way coupling model is qualitatively different because it is only a transient phenomenon.

Unlike $\eta_1$, we see that the other exponents are not scale-dependent. Moreover, $\eta_1'$ differs from $\eta_1$ only in the scale-dependent part. This is a consequence of the fact that this scale-dependence comes specifically from the flow of $T_1$, while all other exponents are not affected by the divergence. Additionally, we see that $z_1$ exhibits logarithmic corrections, corresponding to an intermediate regime between strong and weak dynamic scaling. This is qualitatively similar to the Ising model in four dimensions where the fourth-order terms become marginal, leading to logarithmic corrections to the critical exponents \cite{Tauber2014}. Furthermore, $\nu$ remains the same as in equilibrium, similar to the  $n_2 > n_1$ case. However, in this case an exceptional point is present in the flow of $r_i$ due to a combination of $n_1 = n_2$ and the one-way coupling, resulting in a nominal crossover exponent $\phi = 1$ as well as a phase diagram similar to Fig.~\ref{fig:phasediag}(b) and scaling functions of the same form as Eq.~(\ref{scalingexceptional}).

Finally, we briefly remark on the stability of these fixed points. Since the $\tilde{u}_i$ are not coupled to other parameters in the beta functions, they are stable as in the $O(n)$ model. Furthermore, we find that in the limit of $W_R \to 0$, marginality implies that the stability matrix becomes triangular, so stability can be considered independently for $\tilde{G}_{12_R}$ and $W_R$, and we readily determine flow in $\tilde{G}_{12_R}$ as stable. In the case of $W_R$, marginality leads to weak stability to perturbations towards positive $W_R$ and weak instability to perturbations towards negative $W_R$, where by ``weak'' we mean that the flow is logarithmically slow in $l$.

\section{Conclusion and Outlook}
\label{sec:outlook}

In this work, we have considered the effect that $O(n)$ symmetries have on the critical behavior of nonequilibrium multicritical points described by two nonreciprocally coupled Ising-like order parameters using a perturbative RG approach. To lowest non-trivial order, we determined the behavior of the NEFPs as a function of $n_1, n_2$, illustrating both similarities with and differences from the behavior of the related equilibrium fixed points, particularly the biconical fixed points. Some of the key features of this behavior are the connection between the stability of one of the NEFPs to the biconical fixed point, the potential existence of NEFPs in the limit of $n_2 = 1, n_1 \to \infty$, and non-perturbative regions of $n_1, n_2$.

We have also investigated the previously neglected case of a one-way nonreciprocal coupling, which becomes non-perturbative in the model considered in Ref.~\cite{Young2020}, and could possibly provide a realization of critical exceptional points \cite{Zelle2023}. We discovered an additional type of perturbative fixed point which can emerge and identified the corresponding nonequilibrium modifications to the critical exponents for the dependent field. Furthermore, we clarified the nature of the non-perturbative behavior initially identified in Ref.~\cite{Young2020}. We showed that this behavior is connected to a breakdown in strong dynamical scaling, where the corresponding parameter that captures the presence/absence of strong dynamical scaling simultaneously becomes marginal. Due to this marginality, the model becomes non-perturbative slowly, allowing for the identification of critical exponents, including logarithmic corrections to the dynamical critical exponent from the breakdown in strong dynamic scaling and non-universal anomalous dimension which depends logarithmically on the momentum scale.

Numerical and experimental realizations of the criticality associated with these fixed points  remain important directions to investigate. By investigating toy models which give rise to these new forms of criticality numerically, the field theoretical predictions may be put on more solid foundations, and the numerical models will give further insight into formulating experimental realizations. For example, in the case of one-way coupling, the behavior of the independent field is already well-understood, which could allow for a simplified model in terms of only the dependent field. Moreover, due to the non-perturbative behavior at some $n_1, n_2$ for both $\sigma = -1, 0$, approaches beyond perturbative RG must be used, such as functional/exact RG methods \cite{Dupuis2021,Boettcher2015} or large $n$ expansions \cite{Moshe2003} for the interesting case where the number of components goes to infinity for one of the fields. This similarly applies to two-dimensional systems, which are non-perturbative for all values of $n_1, n_2$. Furthermore, it is unclear whether a nonequilibrium dynamical version of self-avoiding walks can allow for a realization of the $n_i \to 0$ limit of our model. In all cases, understanding further nonequilibrium implications of the fixed points warrants additional investigation, such as entropy production and violations of time-reversal symmetry \cite{Landi2021,Paz2022} or ageing \cite{Calabrese2005}.

One interesting avenue for the realization of nonreciprocal multicriticality, whether experimentally or numerically, emerges in an alternative formulation of our model. In particular, by mapping the free energy $\mathcal{F}_2 \to -\mathcal{F}_2$ and simultaneously changing $\zeta_2 \to - \zeta_2, T_2 \to - T_2$, the resulting dynamics are described by a free energy $\mathcal{F} = \mathcal{F}_1 + \mathcal{F}_2 + (u_{12}/2) \int_x |\bphi_1|^2 |\bphi_2|^2$ with a negative temperature $T_2$. Thus, the nonreciprocity is fully encoded in the two effective temperatures. The concept of negative temperatures has been the subject of extensive theoretical debate \cite{Mosk2005,Rapp2010,Carr2013b,Puglisi2017,Baldovin2021} and experimental investigation \cite{Purcell1951,Braun2013,Gauthier2019,MarquesMuniz2023}, and the current work can provide insight into how phase transitions in these systems are modified when different aspects of the system experience different temperatures, such as the two-temperature models studied in Refs.~\cite{Tauber1997,Tauber2002,Santos2002,Akkineni2004}, which investigated scenarios similar to $\sigma = 0$ but not $\sigma = -1$.

Additionally, several open questions remain concerning the possible forms of universality that can occur in the presence of nonreciprocal coupling. 
For example, a complex order parameter with $U(1)$ symmetry is equivalent to a real two-component order parameter with $SO(2)$ symmetry. Although rotations leave the order parameter invariant, reflections (complex conjugation) do not, thereby distinguishing it from the $O(2)$ group we considered. While this distinction is absent for an equilibrium model described by a real free energy, there is a greater range of nonequilibrium dynamics possible, which has been previously investigated in Refs.~\cite{Risler2004, Risler2005, Sieberer2013, Sieberer2014, Tauber2014a,Altman2015}. Specifically, in terms of a complex order parameter, several parameters in the action become complex which would otherwise be real for an $O(2)$ symmetry.
While, for a single order parameter, the RG fixed points for a $U(1)$ symmetry are all the same as for an $O(2)$ order parameter \cite{Tauber2014a}, this situation might be modified in the presence of a multicritical point with nonreciprocal couplings. $SO(n)$ models for $n>2$ also introduce new dynamics not present for $O(n)$, but the corresponding terms in the action are all irrelevant in the sense of RG, so $n=2$ represents a unique case. Finally, the results presented in this work and others \cite{Pichler2015,Lodahl2017,Nassar2020,Fruchart2021,Bowick2022,Shankar2022,Osat2023,Wang2023,Avni2023, Xue2025} indicate the need to study the effect of nonreciprocal couplings in other types of critical systems, such as quantum generalizations and realizations of the models studied here or generalized multicritical points formed at the intersections between other types of phase transitions in one order parameter (e.g.~model B dynamics, which describe a conserved order parameter with diffusive relaxation).

\begin{acknowledgments}
We thank I.~Boettcher, A. Chiocchetta, S.~Diehl, and M.C.~Marchetti for helpful discussions. J.T.Y. was supported by the NIST NRC Postdoctoral Research Associateship Program, NIST, and the NWO Talent Programme (project number VI.Veni.222.312), which is (partly) financed by the Dutch Research Council (NWO). A.V.G. was supported in part by the NSF QLCI (Award No. OMA-2120757), AFOSR MURI, NSF STAQ program, DoE ASCR Quantum Testbed Pathfinder program (Award No. DE-SC0024220), ONR MURI, DARPA SAVaNT ADVENT, ARL (W911NF-24-2-0107), and NQVL:QSTD:Pilot:FTL. A.V.G. also acknowledges support from the U.S. Department of Energy, Office of Science, National Quantum Information Science Research Centers, Quantum Systems Accelerator (Award No. DE-SCL0000121) and from the U.S. Department of Energy, Office of Science, Accelerated Research in Quantum Computing, Fundamental Algorithmic Research toward Quantum Utility (FAR-Qu).. M.M. acknowledges support from the National Science Foundation under the NSF CAREER Award (DMR-2142866) as well as the NSF grant PHY-2112893. This research was supported in part by the National Science Foundation under Grant No.~NSF PHY-1748958.
\end{acknowledgments}

\section*{Data Availability}
The numerical values of the solutions to Eq.~(\ref{eq:beta}) for Figs.~3 and 6 are openly available \cite{DataOn}.

\appendix

\section{\boldmath $Z$ \unboldmath factors}
\label{sec:Z}

In this section, we present the $Z$ factors used for the renormalization discussed in the main text. These are identified using the minimal subtraction procedure using standard techniques. More comprehensive details concerning the evaluation of the integrals for the corresponding diagrams can be found in Ref.~\cite{Young2020}.

First, we consider the renormalization of $r_1, r_2$, corresponding to Fig.~\ref{fig:loops}(a,b). To lowest order, the location of the multicritical point is shifted to
\begin{subequations}
\begin{multline}
r_{1_c} = -(n_1+2) u_1 T_1 \int \frac{d^d \mathbf{p}}{(2 \pi)^d} \frac{1}{D_1 \mathbf{p}^2} \\
- n_2 u_{12} T_2 \int \frac{d^d \mathbf{p}}{(2 \pi)^d} \frac{1}{D_2 \mathbf{p}^2},
\end{multline}
\begin{multline}
r_{2_c} = -(n_2+2) u_2 T_2 \int \frac{d^d \mathbf{p}}{(2 \pi)^d} \frac{1}{D_2 \mathbf{p}^2} \\
- \sigma n_1 u_{12} T_1 \int \frac{d^d \mathbf{p}}{(2 \pi)^d} \frac{1}{D_1 \mathbf{p}^2}.
\end{multline}
\end{subequations}
Defining an additive renormalized mass term $\overline{r}_i = r_i - r_{i_c}$, we determine the $Z$ factors for $\overline{r}_i$. These are
\begin{subequations}
\begin{equation}
Z_{\overline{r}_1} = 1 - (n_1+2) \frac{\tilde{u}_{1_R}}{\epsilon} - n_2 v_R w_R^{-1} \frac{\tilde{u}_{12_R}}{\epsilon} \frac{\overline{r}_{2_R}}{\overline{r}_{1_R}},
\end{equation}
\begin{equation}
Z_{\overline{r}_2} = 1 - (n_2+2) \frac{\tilde{u}_{2_R}}{\epsilon} - \sigma n_1 w_R \frac{\tilde{u}_{12_R}}{\epsilon} \frac{\overline{r}_{1_R}}{\overline{r}_{2_R}},
\end{equation}
\end{subequations}
From this point on, we simply write $\overline{r}_i$ as $r_i$.

Next, we consider the one-loop corrections to $u_1, u_2$, corresponding to Fig.~\ref{fig:loops}(c,d), and $u_{12}$, corresponding to Fig.~\ref{fig:loops}(e-h). The resulting $Z$ factors are
\begin{subequations}
\begin{equation}
Z_{\tilde{u}_1} = 1 - (n_1 + 8) \frac{\tilde{u}_{1_R}}{\epsilon} - \sigma n_2 v_R \frac{\tilde{u}_{12_R}^2}{\tilde{u}_{1_R} \epsilon},
\end{equation}
\begin{equation}
Z_{\tilde{u}_2} = 1 - (n_2 + 8) \frac{\tilde{u}_{2_R}}{\epsilon} - \sigma n_1 v_R \frac{\tilde{u}_{12_R}^2}{\tilde{u}_{2_R} \epsilon},
\end{equation}
\begin{equation}
Z_{\tilde{u}_{12}} = 1 - 4 \frac{v_R+\sigma w_R}{1+w_R} \frac{\tilde{u}_{12_R}}{\epsilon} - (n_1+2) \frac{\tilde{u}_{1_R}}{\epsilon} - (n_2+2) \frac{\tilde{u}_{2_R}}{\epsilon}.
\end{equation}
\end{subequations}

We now consider the two-loop corrections for terms which are not renormalized at one loop. First, we consider the renormalization of $\zeta_i, D_i$, corresponding to Fig.~\ref{fig:loops}(i-k). The resulting $Z$ factors are
\begin{subequations}
\begin{equation}
\begin{aligned}
Z_{\zeta_1} = 1 &+ C'_\zeta (n_1+2) \frac{\tilde{u}_{1_R}^2}{2 \epsilon} + n_2 v_R^2 G_\zeta(w_R) \frac{\tilde{u}_{12_R}^2}{2 \epsilon} \\ 
& + 2 \sigma n_2 v_R H_\zeta(w_R) \frac{\tilde{u}_{12_R}^2}{2 \epsilon},
\end{aligned}
\end{equation}
\begin{equation}
\begin{aligned}
Z_{\zeta_2} = 1 &+ C'_\zeta (n_2+2) \frac{\tilde{u}_{2_R}^2}{2 \epsilon} + n_1 G_\zeta(w_R^{-1}) \frac{\tilde{u}_{12_R}^2}{2 \epsilon} \\ 
& + 2 \sigma n_1 v_R H_\zeta(w_R^{-1}) \frac{\tilde{u}_{12_R}^2}{2 \epsilon},
\end{aligned}
\end{equation}
\begin{equation}
\begin{aligned}
Z_{D_1} = 1 &+ C'_D (n_1+2) \frac{\tilde{u}_{1_R}^2}{2 \epsilon} + n_2 v_R^2 G_D(w_R) \frac{\tilde{u}_{12_R}^2}{2 \epsilon} \\ 
& + 2 \sigma n_2 v_R H_D(w_R) \frac{\tilde{u}_{12_R}^2}{2 \epsilon},
\end{aligned}
\end{equation}
\begin{equation}
\begin{aligned}
Z_{D_2} = 1 &+ C'_D (n_2+2) \frac{\tilde{u}_{2_R}^2}{2 \epsilon} + n_1 G_D(w_R^{-1}) \frac{\tilde{u}_{12_R}^2}{2 \epsilon} \\ 
& + 2 \sigma n_1 v_R H_D(w_R^{-1}) \frac{\tilde{u}_{12_R}^2}{2 \epsilon},
\end{aligned}
\end{equation}
\end{subequations}
where we have defined the constants and functions
\begin{subequations}
\begin{align}
C'_\zeta &= 3 \log(4/3), & C'_D &=1/2, \\
G_\zeta(w) &= \log \left(\frac{(1+w)^2}{w(2+w)}\right), & G_D(w) &= \frac{1}{2 + 3w + w^2}, \\
H_\zeta(w) & = \frac{1}{w} \log\left(\frac{2+2w}{2+w}\right), & H_D(w) &= \frac{3w + w^2}{8+12w + 4w^2},
\end{align}
\end{subequations}
which were introduced in the main text for the flow functions.

Finally, we consider the two-loop corrections to $\zeta_i T_i$, corresponding to Fig.~\ref{fig:loops}(l,m). For convenience, we remove the contributions from $Z_{\zeta_i}$ using the above expressions, leaving only $Z_{T_i}$. The resulting $Z$ factors are
\begin{subequations}
\begin{equation}
Z_{T_1} = 1 - n_2 F(w_R) v_R(v_R-\sigma) \frac{\tilde{u}_{12_R}^2}{2 \epsilon},
\end{equation}
\begin{equation}
Z_{T_1} = 1 + n_1 F(w_R^{-1}) \sigma(v_R-\sigma) \frac{\tilde{u}_{12_R}^2}{2 \epsilon},
\end{equation}
\end{subequations}
where the function
\begin{equation}
F(w) = -\frac{2}{w} \log \left(\frac{2+2w}{2+w}\right),
\end{equation}
was introduced in the main text for $\beta_v$.

\section{Relaxation behavior}
\label{sec:relax}

In this section, we investigate the relaxation behavior of the system in the doubly-ordered phase. In contrast to the equilibrium fixed points, which exhibit overdamped relaxational dynamics, the NEFPs exhibit a qualitatively different behavior in the form of underdamped relaxation. 

We consider the system in the doubly-ordered phase where both fields take nonzero expectation values. Due to the rotational symmetry, this necessitates a separation of our fields into condensed modes and Goldstone modes. We  take the $n_i$th component of each field to have a nonzero expectation value $\langle \bphi_i \rangle = M_i$, while the remaining $n_i-1$ are Goldstone modes. To distinguish these, we  refer to the ordered components as $\varphi_i$ and make the change of variables $\varphi_i \to \varphi_i + M_i$, where $\varphi_i$ now represent fluctuations around the order parameter. In addition to the terms in the original action, this transformation introduces new quadratic and linear terms as (including the original $r_1$ and $r_2$ terms too)
\begin{widetext}
\begin{equation}
\begin{aligned}
\int_{\mathbf{x},t}&(r_1 + 3 u_1 M_1^2 + u_{12} M_2^2) \varphi_1 \tilde{\varphi}_1 + (r_2 + 3 u_2 M_2^2 + \sigma u_{12} M_1^2) \varphi_2 \tilde{\varphi}_2 + 2 u_{12} M_1 M_2 \varphi_2 \tilde{\varphi}_1 + 2 \sigma u_{12} M_1 M_2 \varphi_1 \tilde{\varphi}_2 \\
&+ M_1(r_1 + u_1 M_1^2 + u_{12} M_2^2) \tilde{\varphi}_1 + M_2(r_2 + u_2 M_2^2 + \sigma u_{12} M_1^2) \tilde{\varphi}_2 \\
&+\sum_{\alpha}^{n_1-1} (r_1 + u_1 M_1^2 +u_{12} M_2^2)\bphi_1^\alpha \tilde{\bphi}_1^\alpha + \sum_{\alpha}^{n_2-1} (r_2 + u_2 M_2^2 + \sigma u_{12} M_1^2)\bphi_2^\alpha \tilde{\bphi}_2^\alpha.
\end{aligned}
\end{equation}
\end{widetext}
In addition, several cubic terms are also introduced which are not reported for simplicity. We set the vertices $\tilde{\varphi}_1, \tilde{\varphi}_2$ to zero since, by definition, $\varphi_i$ solely represent the fluctuations. This in turn sets $r_1 = - u_1 M_1^2 -u_{12} M_2^2$ and $r_2 = -u_2 M_2^2 - \sigma u_{12} M_1^2$, which eliminate the contributions from the Goldstone modes as expected, since these are massless modes. We include the effect of fluctuations up to order $\mathcal{O}(u)$ to determine the remaining quadratic vertices
\begin{multline}
2 u_{1_R} M_1^2 \varphi_1 \tilde{\varphi}_1 + 2 u_{2_R} M_2^2 \varphi_2 \tilde{\varphi}_2 \\
+ 2 u_{12_R} M_1 M_2 (\varphi_2 \tilde{\varphi}_1 + \sigma \varphi_1 \tilde{\varphi}_2),
\end{multline}
which we have written in terms of the renormalized coupling terms after accounting for fluctuations in the form of counterterms. Although the other parameters are not renormalized at this order, they would likewise take on their renormalized values at higher orders since the ordered phases have the same $Z$ factors.

The resulting quadratic part of the action takes the form
\begin{equation}
\begin{aligned}
S_0 = \int_{\mathbf{x},t}& \sum_i \tilde{\varphi}_i ( \zeta \partial_t - D_i \nabla^2 + R_i)\varphi_i - \zeta_i T_i \tilde{\varphi}_i^2 \\
&+ R_{12}( \varphi_2 \tilde{\varphi}_1 + \sigma \varphi_1 \tilde{\varphi}_2),
\end{aligned}
\end{equation}
where $R_i = 2 u_{i_R} M_i^2$, $R_{12} = 2 u_{12_R} M_1 M_2$, and we have taken advantage of the fact that, by rescaling the fields by constant factors, we can set $\zeta_1 = \zeta_2 = \zeta$ without modifying the remaining terms. 
We identify the poles of the propagators via
\begin{equation}
0 = \sigma R_{12}^2 - (D_1 \mathbf{k}^2 + R_1 + i \zeta \omega)(D_2 \mathbf{k}^2 + R_2 + i \zeta \omega),
\end{equation}
whose roots are
\begin{equation}
-i \zeta \omega = \frac{D_1 + D_2}{2} \mathbf{k}^2 + \frac{R_1 + R_2}{2} \pm \sqrt{\sigma R_{12}^2 + \left(\frac{R_1-R_2}{2}\right)^2}.
\end{equation}
From this, we determine that underdamped relaxation will occur when
\begin{equation}
- \sigma R_{12}^2 > \left(\frac{R_1-R_2}{2}\right)^2.
\end{equation}
At the equilibrium fixed points, where $\sigma = 1$, this condition is impossible, so the relaxation can only be overdamped with no oscillations. This likewise holds for the one-way coupled fixed points since $\sigma = 0$.
In contrast, for the NEFPs, where $\sigma = -1$, this condition can be rewritten as
\begin{equation}\label{condition}
4 u_{12_R}^2 M_1^2 M_2^2 > (u_{1_R} M_1^2-u_{2_R} M_2^2)^2.
\end{equation}
This is trivially satisfied when $u_{1_R} M_1^2 = u_{2_R} M_2^2$, which can always be realized for some parameters in the doubly-ordered phase. Defining $|M_1 M_2| \equiv M^2$ and considering the limit $\mathbf{k} \to 0$, the pole with lowest nonzero decay rate takes the form
\begin{equation}
-i\zeta \omega = 2 M^2 (\sqrt{u_{1_R} u_{2_R}} \pm i u_{12_R} ).
\end{equation}
In fact, this scenario corresponds to when the pole achieves its largest real value relative to its real part. We can thus identify the maximal angle $\theta^*$ formed by the pole relative to the imaginary axis as
\begin{equation}
\theta^* = \tan^{-1} \left(\frac{\sqrt{v_R^*} \tilde{u}_{12_R}^*}{\sqrt{\tilde{u}_{1_R}^* \tilde{u}_{2_R}^*}}\right),
\end{equation}
where we have replaced $u_R$ values with the fixed point values of $\tilde{u}_R$, leading to the additional contribution from $v_R^*$. The presence of this factor can be better understood by recasting this equation in terms of
\begin{equation}
\tilde{g}_{12_R} \equiv v_R \tilde{u}_{12_R}, \quad \tilde{g}_{21_R} \equiv \sigma \tilde{u}_{12_R}, \quad \tilde{g}_{i_R} \equiv \tilde{u}_{i_R},
\end{equation}
which corresponds to shifting all features of the nonreciprocity into $g_{12},g_{21}$ and setting $T_{1_R} = T_{2_R}$. The new maximal angle is then
\begin{equation}
\theta^* = \tan^{-1} \left(\frac{\sqrt{- \tilde{g}_{12_R}^* \tilde{g}_{21_R}^*}}{\sqrt{\tilde{g}_{1_R}^* \tilde{g}_{2_R}^*}}\right).
\end{equation}
Hence we see that the two nonreciprocal terms enter this equation in the same fashion and that we must have $\tilde{g}_{12_R}^* \tilde{g}_{21_R}^* < 0$ for underdamped relaxation, which is only possible for $\sigma = -1$.

\section{Marginal fixed points}
\label{app:marginal}

In this section, we present detailed analysis and discussion of the marginal, non-perturbative fixed points of the one-way coupled model at $n_1 = n_2$. We also discuss the stability of the fixed points in the region of $n_1 \approx n_2$.

\subsection{Beta functions and fixed points}

To understand the behavior near $n_1=n_2 = n$, we first note that the $\tilde{u}_i$ couplings flow to their fixed point values quickly compared to the other two parameters, and as such we may ignore their flow. Additionally, given the divergence of $w_R, g_{12_R}$, we  consider the beta functions of
\begin{equation}
    W \equiv 1/w, \qquad \tilde{G}_{12} \equiv W \tilde{g}_{12},
\end{equation}
which allows us to remove the divergent behavior from the beta functions (although not necessarily the flow equations). Noting that $\beta_W = -\beta_w/w_R^2$, the relevant beta functions are
\begin{subequations}
\begin{equation}
    \beta_{\tilde{G}_{12}} = \tilde{G}_{12_R} \left[\left(-1 + 2\frac{n+2}{n+8} \right) \epsilon + 4 \frac{\tilde{G}_{12_R}}{1+ W_R} \right],
\end{equation}
\begin{equation}
\begin{split}
    \beta_W & = n \tilde{G}_{12_R}^2 G(W_R^{-1})/W_R \\
    &= n \tilde{G}_{12_R}^2 W_R^2 + \mathcal{O}(W_R^3),
\end{split}
\end{equation}
\end{subequations}
where we have considered the small $W$ limit of $\beta_W$ in the last line and neglect higher order terms in $W_R$. In terms of these new parameters, we see that $\tilde{G}_{12}$ also quickly flows to its fixed point value
\begin{equation}
    \tilde{G}_{12_R}^*(W_R) = \frac{1+W_R}{4} \left(1 - 2 \frac{n +2}{n+8} \right) \epsilon.
\end{equation}
In the limit of $W_R \to 0$, this quantity approaches a constant $\tilde{G}_{12_R}^* \equiv \tilde{G}_{12_R}^*(0)$. 

Let us now consider the implications of the marginality of $W_R$. Like the logarithmic corrections to criticality in $d=4$ where $\epsilon = 0$, there are logarithmic corrections to the scaling here as well, although not necessarily in the same fashion. The beta function at small $W_R$ implies
\begin{equation}
    \mu \partial_\mu W_R = n {\tilde{G}^{*2}}_{12_R} W_R^2,
\end{equation}
whose solution is given by
\begin{equation}
   W(l) = \frac{W_R}{1 - n {\tilde{G}^{*2}}_{12_R} W_R \log l}, 
\end{equation}
where we have introduced the flowing parameters $W(l)$, $\mu(l) \equiv \mu l$ and fix $W(1) = W_R$. We see that, while the model becomes non-perturbative in the thermodynamic limit, it does so logarithmically in $l$. Thus, for finite systems, the perturbative approach here may nevertheless remain valid in the form of transient criticality.

\subsection{Flow functions and critical exponents}

Next, let us consider how the behavior of $W$modifies the flow functions and thus the critical behavior. In the limit of $W_R \to 0$, these are given by
\begin{subequations}
\begin{equation}
\gamma_{\zeta_1} = -C_\zeta' (n+2) \tilde{u}_{1_R}^2 - n \tilde{G}_{12_R}^2,
\end{equation}
\begin{equation}
\gamma_{D_1} = -C'_D (n+2)\tilde{u}_{1_R}^2 - n  \tilde{G}_{12_R}^2 ,
\end{equation}
\begin{equation}
\gamma_{\tilde{D}_1} = C' (n+2)\tilde{u}_{1_R}^2 - n  \tilde{G}_{12_R}^2 W_R ,
\end{equation}
\begin{equation}
\gamma_{T_1} = - \frac{2 \log 2}{W_R} n \tilde{G}_{12_R}^2,
\end{equation}
\end{subequations}
where we have included the flow for $\tilde{D}_1 = D_1/\zeta_1$ since the leading order terms (in $W_R$) from $D_1, \zeta_1$ cancel, so we include the next order. The flow equations for the second field are unchanged. We can see that, although the flow of $\zeta_1$ and $D_1$ are not affected by $W_R$, the flow of $\tilde{D}_1$ has small corrections due to $W_R$, and the flow of $T_1$ diverges as $W_R \to 0$.

First, let us focus on the flow of $\tilde{D}_1$ since there is no divergence involved. In this case, the corrections to the scaling are analogous to the logarithmic corrections for $d=4$, where the couplings become marginal. Utilizing the method of characteristics, we have
\begin{equation}
 l \partial_l \tilde{D}_1(l) \approx -C' (n+2)\tilde{u}_{1_R}^2 - 1/|\log l|,    
\end{equation} 
which we may readily solve for $l \ll 1$:
\begin{equation}
    \tilde{D}_1(l) \sim \tilde{D}_{1_R} l^{- C'(n+2) \tilde{u}_{1_R}^{*2}} |\log l|.
\end{equation}
Matching $|\mathbf{q}| = q = \mu l$ and noting $\omega_1 \sim \tilde{D}_1(l) q^2$, we have $\omega_1 \sim q^{2 - C'(n+2)\tilde{u}_{1_R}^{*2}} |\log q|$, and hence
\begin{equation}
    \omega_1 \sim q^{z_{O(n)}} |\log q|, \qquad \tau_1 \sim \xi_1^{z_O(n)}/|\log \xi_1|,
\end{equation}
where $\tau_1$ is the correlation time of the first field. Since we have $z_2 = z_{O(n)}$, this indicates the presence of weak dynamic scaling, although the difference is only logarithmic. Additionally, this means that the correlation time for the first field is smaller than that of the second field in the thermodynamic limit due to this logarithmic correction.

Next, let us consider the flow of $T_1$, which  affects the behavior of $\eta_1$. In this case, we have
\begin{equation}
    l \partial_l T_1(l) \approx -2 (\log 2) n^2 \tilde{G}_{12_R}^{*4} |\log l|,
\end{equation}
whose solution is 
\begin{equation}
    T_1(l) \sim l^{- (\log 2) n^2 \tilde{G}_{12_R}^{*4} |\log l|},
\end{equation}
which implies that, in addition to the effective temperature increasing, this increase accelerates at small wavelengths. In terms of the critical exponent $\eta_1$, this means that the exponent is no longer constant but depends on the length scale. 

Let us now turn our attention to the behavior of $r_i$. Here, we see that $\mathcal{R}_{11} = \mathcal{R}_{22}$, while $\mathcal{R}_{12} = n \tilde{G}_{12}^*$, $\mathcal{R}_{21} = 0$ [cf.~Eq.~\ref{eq:Rflow}], and there are no divergences. Interestingly, this corresponds to an exceptional point in the flow equations for $r_i$, and the phase diagram is similar to Fig.~\ref{fig:phasediag}(b). As before, the curve associated with a phase transition in $\bphi_2$ becomes a straight line since it is independent of $\bphi_1$. 

Finally, we consider the stability of this fixed point as well as the qualitative consequences that higher-order terms may have on the critical behavior discussed above. Given the independence of $\beta_{\tilde{u}_i}$ on the other two parameters, we may ignore their contribution to the stability, allowing us to only consider $\tilde{G}_{12}, W$, resulting in the following matrix in the limit of $W_R \to 0$:
\begin{equation}
    \Lambda = \left( \begin{array}{cc}
    4 \tilde{G}_{12}^* & -4 \tilde{G}_{12}^{*2}  \\
    \tilde{G}_{12}^* W_R^2 & \tilde{G}_{12}^{*2} W_R
    \end{array} \right),
\end{equation}
where the top and bottom rows correspond to $\beta_{\tilde{G}_{12}}$ and $\beta_W$, respectively. In the limit of small $W_R$, we see that the bottom left term becomes small faster than the bottom right term, and hence $\Lambda$ becomes increasingly triangular. As such, the stability is purely determined by the diagonal terms, so again we have stability in $\tilde{G}_{12}$ when $\tilde{G}_{12}^* > 0$ and (marginal) stability in $W_R$ since we only need consider perturbations to positive $W_R$. 

When higher-order terms are included, there are two primary possibilities. First, $W_R$ no longer becomes marginal anywhere, and there is always a strong dynamic scaling fixed point. In this case, none of the above analysis applies given the drastic change to the behavior of the fixed points as a function of $n_1, n_2$. Second, $\beta_W \propto W_R^2$ still can occur, but the line where this occurs is shifted away from $n_1 = n_2$. While there may be additional terms couplings $W_R$ to the $\tilde{u}_R$ parameters we  nevertheless have the same triangular form of the stability matrix since the off-diagonal terms in the $\beta_W$ row are higher-order in $W_R$, and the stability is determined purely by $\partial_{W_R} \beta_{W}$ and thus still marginal. 
While the shift in the locus of marginality means that $W$ at $n_1 = n_2$ is no longer marginal, if the shift is not too large, it may nevertheless appear nearly marginal for finite system sizes, leading to similar behavior in the criticality. 

We can also utilize this approach to determine stability in the vicinity of $n_1 = n_2$ by defining $\delta_n = n_1 - n_2, \overline{n} \equiv (n_1+n_2)/2$, resulting in the stability matrix of the coupled fixed point
\begin{equation}
    \Lambda = \left( \begin{array}{cc}
    \frac{4 - \overline{n}}{8 + \overline{n}} & - \frac{(\overline{n}-4)^2}{4 (8+\overline{n})^2} \\
     C'^2 \frac{128 }{\overline{n} (4 - \overline{n}) (\overline{n}+8)^3} \delta n^2
    & -C' \frac{4 - \overline{n}}{(8 + \overline{n})^3} \delta n \\
    \end{array}
    \right),
\end{equation}
which exhibits a similar triangular form in the limit of $\delta n \to 0$. Thus for $\overline{n} < 4$, we see that stability is achieved only for $\delta n < 0$, i.e., $n_1 < n_2$. We remark that the divergence in the stability matrix near $\overline{n} = 4$ is a consequence of the fact that there is a discontinuity in the fixed point value of $W_R^*$ when moving to higher $n_2$ from the stable fixed point region, where it jumps from $0$ to just below $-1/2$ since the flow is not well-defined for $-1/2 < W_R < 0$ due to the function $G(1/W_R)$.

\section{Fixed points and critical exponents}

\label{sec:tables}

In this section, we present the numerical fixed-point values of $\tilde{u}_R^*, v_R^*, w_R^*$ in Table \ref{tab:fixed} and their corresponding critical behaviors in Table \ref{tab:exp} for the stable NEFPs with $n_1,n_2 \leq 16$. Analytic values for $n_1 = n_2 = n$ can be found in the main text in Eqs.~(\ref{eq:equalfixed},\ref{eq:equalexp}) for the fixed point values and critical exponents, respectively. Additionally, we report the fixed point values of $g_{12_R}^*, w_R^*$ and the non-trivial critical exponent shifts $\eta_1-\eta_{O(n_1)}, \eta_1'-\eta_{O(n_1)}$ for the stable one-way coupling fixed points in Table \ref{tab:oneway}.

\onecolumngrid

\begin{table*}[h!]
\caption{Fixed point values $\tilde{u}^*_{1_R}, \tilde{u}^*_{2_R}, \tilde{u}^*_{12_R}, v_R^*, w_R^*$ for values of $n_1, n_2$ where the $u_{12_R}^* < 0$ NEFP is stable. \label{tab:fixed}}
\centering
\def\arraystretch{1}
\setlength\tabcolsep{2.4mm}
\begin{tabular}{ccccccccc}
\toprule
\toprule
$n_1$ & $n_2$ & $\tilde{u}_{1_R}^*$ & $\tilde{u}_{2_R}^*$ & $\tilde{u}_{12_R}^*$ 
& $v_R^*$ & $w_R^*$ \\
\midrule
1 & 1 & $0.17\epsilon$ & $0.17\epsilon$ & $-0.29\epsilon$ & $1.0$ & $1.0$ \\
\midrule
1 & 2 & $0.13 \epsilon$ & $0.11 \epsilon$ & $-0.13 \epsilon$ & $0.66$ & $1.4$ \\
\midrule
1 & 3 & $0.12 \epsilon$ & $0.095 \epsilon$ & $-0.086\epsilon$ & $0.57$ & $1.9$ \\
\midrule
1 & 4 & $0.12 \epsilon$ & $0.085 \epsilon$ & $-0.058 \epsilon$ & $0.58$ & $2.7$ \\
\midrule
1 & 5 & $0.12 \epsilon$ & $0.078 \epsilon$ & $-0.040 \epsilon$ & $0.65$ & $3.9$ \\
\midrule
1 & 6 & $0.11\epsilon$ & $0.072 \epsilon$ & $-0.027 \epsilon$ & $0.78$ & $5.9$\\
\midrule
1 & 7 & $0.11 \epsilon$ & $0.067 \epsilon$ & $-0.018 \epsilon$ & $1.0$ & $9.1$ \\
\midrule
1 & 8 & $0.11 \epsilon$ & $0.063 \epsilon$ & $-0.011 \epsilon$ & $1.4$ & $15$\\
\midrule
1 & 9 & $0.11\epsilon$ & $0.059 \epsilon$ & $-0.0048 \epsilon$ & $2.7$ & $33$ \\
\midrule
2 & 1 & $0.21 \epsilon$ & $0.29 \epsilon$ & $-0.39 \epsilon$ & $1.6$ & $0.75$\\
\midrule
2 & 2 & $0.12\epsilon$ & $0.12\epsilon$ & $-0.13\epsilon$ & $1.0$ & $1.0$ \\
\midrule
2 & 3 & $0.11 \epsilon$ & $0.099 \epsilon$ & $-0.075 \epsilon$ & $0.79$ & $1.2$ \\
\midrule
2 & 4 & $0.11\epsilon$ & $0.086\epsilon$ & $-0.046 \epsilon$ & $0.71$ & $1.5$ \\
\midrule
2 & 5 & $0.10 \epsilon$ & $0.078 \epsilon$ & $-0.026 \epsilon$ & $0.75$ & $2.1$ \\
\midrule
2 & 6 & $0.10 \epsilon$ & $0.072 \epsilon$ & $-0.010 \epsilon$ & $1.1$ & $4.2$ \\
\midrule
3 & 1 & $0.26\epsilon$ & $0.46 \epsilon$ & $-0.48 \epsilon$ & $2.1$ & $0.64$ \\
\midrule
3 & 2 & $0.11 \epsilon$ & $0.13 \epsilon$ & $-0.11 \epsilon$ & $1.3$ & $0.86$ \\
\midrule
3 & 3 & $0.10 \epsilon$ & $0.10 \epsilon$ & $-0.058 \epsilon$ & $1.0$ & $1.0$ \\
\midrule
3 & 4 & $0.094\epsilon$ & $0.086\epsilon$ & $-0.030\epsilon$ & $0.85$ & $1.2$ \\
\midrule
3 & 5 & $0.091\epsilon$ & $0.077\epsilon$ & $-0.0046\epsilon$ & $1.1$ & $2.1$ \\
\midrule
4 & 1 & $0.30\epsilon$ & $0.64\epsilon$ & $-0.55\epsilon$ & $2.6$ & $0.59$ \\
\bottomrule
\bottomrule
\end{tabular}
\quad
\begin{tabular}{ccccccccc}
\toprule
\toprule
$n_1$ & $n_2$ & $\tilde{u}_{1_R}^*$ & $\tilde{u}_{2_R}^*$ & $\tilde{u}_{12_R}^*$ 
& $v_R^*$ & $w_R^*$ \\
\midrule
4 & 2 & $0.10\epsilon$ & $0.14\epsilon$ & $-0.087\epsilon$ & $1.6$ & $0.78$ \\
\midrule
4 & 3 & $0.089\epsilon$ & $0.099\epsilon$ & $-0.041\epsilon$ & $1.2$ & $0.90$ \\
\midrule
5 & 1 & $0.33\epsilon$ & $0.84\epsilon$ & $-0.61\epsilon$ & $3.0$ & $0.55$ \\
\midrule
5 & 2 & $0.093\epsilon$ & $0.14\epsilon$ & $-0.071\epsilon$ & $1.9$ & $0.74$ \\
\midrule
5 & 3 & $0.079\epsilon$ & $0.095\epsilon$ & $-0.024\epsilon$ & $1.5$ & $0.88$ \\
\midrule
6 & 1 & $0.36\epsilon$ & $1.1\epsilon$ & $-0.65\epsilon$ & $3.5$ & $0.53$ \\
\midrule
6 & 2 & $0.084\epsilon$ & $0.13\epsilon$ & $-0.057\epsilon$ & $2.3$ & $0.73$ \\
\midrule
7 & 1 & $0.39\epsilon$ & $1.3\epsilon$ & $-0.69\epsilon$ & $4.0$ & $0.51$ \\
\midrule
7 & 2 & $0.076\epsilon$ & $0.13\epsilon$ & $-0.045\epsilon$ & $2.7$ & $0.72$ \\
\midrule
8 & 1 & $0.41\epsilon$ & $1.5\epsilon$ & $-0.71\epsilon$ & $4.4$ & $0.50$ \\
\midrule
8 & 2 & $0.069\epsilon$ & $0.12\epsilon$ & $-0.035\epsilon$ & $3.1$ & $0.74$ \\
\midrule
9 & 1 & $0.42\epsilon$ & $1.7\epsilon$ & $-0.73\epsilon$ & $4.9$ & $0.49$ \\
\midrule
9 & 2 & $0.063\epsilon$ & $0.12\epsilon$ & $-0.023\epsilon$ & $3.8$ & $0.81$ \\
\midrule
10 & 1 & $0.43\epsilon$ & $1.8\epsilon$ & $-0.73\epsilon$ & $5.4$ & $0.48$ \\
\midrule
11 & 1 & $0.43\epsilon$ & $2.0\epsilon$ & $-0.73\epsilon$ & $5.9$ & $0.48$ \\
\midrule
12 & 1 & $0.44\epsilon$ & $2.2\epsilon$ & $-0.73\epsilon$ & $6.3$ & $0.47$ \\
\midrule
13 & 1 & $0.43\epsilon$ & $2.3\epsilon$ & $-0.72\epsilon$ & $6.8$ & $0.47$ \\
\midrule
14 & 1 & $0.43\epsilon$ & $2.4\epsilon$ & $-0.70\epsilon$ & $7.3$ & $0.46$ \\
\midrule
15 & 1 & $0.42\epsilon$ & $2.5\epsilon$ & $-0.69\epsilon$ & $7.8$ & $0.46$ \\
\midrule
16 & 1 & $0.41\epsilon$ & $2.6\epsilon$ & $-0.67\epsilon$ & $8.3$ & $0.46$ \\
\midrule
$\to \infty$ & 1 & $ 7.4 \epsilon/n_1$ & $2.4 \epsilon$ & $-9.9 \epsilon/n_1$ & $.49 n_1$ & $0.43$ \\
\bottomrule
\bottomrule
\end{tabular}
\end{table*}

\begin{table*}[h!]
\caption{Critical exponents for values of $n_1, n_2$ where the $u_{12_R}^* < 0$ NEFP is stable. \label{tab:exp}}
\def\arraystretch{.5}
\setlength\tabcolsep{2.4mm}
\begin{tabular}{ccccccccccc}
\toprule
\toprule
$n_1$ & $n_2$ & \multicolumn{2}{c}{$\nu^{-1}-2$} & $\eta_1$ & $\eta_2$ & $\eta_1'$ 
& $\eta_2'$ & $\gamma_T$ & $z-2$ & $\theta^*/(\pi/2)$ \\
\midrule
1 & 1 & \multicolumn{2}{c}{$(-0.50 \pm 0.29i)\epsilon$} & $-0.068 \epsilon^2$ & $-0.068 \epsilon^2$ & $0.028 \epsilon^2$ & $0.028 \epsilon^2$ & $-0.096 \epsilon^2$ & $0.020 \epsilon^2$ & $0.67$ \\
\midrule
1 & 2 & \multicolumn{2}{c}{$(-0.42 \pm 0.15 i)\epsilon$} & $-0.00062 \epsilon^2$ & $0.0055 \epsilon^2$ & $0.019 \epsilon^2$ & $0.025 \epsilon^2$ & $-0.019 \epsilon^2$ & $0.017 \epsilon^2$ & $0.47$ \\
\midrule
1 & 3 & \multicolumn{2}{c}{$(-0.42 \pm 0.099i)\epsilon$} & $0.0096\epsilon^2$ & $0.015 \epsilon^2$ & $0.018 \epsilon^2$ & $0.023 \epsilon^2$ & $-0.0084 \epsilon^2$ & $0.017 \epsilon^2$ & $0.35$ \\
\midrule
1 & 4 & \multicolumn{2}{c}{$(-0.43 \pm 0.042i)\epsilon$} & $0.014\epsilon^2$ & $0.018 \epsilon^2$ & $0.018 \epsilon^2$ & $0.022 \epsilon^2$ & $-0.0042 \epsilon^2$ & $0.016 \epsilon^2$ & $0.26$ \\
\midrule
1 & 5 & $-0.51\epsilon$ & $-0.38 \epsilon$ & $0.016\epsilon^2$ & $0.019 \epsilon^2$ & $0.018 \epsilon^2$ & $0.022 \epsilon^2$ & $-0.0022 \epsilon^2$ & $0.016 \epsilon^2$ & $0.21$\\
\midrule
1 & 6 & $-0.56 \epsilon$ & $-0.36 \epsilon$ & $0.017\epsilon^2$ & $0.020 \epsilon^2$ & $0.018 \epsilon^2$ & $0.21 \epsilon^2$ & $-0.0012 \epsilon^2$ & $0.015 \epsilon^2$ & $0.16$ \\
\midrule
1 & 7 & $-0.59 \epsilon$ & $-0.35 \epsilon$ & $0.018 \epsilon^2$ & $0.020 \epsilon^2$ & $0.018 \epsilon^2$ & $0.020 \epsilon^2$ & $-0.00059 \epsilon^2$ & $0.015 \epsilon^2$ & $0.13$ \\
\midrule
1 & 8 & $ -0.62 \epsilon$ & $-0.34 \epsilon$ & $0.018 \epsilon^2$ & $0.019\epsilon^2$ & $0.018 \epsilon^2$ & $0.020 \epsilon^2$ & $-0.00026\epsilon^2$ & $0.014 \epsilon^2$ & $0.096$ \\
\midrule
1 & 9 & $-0.65 \epsilon$ & $-0.34 \epsilon$ & $0.018 \epsilon^2$ & $0.019 \epsilon^2$ & $0.018 \epsilon^2$ & $0.019 \epsilon^2$ & $-0.000084  \epsilon^2$ & $0.014 \epsilon^2$ & $0.062$ \\
\midrule
2 & 1 & \multicolumn{2}{c}{$(-0.87 \pm 0.70 i) \epsilon$} & $-0.30 \epsilon^2$ & $-0.41 \epsilon^2$ & $0.099\epsilon^2$ & $ -0.010 \epsilon^2$ & $-0.40 \epsilon^2$ & $0.052 \epsilon^2$ & $0.70$  \\
\midrule
2 & 2 & \multicolumn{2}{c}{$(-0.50 \pm 0.25i)\epsilon$} & $-0.0099 \epsilon^2$ & $-0.0099 \epsilon^2$ & $0.026 \epsilon^2$ & $0.026 \epsilon^2$ & $-0.036 \epsilon^2$ & $0.019 \epsilon^2$ & $0.50$ \\
\midrule
2 & 3 & \multicolumn{2}{c}{$(-0.47 \pm 0.16 i)\epsilon$} & $0.0093\epsilon^2$ & $0.012 \epsilon^2$ & $0.022 \epsilon^2$ & $0.024 \epsilon^2$ & $-0.012 \epsilon^2$ & $0.017\epsilon^2$ & $0.36$ \\
\midrule
2 & 4 & \multicolumn{2}{c}{$(-0.47 \pm 0.10i)\epsilon$} & $0.016 \epsilon^2$ & $0.018 \epsilon^2$ & $0.021 \epsilon^2$ & $0.022 \epsilon^2$ & $-0.0049 \epsilon^2$ & $0.016\epsilon^2$ & $0.25$ \\
\midrule
2 & 5 & \multicolumn{2}{c}{$(-0.48 \pm 0.020 i)\epsilon$} & $0.018 \epsilon^2$ & $0.020 \epsilon^2$ & $0.020 \epsilon^2$ & $0.021 \epsilon^2$ & $-0.0017 \epsilon^2$ & $0.015 \epsilon^2$ & $0.16$ \\
\midrule
2 & 6 & $-0.56\epsilon$ & $-0.41 \epsilon$ & $0.020 \epsilon^2$ & $0.020 \epsilon^2$ & $0.020 \epsilon^2$ & $0.021 \epsilon^2$ & $-0.00038 \epsilon^2$ & $0.015 \epsilon^2$ & $0.081$ \\
\midrule
3 & 1 & \multicolumn{2}{c}{$(-1.3 \pm 1.2 i)\epsilon$} & $-0.72 \epsilon^2$ & $-1.1 \epsilon^2$ & $0.26 \epsilon^2$ & $-0.16 \epsilon^2$ & $-0.98 \epsilon^2$ & $0.16 \epsilon^2$ & $0.71$ \\
\midrule
3 & 2 & \multicolumn{2}{c}{$(-0.55 \pm 0.30i)\epsilon$} & $-0.011 \epsilon^2$ & $-0.017 \epsilon^2$ & $0.031 \epsilon^2$ & $0.025 \epsilon^2$ & $-0.042 \epsilon^2$ & $0.021 \epsilon^2$ & $0.5$ \\
\midrule
3 & 3 & \multicolumn{2}{c}{$(-0.50 \pm 0.17i)\epsilon$} & $0.012 \epsilon^2$ & $0.012 \epsilon^2$ & $0.023 \epsilon^2$ & $0.023 \epsilon^2$ & $-0.12 \epsilon^2$ & $0.017 \epsilon^2$ & $0.33$ \\
\midrule
3 & 4 & \multicolumn{2}{c}{$(-0.49 \pm 0.093i)\epsilon$} & $0.018 \epsilon^2$ & $0.019 \epsilon^2$ & $0.021 \epsilon^2$ & $0.022 \epsilon^2$ & $-0.003 \epsilon^2$ & $0.016 \epsilon^2$ & $0.19$ \\
\midrule
3 & 5 & $-0.53\epsilon$ & $-0.46\epsilon$ & $0.021 \epsilon^2$ & $0.021 \epsilon^2$ & $0.021 \epsilon^2$ & $0.021 \epsilon^2$ & $-0.0001 \epsilon^2$ & $0.015 \epsilon^2$ & $0.038$ \\
\midrule
4 & 1 & \multicolumn{2}{c}{$(-1.9 \pm 1.8i)\epsilon$} & $-1.4 \epsilon^2$ & $-2.4 \epsilon^2$ & $0.55 \epsilon^2$ & $-0.49 \epsilon^2$ & $-1.9 \epsilon^2$ & $0.37 \epsilon^2$ & $0.71$ \\
\midrule
4 & 2 & \multicolumn{2}{c}{$(-0.58 \pm 0.31i)\epsilon$} & $-0.0085 \epsilon^2$ & $-0.018 \epsilon^2$ & $0.033 \epsilon^2$ & $0.023 \epsilon^2$ & $-0.041 \epsilon^2$ & $0.022 \epsilon^2$ & $0.48$ \\
\midrule
4 & 3 & \multicolumn{2}{c}{$(-0.51 \pm 0.16i)\epsilon$} & $0.015 \epsilon^2$ & $0.014 \epsilon^2$ & $0.023 \epsilon^2$ & $0.022 \epsilon^2$ & $-0.0083 \epsilon^2$ & $0.017 \epsilon^2$ & $0.29$ \\
\midrule
5 & 1 & \multicolumn{2}{c}{$(-2.4 \pm 2.4i)\epsilon$} & $-2.2 \epsilon^2$ & $-4.2 \epsilon^2$ & $0.96 \epsilon^2$ & $-1.0 \epsilon^2$ & $-3.2 \epsilon^2$ & $0.72 \epsilon^2$ & $0.70$ \\
\midrule
5 & 2 & \multicolumn{2}{c}{$(-0.60 \pm 0.31i)\epsilon$} & $-0.0047 \epsilon^2$ & $-0.015 \epsilon^2$ & $0.033 \epsilon^2$ & $0.022 \epsilon^2$ & $-0.037 \epsilon^2$ & $0.022 \epsilon^2$ & $0.46$ \\
\midrule
5 & 3 & \multicolumn{2}{c}{$(-0.52 \pm 0.11i)\epsilon$} & $0.018 \epsilon^2$ & $0.018 \epsilon^2$ & $0.022 \epsilon^2$ & $0.022 \epsilon^2$ & $-0.004 \epsilon^2$ & $0.016 \epsilon^2$ & $0.21$ \\
\midrule
6 & 1 & \multicolumn{2}{c}{$(-3.0 \pm 3.0i)\epsilon$} & $-3.3 \epsilon^2$ & $-6.6 \epsilon^2$ & $1.5 \epsilon^2$ & $-1.8 \epsilon^2$ & $-4.8 \epsilon^2$ & $1.2 \epsilon^2$ & $0.70$ \\
\midrule
6 & 2 & \multicolumn{2}{c}{$(-0.60 \pm 0.29i)\epsilon$} & $-0.00068 \epsilon^2$ & $-0.011 \epsilon^2$ & $0.031 \epsilon^2$ & $0.021 \epsilon^2$ & $-0.032 \epsilon^2$ & $0.022 \epsilon^2$ & $0.44$ \\
\midrule
7 & 1 & \multicolumn{2}{c}{$(-3.6 \pm 3.6i)\epsilon$} & $-4.5 \epsilon^2$ & $-9.5 \epsilon^2$ & $2.2 \epsilon^2$ & $-2.8 \epsilon^2$ & $-6.7 \epsilon^2$ & $1.9 \epsilon^2$ & $0.70$ \\
\midrule
7 & 2 & \multicolumn{2}{c}{$(-0.60 \pm 0.26i)\epsilon$} & $0.0032 \epsilon^2$ & $-0.0051 \epsilon^2$ & $0.029 \epsilon^2$ & $0.021 \epsilon^2$ & $-0.026 \epsilon^2$ & $0.021 \epsilon^2$ & $0.41$ \\
\midrule
8 & 1 & \multicolumn{2}{c}{$(-4.2 \pm 4.2i)\epsilon$} & $-5.9 \epsilon^2$ & $-13 \epsilon^2$ & $3.0 \epsilon^2$ & $-3.9 \epsilon^2$ & $-8.9 \epsilon^2$ & $2.7 \epsilon^2$ & $0.70$ \\
\midrule
8 & 2 & \multicolumn{2}{c}{$(-0.59 \pm 0.22i)\epsilon$} & $0.0069 \epsilon^2$ & $0.0012 \epsilon^2$ & $0.027 \epsilon^2$ & $0.021 \epsilon^2$ & $-0.02 \epsilon^2$ & $0.019 \epsilon^2$ & $0.37$ \\
\midrule
9 & 1 & \multicolumn{2}{c}{$(-4.8 \pm 4.8i)\epsilon$} & $-7.4 \epsilon^2$ & $-16 \epsilon^2$ & $3.8 \epsilon^2$ & $-5.2 \epsilon^2$ & $-11 \epsilon^2$ & $3.6 \epsilon^2$ & $0.69$ \\
\midrule
9 & 2 & \multicolumn{2}{c}{$(-0.58 \pm 0.15i)\epsilon$} & $0.011 \epsilon^2$ & $0.009 \epsilon^2$ & $0.023 \epsilon^2$ & $0.021 \epsilon^2$ & $-0.012 \epsilon^2$ & $0.017 \epsilon^2$ & $0.31$ \\
\midrule
10 & 1 & \multicolumn{2}{c}{$(-5.3 \pm 5.4i)\epsilon$} & $-8.9 \epsilon^2$ & $-20 \epsilon^2$ & $4.7 \epsilon^2$ & $-6.6 \epsilon^2$ & $-14 \epsilon^2$ & $4.6 \epsilon^2$ & $0.69$ \\
\midrule
11 & 1 & \multicolumn{2}{c}{$(-5.8 \pm 5.9i)\epsilon$} & $-10 \epsilon^2$ & $-24 \epsilon^2$ & $5.6 \epsilon^2$ & $-8.0 \epsilon^2$ & $-16 \epsilon^2$ & $5.6 \epsilon^2$ & $0.69$ \\
\midrule
12 & 1 & \multicolumn{2}{c}{$(-6.3 \pm 6.3i)\epsilon$} & $-12 \epsilon^2$ & $-28 \epsilon^2$ & $6.4 \epsilon^2$ & $-9.4 \epsilon^2$ & $-18 \epsilon^2$ & $6.6 \epsilon^2$ & $0.69$ \\
\midrule
13 & 1 & \multicolumn{2}{c}{$(-6.7 \pm 6.7i)\epsilon$} & $-13 \epsilon^2$ & $-31 \epsilon^2$ & $7.2 \epsilon^2$ & $-11. \epsilon^2$ & $-20 \epsilon^2$ & $7.5 \epsilon^2$ & $0.69$ \\
\midrule
14 & 1 & \multicolumn{2}{c}{$(-7.1 \pm 7.1i)\epsilon$} & $-14 \epsilon^2$ & $-34 \epsilon^2$ & $8.0 \epsilon^2$ & $-12. \epsilon^2$ & $-22 \epsilon^2$ & $8.5 \epsilon^2$ & $0.69$ \\
\midrule
15 & 1 & \multicolumn{2}{c}{$(-7.4 \pm 7.4i)\epsilon$} & $-15 \epsilon^2$ & $-37 \epsilon^2$ & $8.7 \epsilon^2$ & $-13. \epsilon^2$ & $-24 \epsilon^2$ & $9.3 \epsilon^2$ & $0.69$ \\
\midrule
16 & 1 & \multicolumn{2}{c}{$(-7.7 \pm 7.7i)\epsilon$} & $-16 \epsilon^2$ & $-40 \epsilon^2$ & $9.3 \epsilon^2$ & $-14 \epsilon^2$ & $-26 \epsilon^2$ & $10 \epsilon^2$ & $0.68$ \\
\midrule
$\to \infty$ & 1 & \multicolumn{2}{c}{$(-7.3 \pm 6.9i)\epsilon$} & $-11 \epsilon^2$ & $-30 \epsilon^2$ & $6.7 \epsilon^2$ & $-12 \epsilon^2$ & $-18 \epsilon^2$ & $9.0 \epsilon^2$ & $0.65$ \\
\bottomrule
\bottomrule
\end{tabular}
\end{table*}

\begin{table*}[h!]
\caption{Fixed-point values $g_{12_R}^*, w_R^*$ and non-trivial critical exponent shifts for the stable one-way coupled fixed points. \label{tab:oneway}}
\def\arraystretch{1}
\setlength\tabcolsep{2.4mm}
\begin{tabular}{ccccccccccc}
\toprule
\toprule
$n_1$ & $n_2$ & $w_R^*$ & $g_{12_R}^*$ & $\eta_1 - \eta_{O(n_1)}$ & $\eta_1'-\eta_{O(n_1)}$ & $\gamma_{T_1}$\\
\midrule
1 & 2 & $0.52\epsilon$ & 6.9 & $-0.084\epsilon^2$ & $0.0079\epsilon^2$ & $-0.092\epsilon^2$ \\
\midrule
1 & 3 & $0.27\epsilon$ & 4.1 & $-0.048\epsilon^2$ & $0.007\epsilon^2$ & $-0.055\epsilon^2$ \\
\midrule
1 & 4 & $0.16\epsilon$ & 2.8 & $-0.028\epsilon^2$ & $0.0055\epsilon^2$ & $-0.033\epsilon^2$ \\
\midrule
1 & 5 & $0.096\epsilon$ & 2. & $-0.015\epsilon^2$ & $0.0039\epsilon^2$ & $-0.019\epsilon^2$ \\
\midrule
1 & 6 & $0.056\epsilon$ & 1.3 & $-0.007\epsilon^2$ & $0.0024\epsilon^2$ & $-0.0094\epsilon^2$ \\
\midrule
1 & 7 & $0.03\epsilon$ & 0.8 & $-0.0027\epsilon^2$ & $0.0012\epsilon^2$ & $-0.004\epsilon^2$ \\
\midrule
1 & 8 & $0.014\epsilon$ & 0.37 & $-0.00078\epsilon^2$ & $0.0005\epsilon^2$ & $-0.0013\epsilon^2$ \\
\midrule
1 & 9 & $0.0053\epsilon$ & 0.084 & $-0.00013\epsilon^2$ & $0.00011\epsilon^2$ & $-0.00024\epsilon^2$ \\
\midrule
2 & 3 & $0.29\epsilon$ & 6.9 & $-0.037\epsilon^2$ & $0.0035\epsilon^2$ & $-0.041\epsilon^2$ \\
\midrule
2 & 4 & $0.096\epsilon$ & 2.8 & $-0.01\epsilon^2$ & $0.002\epsilon^2$ & $-0.012\epsilon^2$ \\
\midrule
2 & 5 & $0.034\epsilon$ & 1.2 & $-0.0022\epsilon^2$ & $0.00081\epsilon^2$ & $-0.003\epsilon^2$ \\
\midrule
2 & 6 & $0.0092\epsilon$ & 0.29 & $-0.00025\epsilon^2$ & $0.00017\epsilon^2$ & $-0.00042\epsilon^2$ \\
\midrule
3 & 4 & $0.044\epsilon$ & 2.8 & $-0.0021\epsilon^2$ & $0.00041\epsilon^2$ & $-0.0025\epsilon^2$ \\
\midrule
3 & 5 & $0.0018\epsilon$ & 0.038 & $-8.2 \times 10^{-6}\epsilon^2$ & $7.8 \times 10^{-6}\epsilon^2$ & $-0.000016\epsilon^2$ \\
\bottomrule
\bottomrule
\end{tabular}
\end{table*}
\FloatBarrier
\twocolumngrid

\bibliography{NonEqOn,article-driven-dissipative,DataOn}

\end{document}